\documentclass[aps,prd,reprint,superscriptaddress]{revtex4-1}
\usepackage{amsmath}
\usepackage{amssymb}
\usepackage{graphicx}
\usepackage[caption=false]{subfig}
\usepackage{enumitem}
\usepackage{color}
\usepackage[percent]{overpic}
\usepackage{bm}
\usepackage{rotating}
\usepackage{hyperref}
\usepackage{cancel}
\usepackage{comment}
\usepackage{braket}
\usepackage{wasysym}
\usepackage{mathtools}

 \newcommand\rwaeq{\stackrel{\mathclap{\normalfont\mbox{\tiny RWA}}}{=}}
  \newcommand\fulleq{\stackrel{\mathclap{\normalfont\mbox{\scriptsize Full}}}{=}}
\newcommand{\ketbra}[2]{\left| #1 \vphantom{#2}\right>\!\!\left< #2\vphantom{#1}\right|}
\newcommand{\abs}[1]{\left| #1 \right|} 

\newcommand{\ii}{\mathrm{i}}

\newcommand{\dd}{\mathrm{d}}
 \newcommand{\expect}[1]{\left< #1\right>} 

\newcommand{\K}{K}

\newcommand{\px}{+}
\newcommand{\mx}{-}
\newcommand{\Tr}[0]{\text{Tr}}

\begin{document}

\title{Faster-than-light signalling in the rotating-wave approximation}

\author{Nicholas Funai}
\affiliation{Institute for Quantum Computing, University of Waterloo, Waterloo, Ontario, N2L 3G1, Canada}
\affiliation{Department of Applied Mathematics, University of Waterloo, Waterloo, Ontario, N2L 3G1, Canada}
\author{Eduardo Mart\'{i}n-Mart\'{i}nez}
\affiliation{Institute for Quantum Computing, University of Waterloo, Waterloo, Ontario, N2L 3G1, Canada}
\affiliation{Department of Applied Mathematics, University of Waterloo, Waterloo, Ontario, N2L 3G1, Canada}
\affiliation{Perimeter Institute for Theoretical Physics, 31 Caroline St N, Waterloo, Ontario, N2L 2Y5, Canada}


\begin{abstract}
We present new results on the causality violations introduced by the rotating wave approximation commonly used in quantum optics and high-energy physics. We find that the causality violations and faster-than-light signalling induced by the approximation have `fat tails', i.e., they are polynomially decaying with the distance from the light-cone of the emitter. Furthermore, we also show that the fundamental problems with the incompatibility between the approximation and relativity are not cured even in the long interaction time regime (where the approximation is often taken). This renders the approximation unsuitable for any regime where we are concerned about relativistic causality and information transmission via the electromagnetic field. 

\end{abstract}

\maketitle

\section{Introduction}

Early on in the study of quantized light and its interaction with matter it became evident that there was a need for approximations that would allow one to reduce the inherent complexity of quantum electrodynamical approaches to the light-matter interaction. It was therefore quickly determined that certain approximations could be made to facilitate predictions and intuition in most experimentally achievable regimes. The first of these intuitive steps would be to replace the minimal coupling model by the dipole model \cite{1931AnP...401..273G} and a larger step would involve the study of scalar field interactions rather than electromagnetic field interactions, together with the reduction of the matter systems (typically atoms) to two-level quantum systems. We know that such approximations can capture most of the relevant features of the light-matter interaction where angular momentum exchange is not involved \cite{PhysRevD.94.064074,Pablo}. This motivated the study of the simpler  Unruh-DeWitt type couplings in scalar field theories as a simplified, yet effective, form of the light-matter interaction. Concretely, this interaction is described by the following interaction picture Hamiltonian:
\begin{align}
\hat{H}_{\text{I}}^{\vphantom{\textsc{rwa}}}&=\left(e^{\ii\Omega t}\hat{\sigma}^{+}+e^{-\ii\Omega t}\hat{\sigma}^{-}\right)\hat{\phi}(\bm{y},t),\label{eq1}
\end{align}
where $\hat{\sigma}^{\pm}$ are the raising and lowering operators of the detector and $\Omega$ is the frequency of the energy gap between the ground and excited state of the detector.

Within certain disciplines, e.g. quantum optics, this model can still be unnecessarily complex for the purpose of modelling observable experiments; therefore, it is common practice to implement further approximations to the already simplified Unruh-DeWitt model. The two most popular ones are the single-mode (or few-mode) approximation (SMA) and  the rotating wave approximation (RWA). They are used extensively in quantum optics \cite{Scully}, but they are also a fundamental part of some scattering theory techniques in high-energy physics such as  Fermi's golden rule \cite{Dirac1927}. A few-mode approximation has been already shown to be incompatible with a relativistic description of the light-matter interaction and leads, unsurprisingly, to faster-then-light predictions that are not present in the fully covariant Unruh-DeWitt model \cite{PhysRevA.89.022330,PhysRevD.92.104019}.

We will turn our attention in this manuscript to the RWA instead. This approximation, (implicitly) employed early on by Fermi himself (eq. 51 of \cite{RevModPhys.4.87}), can be illustrated by first expanding the field in \eqref{eq1} into plane wave modes for a 3+1D flat spacetime:
\begin{align}
\begin{split}
&\left(e^{\ii\Omega t}\hat{\sigma}^{+}+e^{-\ii\Omega t}\hat{\sigma}^{-}\right)\hat{\phi}(\bm{y},t)=\int\frac{\dd^{3}\bm{k}}{(2\pi)^{3/2}\sqrt{2\omega}}\\
&\bigg(e^{-\ii(\omega-\Omega)t+\ii\bm{k}\cdot\bm{y}}\hat{a}^{\vphantom{\dagger}}_{\bm{k}}\hat{\sigma}^{+}+e^{\ii(\omega-\Omega)t-\ii\bm{k}\cdot\bm{y}}\hat{a}_{\bm{k}}^{\dagger}\hat{\sigma}^{-}\\
&+e^{-\ii(\omega+\Omega)t+\ii\bm{k}\cdot\bm{y}}\hat{a}^{\vphantom{\dagger}}_{\bm{k}}\hat{\sigma}^{-}+e^{\ii(\omega+\Omega)t-\ii\bm{k}\cdot\bm{y}}\hat{a}_{\bm{k}}^{\dagger}\hat{\sigma}^{+}\bigg),
\end{split}\label{eq2}
\end{align}
where $\hat{a}_{\bm{k}}^{\vphantom{\dagger}}$ and $\hat{a}_{\bm{k}}^{\dagger}$ are the usual annihilation and creation operators satisfying that all commutators between them are zero except for  $\big[\hat{a}_{\bm{k}}^{\vphantom{\dagger}},\hat{a}_{\bm{k}'}^{\dagger}\big]=\delta^{3}(\bm{k}-\bm{k}')$ . The usual (RWA) argument goes as follows: given that $\omega,\Omega>0$, the terms of the form $e^{\pm\ii(\omega-\Omega)t}$ (rotating terms) oscillate much slower than terms of the form $e^{\pm\ii(\omega+\Omega)t}$ (counter rotating terms). Given that to compute time evolution, the Hamiltonian must ultimately be integrated over time, the RWA involves neglecting the counter rotating terms, reasoning that, once integrated over a long enough time, their highly oscillatory behaviour will be dominated by the relative `stationary' terms of the rotating terms. This leaves
\begin{align}
\begin{split}
&\left(e^{\ii\Omega t}\hat{\sigma}^{+}+e^{-\ii\Omega t}\hat{\sigma}^{-}\right)\hat{\phi}(\bm{y},t) \stackrel{\textsc{rwa}}{\approx}\int\frac{\dd^{3}\bm{k}}{(2\pi)^{3/2}\sqrt{2\omega}}\\
&\bigg(e^{-\ii(\omega-\Omega)t+\ii\bm{k}\cdot\bm{y}}\hat{a}^{\vphantom{\dagger}}_{\bm{k}}\hat{\sigma}^{+}+e^{\ii(\omega-\Omega)t-\ii\bm{k}\cdot\bm{y}}\hat{a}_{\bm{k}}^{\dagger}\hat{\sigma}^{-}\bigg).
\end{split}\label{eq3}
\end{align}
This model is easy to understand even from a classical point of view since it has built-in some (misplaced) intuition coming from the early times of spectroscopy and Kirchhoff's spectroscopy laws \cite{kirchhoff2009abhandlungen}, i.e. the idea that the absorption of 1 photon leads to a single atomic excitation and vice versa and that an atom in its ground state cannot get excited via photon emission. In particular when coupled with the single-mode approximated Hamiltonian this model seems to respect classical notions of energy conservation. However, it would be a misconception to think that the counter-rotating terms do not conserve energy: the Hamiltonians in the Schr\"odinger picture are time independent and therefore they conserve the expectation of energy at all times. One should realize, however, that single-photon states and the excited and ground state of the atoms are not eigenstates of the interaction Hamiltonian and therefore they do not have a well-defined value of energy and hence the total Hamiltonian eigenstates will not include the eigenstates of the free theories (such as the ground state of the atom and the vacuum of the field), so the whole energy conservation argument of the RWA cannot be carried out a-priori to establish the model's Hamiltonian.

In spite of the issues associated with the rotating wave approximation and the single mode approximation, they have been commonly used in condensed matter \cite{0953-4075-46-22-224020} as well as in quantum optics \cite{MEYSTRE1992243} in the form of Jaynes-Cummings models or as rotating-frame approximations. Due to the frequent use of finite size cavities within these communities, the RWA and SMA are in general good approximations when causality and relativistic considerations are not of paramount importance, i.e. extremely long times.

The RWA is indeed very tempting, given the enormous mathematical simplifications it brings with it as well as the coincidence with empirical observations from spectral lines and the seemingly intuitive `conservation of energy' arguments of the model; however, from a relativist's standpoint, it becomes apparent that the RWA modifies a once local interaction into a non-local interaction that, if taken seriously, would enable superluminal signalling. Compagno et al. \cite{Compagno_1989,Compagno_1990} demonstrated this non-locality with electromagnetic (EM) fields, highlighting the importance of the counter-rotating terms interfering with rotating terms in order to maintain causality, which was measured by means of the support of the renormalized stress-energy tensor propagation.

Whilst looking at the interaction terms themselves \eqref{eq3}, Clerk \& Sipe \cite{Clerk1998} demonstrated that the RWA's non-locality stems from the RWA interaction Hamiltonian's non-locality, a result that is independent of the field measurment device used. This non-locality can then be quantified by the standard approach in Relativistic Quantum Information where we consider the actual communication between two particle detectors under the Unruh-DeWitt interaction (see, among others, \cite{PhysRevD.92.104019,PhysRevD.93.104019,PhysRevLett.114.110505,PhysRevLett.114.141103,Jonsson_2018,PhysRevA.89.022330}).

In this paper we will extend and explicitly quantify what was hinted in previous works on the RWA's non-locality and superluminal communication. This will first involve a study of the expectations of energy density and field fluctuations $\expect{\phi^{2}}$, both perturbatively and non-perturbatively, outside of the interactions light-cone extending \cite{Compagno_1989,Compagno_1990} by studying in detail the decay rate of the non-locality. We will then  revisit Clerk \& Sipe's calculation on the Hamiltonian's non-locality, further highlighting its connection to non-localities of observables. Finally we will explore, from the point of view of information theory, the communication of two particle detectors (modelling, e.g., atoms) communicating under RWA in order to explicitly quantify the presence of faster-than-light signalling and its rate of decay with the different scales of the problem. Furthermore, we will analytically show that even for long times, when the RWA is supposed to be ultimately valid, there are polynomially suppressed violations of causality as a receiver increases their distance from the light-cone of the emitter, which means that faster-than-light signalling is therefore always possible under this approximation even for arbitrarily long evolution times. Besides the consideration of communication protocols, another significant novelty in this paper is the extensive numerics included to quantify to what extent the RWA is an acceptable approximation when relativistic considerations are important. 

\section{Rotating wave approximation}
\subsection{Theoretical review}
The object of our study are the acausal non-localities introduced into a reasonable relativistic theory as a consequence of the rotating wave approximation. In addition to determining the exact magnitude of these non-localities we are also interested in if and how the RWA becomes exact in the long time limit.

Our work will be focused on the scalar field in 3+1 D, which in terms of a plane-wave expansion can be written as 
\begin{align}
\hat{\phi}(\bm{x},t)&= \int\frac{\dd^{3}\bm{k}}{(2\pi)^{3/2}\sqrt{2\omega}}\left(e^{-\ii\omega t+\ii\bm{k}\cdot\bm{x}}\hat{a}_{\bm{k}}^{\vphantom{\dagger}}+e^{\ii\omega t-\ii\bm{k}\cdot\bm{x}}\hat{a}_{\bm{k}}^{\dagger}\right)\label{eqphi}
\end{align}
and has $\hat\pi$ as its canonical conjugate momentum, which in terms of plane-wave modes takes the form
\begin{align}
    \hat{\pi}(\bm{x},t)&= -\ii\int\frac{\dd^{3}\bm{k}}{(2\pi)^{3/2}}\sqrt{\frac{\omega}{2}}\left(e^{-\ii\omega t+\ii\bm{k}\cdot\bm{x}}\hat{a}_{\bm{k}}^{\vphantom{\dagger}}-e^{\ii\omega t-\ii\bm{k}\cdot\bm{x}}\hat{a}_{\bm{k}}^{\dagger}\right).\label{eqpi}
\end{align}

To increase the physicality of the model at the same time as avoiding spurious divergences, the interaction used will be a spatially smeared Unruh-DeWitt interaction that, as has been discussed, captures all the fundamental features of the light-matter interaction when exchange of angular momentum between atoms and light is not relevant \cite{PhysRevD.94.064074,Pablo}:
\begin{align}
\hat{H}_{\text{I}}^{\textsc{full}}&=\lambda\chi(t)\left(e^{\ii\Omega t}\hat{\sigma}^{+}+e^{-\ii\Omega t}\hat{\sigma}^{-}\right)\int\dd^{3}\bm{y}\,F(\bm{y})\hat{\phi}(\bm{y},t),
\end{align}
where $\chi(t)$ is a switching function controlling the duration of the interaction and its adiabaticity or suddenness, $\lambda$ is the interaction strength, $\hat{\sigma}^{\pm}$ are the detector raising and lower operators, $\Omega$ is the detector's energy gap and $F(\bm{y})$, which has dimensions of $[L]^{-3}$, is the detector's smearing that can be associated to the wavefunctions of the excited and ground state of the atom being modelled \cite{PhysRevD.94.064074,Pablo}. We will assume that the smearing function is only non-negligible for a length scale $R$ (the size of the atom). For convenience, in this paper we will rewrite this smearing function in terms of a dimensionless smearing as follows
\begin{equation}
    F(\bm{y})=\frac{1}{R^3}G\left(\frac{\bm y}{R}\right),
\end{equation}
where the dimensionless function $G(\bm\zeta)$ is localized around $|\bm \zeta|\lesssim 1$.

As shown in \eqref{eq2} the Unruh-DeWitt interaction has `rotating terms' $e^{\pm\ii(\omega-\Omega)t}$ and `counter-rotating terms' $e^{\pm\ii(\omega+\Omega)t}$. Given that $\omega>0$ and $\Omega>0$ then $e^{\pm\ii(\omega+\Omega)t}$ oscillates at least as quickly as $e^{\pm\ii\Omega t}$, therefore any integral over time (as required for the unitary time evolution operator) will bound these counter-rotating terms by $\tilde{\chi}(\Omega)$, the Fourier transform of $\chi(t)$. On the other hand, for contributions from modes where $\omega\approx \Omega$, the rotating terms hardly oscillate. Therefore the `resonant' rotating terms are roughly $\int\dd t\,\chi(t)$ large, where $\chi(t)\geq 0$. Therefore if $\chi(t)$ is on for long times these rotating terms should easily dominate over the counter-rotating terms. With this progression of logic the RWA would be justified and we could approximately have 
\begin{align}
\begin{split}
\hat{H}_{\text{I}}^{\textsc{rwa}}&=\lambda\chi(t)\int\dd^{3}\bm{y}\,\frac{1}{R^3}G\left(\frac{\bm y}{R}\right)
\int\frac{\dd^{3}\bm{k}}{(2\pi)^{3/2}\sqrt{2\omega}}\\
&\bigg(e^{-\ii(\omega-\Omega)t+\ii\bm{k}\cdot\bm{y}}\hat{a}^{\vphantom{\dagger}}_{\bm{k}}\hat{\sigma}^{+}+e^{\ii(\omega-\Omega)t-\ii\bm{k}\cdot\bm{y}}\hat{a}_{\bm{k}}^{\dagger}\hat{\sigma}^{-}\bigg),
\end{split}\label{eq7}
\end{align}
where the approximation is expected to work in the very long time limit.

\subsection{Hamiltonian non-locality}\label{sec2b}
In order to see the non-locality of the RWA Hamiltonian it is useful to express the creation and annihilation operators in terms of the local operators \eqref{eqphi} and \eqref{eqpi}. Namely, 
\begin{align}
\hat{a}_{\bm{k}}^{\vphantom{\dagger}}&=\int\frac{\dd^{3}\bm{y}}{\sqrt{2}(2\pi)^{3/2}}e^{\ii\omega t-\ii\bm{k}\cdot\bm{y}}\left(\sqrt{\omega}\hat{\phi}(\bm{y},t)+\frac{\ii}{\sqrt{\omega}}\hat{\pi}(\bm{y},t)\right),\\
\hat{a}_{\bm{k}}^{\dagger}&=\int\frac{\dd^{3}\bm{y}}{\sqrt{2}(2\pi)^{3/2}}e^{-\ii\omega t+\ii\bm{k}\cdot\bm{y}}\left(\sqrt{\omega}\hat{\phi}(\bm{y},t)-\frac{\ii}{\sqrt{\omega}}\hat{\pi}(\bm{y},t)\right).
\end{align}
Consequently by substitution into \eqref{eq7} we obtain
\begin{align}
\begin{split}
\hat{H}_{\text{I}}^{\textsc{rwa}}&=\lambda\chi(t)\int\dd^{3}\bm{y}\,\frac{1}{R^3}G\left(\frac{\bm y}{R}\right)\int\frac{\dd^{3}\bm{z}}{2}\hat{\phi}(\bm{z},t)\int\frac{\dd^{3}\bm{k}}{(2\pi)^{3}}\\
&\left(e^{\ii\Omega t}e^{\ii\bm{k}\cdot(\bm{y}-\bm{z})}\hat{\sigma}^{+}+e^{-\ii\Omega t}e^{-\ii\bm{k}\cdot(\bm{y}-\bm{z})}\hat{\sigma}^{-}\right)\\
&+\ii\lambda\chi(t)\int\dd^{3}\bm{y}\,\frac{1}{R^3}G\left(\frac{\bm y}{R}\right)\int\frac{\dd^{3}\bm{z}}{2}\hat{\pi}(\bm{z},t)\int\frac{\dd^{3}\bm{k}}{(2\pi)^{3}\omega}\\
&\left(e^{\ii\Omega t}e^{\ii\bm{k}\cdot(\bm{y}-\bm{z})}\hat{\sigma}^{+}-e^{-\ii\Omega t}e^{-\ii\bm{k}\cdot(\bm{y}-\bm{z})}\hat{\sigma}^{-}\right). 
\end{split}\label{eq10}
\end{align}
This analysis is analogous to the one performed by  Clerk \& Sipe in \cite{Clerk1998}. We can extend those results further and determine the exact polynomial decay of the non-locality.  We can do so considering that (as shown in appendix \ref{seca1a}) we can write
\begin{align}
\int\dd^{3}\bm{k}\,e^{\ii\bm{k}\cdot(\bm{y}-\bm{z})}&=(2\pi)^{3}\delta(\bm{y}-\bm{z}),\label{eq12_0}\\
\int\dd^{3}\bm{k}\frac{e^{\ii\bm{k}\cdot(\bm{y}-\bm{z})}}{\omega}&=\frac{4\pi}{\abs{\bm{y}-\bm{z}}^{2}},\label{eq12}
\end{align}
which can be appropriately substituted into \eqref{eq10} such that we are left with
\begin{align}
\begin{split}
\hat{H}_{\text{I}}^{\textsc{rwa}}&=\frac{\lambda\chi(t)}{2}\int\frac{\dd^{3}\bm{y}}{R^3}G\left(\frac{\bm y}{R}\right)\bigg[\left(e^{\ii\Omega t}\hat{\sigma}^{+}+e^{-\ii\Omega t}\hat{\sigma}^{-}\right)\hat{\phi}(\bm{y},t)\\
&-\frac{2\ii}{(2\pi)^{2}}\left(e^{\ii\Omega t}\hat{\sigma}^{+}-e^{-\ii\Omega t}\hat{\sigma}^{-}\right)\int\dd^{3}\bm{z}\frac{\hat{\pi}(\bm{z},t)}{\abs{\bm{y}-\bm{z}}^{2}}\bigg].
\end{split}\label{eq13}
\end{align}
By means of backward substitution \eqref{eq13} has shown us that implementation of the RWA introduces a polynomial non-locality into the Hamiltonian; and, rather importantly, this non-locality has no indication of improving for long-time limits.

\section{Non-localities in field observables}

Whilst we have seen how the RWA Hamiltonian contains a $1/r^{2}$ non-locality, we would like to know how exactly this non-locality translates into possible faster-than-light behaviour of  field observables. For example, we would like to know if any coincidental cancellations may improve the decay rate of the non-locality in any parameter regimes either for short or long interaction times.

\subsection{Non-perturbative very short time regimes}\label{RWA:sec3}

\subsubsection{Time evolution}\label{RWA:sec3a}

We begin our considerations of field observables in the very short time regime. In particular we will consider that our switching function is proportional to a delta distribution (see, e.g.,\cite{Hotta08}), i.e., $\chi(t)=\eta\delta(t)$, where $\eta$ is a timescale quantifying the intensity of the delta-kick. This switching can be thought of as the limit of Gaussian switching when the interaction time is taken to be very short as compared to all other scales in the problem, but the overall intensity of the coupling over time is kept constant (see e.g., appendix D of \cite{PhysRevD.95.105009}). Picking this particular switching will allows us to use non-perturbative tools. Our expectation is to measure the magnitude of RWA's shortcomings when the interaction time is extremely short, i.e. the opposite of the RWA's validity criterion. We are also interested to see how exactly the interaction Hamiltonian's non-locality translates onto the time evolution operator's non-locality. 

The RWA interaction Hamiltonian, for the $\delta$-switching case takes the form
\begin{align}
\begin{split}
\hat{H}_{\text{I}}^{\textsc{rwa}}&=\tilde{\lambda}\delta(t)\int\dd^{3}\bm{y}\,\frac{1}{R^{3}}G\left(\frac{\bm{y}}{R}\right)\int\frac{\dd^{3}\bm{k}}{(2\pi)^{3/2}\sqrt{2\omega}}\\
&\left(e^{-\ii(\omega-\Omega)t+\ii\bm{k}\cdot\bm{y}}\hat{a}_{\bm{k}}^{\vphantom{\dagger}}\hat{\sigma}^{+}+e^{\ii(\omega-\Omega)t-\ii\bm{k}\cdot\bm{y}}\hat{a}_{\bm{k}}^{\dagger}\hat{\sigma}^{-}\right),
\end{split}
\end{align}
where we recall $R$ is the characteristic length of our smearing function and $\tilde\lambda\coloneqq\lambda\eta$ is the overall interaction strength. In order to compress notation we also define
\begin{align}
\tilde{F}(\bm{k})&\coloneqq \int \dd^{3}\bm{y}\,\frac{1}{R^{3}}G\left(\frac{\bm{y}}{R}\right)e^{\ii\bm{k}\cdot\bm{y}},\label{eqq16}\\
\hat{\alpha}(t)&\coloneqq \tilde{\lambda}\int\frac{\dd^{3}\bm{k}}{(2\pi)^{3/2}\sqrt{2\omega}}\tilde{F}(\bm{k})e^{-\ii\omega t}\hat{a}_{\bm{k}}^{\vphantom{\dagger}}.\label{eq17}
\end{align}
This allows us to write the interaction Hamiltonian in a very compact form
\begin{align}
\hat{H}_{\text{I}}^{\textsc{rwa}}&=\delta(t)\left(\hat{\alpha}(t)\hat{\sigma}^{+}(t)+\hat{\alpha}^{\dagger}(t)\hat{\sigma}^{-}(t)\right),\label{eq18}
\end{align}
where $\hat{\sigma}^{\pm}(t)=e^{\pm\ii\Omega t}\hat{\sigma}^{\pm}$. Observe that $\hat{\alpha}(t)$ consists of the sum of annihilation operators \eqref{eq17}, i.e. $\hat{\alpha}(t)$ acts similarly to an annihilation operator (annihilating the same vacuum as all of the $\hat a^{\vphantom{\dagger}}_{\bm k}$). This allows us to think of \eqref{eq18} as a sort of  Jaynes-Cummings model, albeit where  $\hat{\alpha}(t)$ and its adjoint do not satisfy canonical commutation relationships.

Taking advantage of the $\delta$-switching we can evaluate the  time evolution operator,
\begin{align}
\hat{U}=\mathcal{T}\exp\left(-\ii\!\int \dd t\,\hat{H}_{\text{I}}^{\textsc{rwa}}(t)\right)\!=\exp\left[-\ii\left(\hat{\alpha}\hat{\sigma}^{+}\!\!+\!\hat{\alpha}^{\dagger}\hat{\sigma}^{-}\right)\right],
\end{align}
where $\hat{\alpha}$ and $\hat{\sigma}^{\pm}$ are evaluated at $t=0$. As shown in appendix \ref{seca2} the exponential above can be expanded and simplified when acting on the vacuum  state:
\begin{align}
\begin{split}
\hat{U}\ket{\varphi}\ket{0}&=\left[
\vphantom{\hat{\Pi}_{g}\otimes\hat{\openone}+\hat{\Pi}_{e}\otimes\hat{\openone}\cos  K-\ii\frac{\hat{\sigma}^{-}\otimes\hat{\alpha}^{\dagger}(0)}{K}\sin  K}\right.
\hat{\Pi}_{g}\otimes\hat{\openone}+\hat{\Pi}_{e}\otimes\hat{\openone}\cos  K\\
&
\left.\vphantom{\hat{\Pi}_{g}\otimes\hat{\openone}+\hat{\Pi}_{e}\otimes\hat{\openone}\cos  K-\ii\frac{\hat{\sigma}^{-}\otimes\hat{\alpha}^{\dagger}(0)}{K}\sin  K}
-\ii\frac{\hat{\sigma}^{-}\otimes\hat{\alpha}^{\dagger}(0)}{K}\sin  K
\right]\ket{\varphi}\ket{0},
\end{split}\label{eq21}
\end{align}
where $\ket{\varphi}$ is the initial detector state and
\begin{align}
K^{2}\hat{\openone} \coloneqq \left[\hat{\alpha}(0),\hat{\alpha}^{\dagger}(0)\right]=\tilde{\lambda}^{2}\int\frac{\dd^{3}\bm{k}}{(2\pi)^{3}2\omega}\big|{\tilde{F}(\bm{k})}\big|^{2}\hat\openone,
\end{align}
i.e. $K\geq 0$. $\hat{\Pi}_{g,e}$ are the projection operators onto the detector ground and excited state respectively. 

Note that in the RWA, \eqref{eq21} yields 0 and 1 field excitations, conditional on the initial state of the detector. This is particularly interesting when comparing with the non-approximated full interaction Hamiltonian, which has the form
\begin{align}
\hat{H}_{\text{I}}^{\textsc{full}}&=\tilde{\lambda}\delta(t)\hat{\sigma}_{x}(t)\int\dd^{3}\bm{y}\,\frac{1}{R^{3}}G\left(\frac{\bm{y}}{R}\right)\int\frac{\dd^{3}\bm{k}}{(2\pi)^{3/2}\sqrt{2\omega}}\nonumber\\
&\left(e^{-\ii\omega t+\ii\bm{k}\cdot\bm{y}}\hat{a}_{\bm{k}}^{\vphantom{\dagger}}+e^{\ii\omega t-\ii\bm{k}\cdot\bm{y}}\hat{a}_{\bm{k}}^{\dagger}\right)\\
&=\delta(t)\hat{\sigma}_{x}(t)\left(\hat{\alpha}(t)+\hat{\alpha}^{\dagger}(t)\right),\label{eq28}
\end{align}
where $\hat{\sigma}_{x}(t)=e^{\ii\Omega t}\hat{\sigma}^{+}+e^{-\ii\Omega t}\hat{\sigma}^{-}$. This particular Hamiltonian allows for terms of the form $\hat{\alpha}^{\dagger}\hat{\sigma}^{+}$, i.e. emission of a field excitation via a detector excitation. The corresponding time evolution operator then becomes
\begin{align}
\hat{U}&=\mathcal{T}\exp\left(-\ii\int\hat{H}_{\text{I}}\dd t\right)\nonumber\\
&=\ketbra{\px}{\px}\otimes\exp\left(-\ii\left(\hat{\alpha}+\hat{\alpha}^{\dagger}\right)\right)\nonumber\\
&+\ketbra{\mx}{\mx}\otimes\exp\left(\ii\left(\hat{\alpha}+\hat{\alpha}^{\dagger}\right)\right),\label{eq24p}
\end{align}
where $\ket{\px}$ and $\ket{\mx}$ are the $\pm$ eigenstates of $\hat{\sigma}_{x}$ and $\hat{\alpha},\hat{\alpha}^{\dagger}$ are evaluated at $t=0$. In contrast to the \eqref{eq21}, the full interaction time evolution operator generates phase-space displacements conditioned to the state of the detector, which applied to the vacuum state generates superpositions of coherent states and therefore states with multiple field excitations. This is in stark contrast with the RWA where only single-photon excitations are produced. 

However, note that if $\tilde{\lambda}$ is very small then the coherent state displacements in \eqref{eq24p} approximate zero and one excitation states, but as the coupling increases the approximation becomes exponentially worse. In fact the final states produced by these two Hamiltonians, $\ket{\psi_{\textsc{rwa}}}$ and $\ket{\psi_{\text{Full}}}$ have the following overlap:
\begin{align}
    \braket{\psi_{\textsc{rwa}}|\psi_{\text{Full}}}&=
    \braket{\varphi|\hat{\Pi}_{g}|\varphi}e^{-K^{2}/2}\\
    &+\braket{\varphi|\hat{\Pi}_{e}|\varphi}e^{-K^{2}/2}(\cos K+K\sin K)\nonumber
\end{align}
which, regardless of the detector's initial state, goes to zero exponentially fast on $\tilde\lambda$.

\subsubsection{Faster-than-light effects in physical observables}

In this subsection we will focus on the energy deposited in the field and in the amplitude of the field. In particular we will evaluate the expectation of the stress-energy density and the square of the field amplitude. The expectation values we are interested in finding correspond to the operators
\begin{widetext}
\begin{align}
\begin{split}
:\hat{T}_{\mu\nu}(\bm{x},t):&=\int\frac{\dd^{3}\bm{k}\dd^{3}\bm{k}'}{(2\pi)^{3}\sqrt{4\omega\omega'}}\left(k_{\mu}k'_{\nu}-\frac{\eta_{\mu\nu}}{2}k_{\gamma}k'^{\gamma}\right)\left[e^{-\ii(\omega-\omega')t+\ii(\bm{k}-\bm{k}')\cdot\bm{x}}\hat{a}_{\bm{k}'}^{\dagger}\hat{a}_{\bm{k}}^{\vphantom{\dagger}}
+e^{\ii(\omega-\omega')t-\ii(\bm{k}-\bm{k}')\cdot\bm{x}}\hat{a}_{\bm{k}}^{\dagger}\hat{a}_{\bm{k}'}^{\vphantom{\dagger}}\right.\\
&\left.-e^{-\ii(\omega+\omega')t+\ii(\bm{k}+\bm{k}')\cdot\bm{x}}\hat{a}_{\bm{k}'}^{\vphantom{\dagger}}\hat{a}_{\bm{k}}^{\vphantom{\dagger}}-e^{\ii(\omega-\omega')t-\ii(\bm{k}-\bm{k}')\cdot\bm{x}}\hat{a}_{\bm{k}'}^{\dagger}\hat{a}_{\bm{k}}^{\dagger}
\right],
\end{split}\label{eqtmn}\\
\begin{split}
:\phi^{2}(\bm{x},t):&=\int\frac{\dd^{3}\bm{k}\dd^{3}\bm{k}'}{(2\pi)^{3}\sqrt{4\omega\omega'}}\left[e^{-\ii(\omega-\omega')t+\ii(\bm{k}-\bm{k}')\cdot\bm{x}}\hat{a}_{\bm{k}'}^{\dagger}\hat{a}_{\bm{k}}^{\vphantom{\dagger}}
+e^{\ii(\omega-\omega')t-\ii(\bm{k}-\bm{k}')\cdot\bm{x}}\hat{a}_{\bm{k}}^{\dagger}\hat{a}_{\bm{k}'}^{\vphantom{\dagger}}\right.\\
&\left.+e^{-\ii(\omega+\omega')t+\ii(\bm{k}+\bm{k}')\cdot\bm{x}}\hat{a}_{\bm{k}'}^{\vphantom{\dagger}}\hat{a}_{\bm{k}}^{\vphantom{\dagger}}+e^{\ii(\omega-\omega')t-\ii(\bm{k}-\bm{k}')\cdot\bm{x}}\hat{a}_{\bm{k}'}^{\dagger}\hat{a}_{\bm{k}}^{\dagger}
\right],
\end{split}\label{eqphi2}
\end{align}
where $k_{\gamma}=(\omega,-\bm{k})$ is a 4-vector. When considering a normalised initial detector state $\ket{\varphi}=a_{g}\ket{g}+a_{e}\ket{e}$ tensored with an initial vacuum field state, i.e. $\ket{\varphi}\otimes\ket{0}$, the final (RWA evolved) state is described by \eqref{eq21}. This in turn leads to

\begin{align}
\begin{split}
\expect{:\hat{T}_{\mu\nu}(\bm{x},t):}_{\textsc{rwa}}&=\tilde{\lambda}^{2}\frac{\abs{a_{e}}^{2}\sin^{2}K}{K^{2}}\int\frac{\dd^{3}\bm{k}\dd^{3}\bm{k}'}{(2\pi)^{6}4\omega\omega'}\left(k_{\mu}k'_{\nu}-\frac{\eta_{\mu\nu}}{2}k_{\gamma}k'^{\gamma}\right)\\
&\times\left[e^{-\ii(\omega-\omega')t+\ii(\bm{k}-\bm{k}')\cdot\bm{x}}\tilde{F}(\bm{k}')\tilde{F}^{*}(\bm{k})+e^{\ii(\omega-\omega')t-\ii(\bm{k}-\bm{k}')\cdot\bm{x}}\tilde{F}(\bm{k})\tilde{F}^{*}(\bm{k}')\right],
\end{split}\label{eq25}\\
\begin{split}
\expect{:\phi^{2}(\bm{x},t):}_{\textsc{rwa}}&=\tilde{\lambda}^{2}\abs{a_{e}}^{2}\frac{\sin^{2}K}{K^{2}}\frac{1}{2(2\pi)^{6}}\abs{\int\frac{\dd^{3}\bm{k}}{\omega}e^{\ii\omega t-\ii\bm{k}\cdot\bm{x}}\tilde{F}(\bm{k})}^{2}.
\end{split}\label{eq26}
\end{align}
These expressions are non-zero only if $a_{e}\neq 0$, that is the detector must be excited in order to deposit energy in the field. In contrast the final state following the non-approximated interaction Hamiltonian leads to

\begin{align}
\begin{split}
\expect{:\hat{T}_{\mu\nu}(\bm{x},t):}_{\text{Full}}&=\tilde{\lambda}^{2}\int\frac{\dd^{3}\bm{k}\dd^{3}\bm{k}'}{(2\pi)^{6}4\omega\omega'}
\left(k_{\mu}k'_{\nu}-\frac{\eta_{\mu\nu}}{2}k_{\gamma}k'^{\gamma}\right)\\
&\left(e^{\ii\omega t-\ii\bm{k}\cdot\bm{x}}\tilde{F}(\bm{k})+e^{-\ii\omega t+\ii\bm{k}\cdot\bm{x}}\tilde{F}^{*}(\bm{k})\right)\left(e^{\ii\omega' t-\ii\bm{k}'\cdot\bm{x}}\tilde{F}(\bm{k}')+e^{-\ii\omega' t+\ii\bm{k}'\cdot\bm{x}}\tilde{F}^{*}(\bm{k}')\right),
\end{split}\label{eq30}\\
\begin{split}
\expect{:\hat{\phi}^{2}(\bm{x},t):}_{\text{Full}}
&=-\frac{\tilde{\lambda}^{2}}{4(2\pi)^{6}}\left[\int\frac{\dd^{3}\bm{k}}{\omega}\left(e^{\ii\omega t-\ii\bm{k}\cdot\bm{x}}\tilde{F}(\bm{k})-e^{-\ii\omega t+\ii\bm{k}\cdot\bm{x}}\tilde{F}^{*}(\bm{k})\right)\right]^{2}.
\end{split}\label{eq31}
\end{align}
\end{widetext}

Note that these results are independent of the detector energy gap $\Omega$ (as we would expect from a delta switching).  Let us highlight again a relevant qualitative difference between the RWA case and the full model: in the case of the full Hamiltonian, the results are also independent of the the initial detector state unlike in the RWA. Indeed, inspection of \eqref{eq25} and \eqref{eq26} vs \eqref{eq30} and \eqref{eq31} respectively demonstrates two important points. Firstly, the RWA expectation values are dependent on the detector being excited. The RWA does not permit spontaneous excitation of a detector via a field excitation emission. Secondly, the RWA expectation values take the form $\varphi_{i}^{\vphantom{\dagger}}\varphi_{j}^{\dagger}+\varphi_{i}^{\dagger}\varphi_{j}^{\vphantom{\dagger}}$, whilst the full model expectations take the form $(\varphi_{i}^{\vphantom{\dagger}}+\varphi_{j}^{\dagger})(\varphi_{i}^{\dagger}+\varphi_{j}^{\vphantom{\dagger}})$, which is a direct reflection of the differences in the Hamiltonians \eqref{eq18} and \eqref{eq28}.

\subsubsection{Numerical Results}

Using the results above we consider the situation of a spherically symmetric detector smearing 
\begin{align}
G(\bm{\zeta})=\begin{cases}
1 &\text{if $\abs{\bm{\zeta}}<1$}\\
0&\text{otherwise.}
\end{cases}\label{eq31f}
\end{align}
I.e. a ``hard sphere'' extending to radius $R$. We will use the lengthscale $R$ as a reference scale to adimensionalize all the dimensionful parameters in our setup. We will study numerically  the expectation values of the renormalized energy density, and $:\hat{\phi}^{2}:$  at $t=0^{+}$, i.e., immediately following the $\delta$-coupling interaction. Note that (for the purpose of the RWA to yield a non-trivial result) the detector is assumed to be initially excited.

\begin{figure}[!h]
\includegraphics[width=0.9\columnwidth]{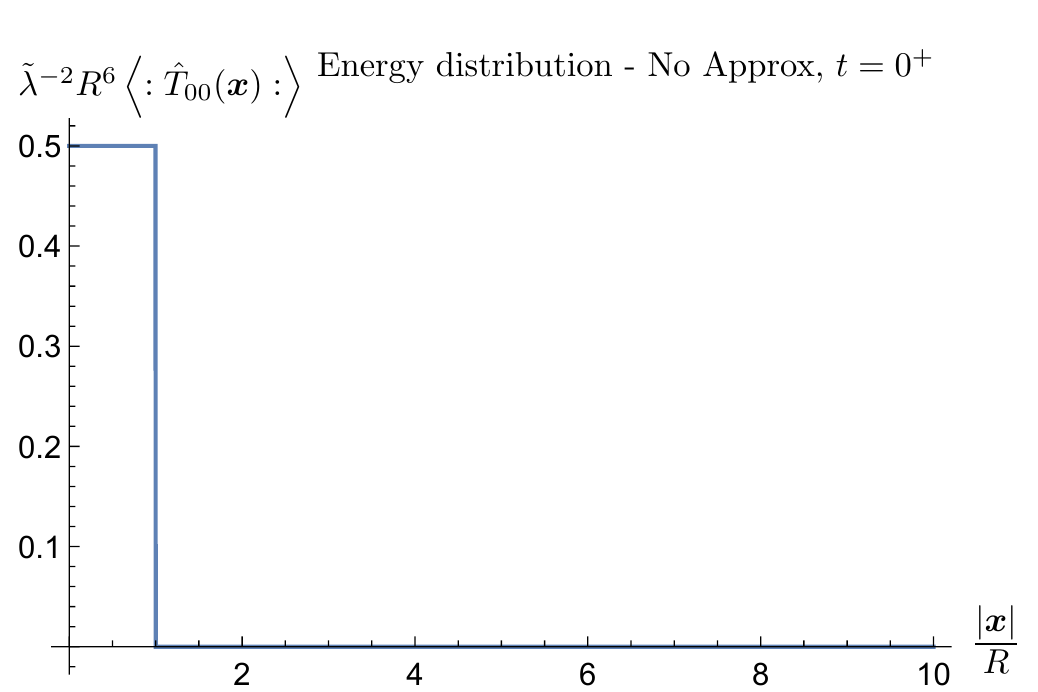}
\caption{Energy density distribution immediately following $\delta$-coupling interaction with no approximations. Here $G(\bm{\zeta})=\Theta(1-\bm{\zeta})$, i.e. a hard sphere. Note that the interaction has no non-local field consequences. }\label{fig2}
\end{figure}

In figure \ref{fig2} the renormalised energy density, for the full interaction model, is plotted as a function of distance from the detector's distribution centre at time $t=0^{+}$, i.e. just after the $\delta$-coupling interaction has taken place. As expected from a local relativistic theory there is no acausal propagation or perturbations of energy beyond the support of the detector distribution (and their support remains always strictly inside the lightcone of the detector). 

\begin{figure}[!h]
\includegraphics[width=0.9\columnwidth]{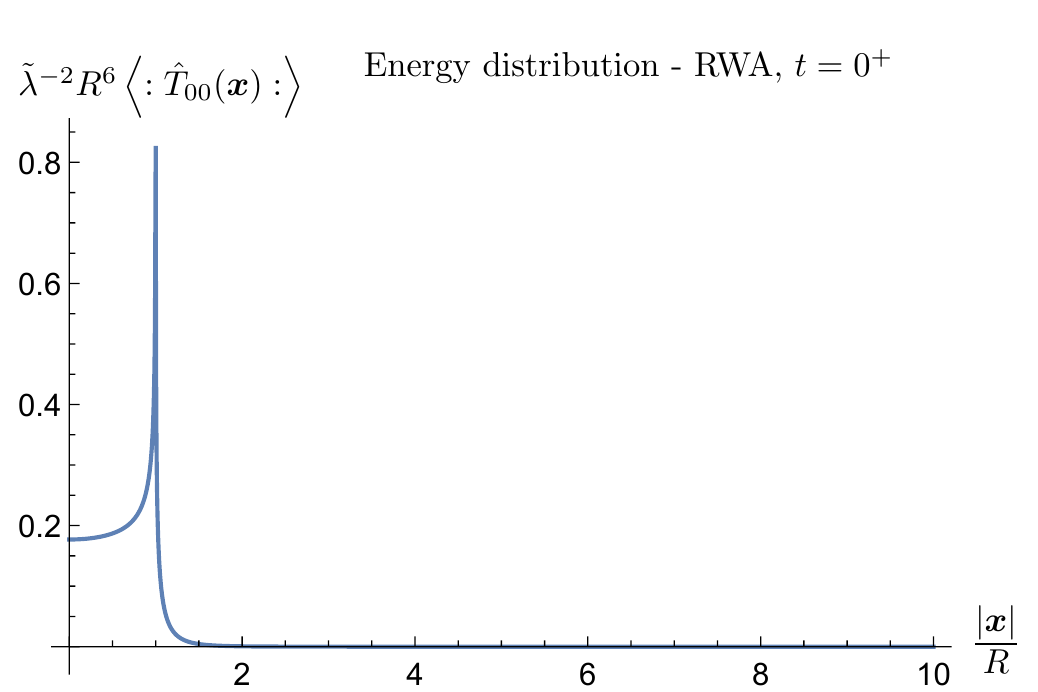}
\caption{Energy density distribution immediately following $\delta$-coupling interaction under the RWA. Here $G(\bm{\zeta})=\Theta(1-\bm{\zeta})$, i.e. spherically symmetric with a sudden cutoff. The spike at $\abs{\bm{x}}=R$ is a consequence of the $F(\bm{x})$ discontinuity. Note the polynomially decaying tail for $\abs{\bm{x}}>R$.  }\label{fig1}
\end{figure}

\begin{figure}[!h]
\includegraphics[width=0.9\columnwidth]{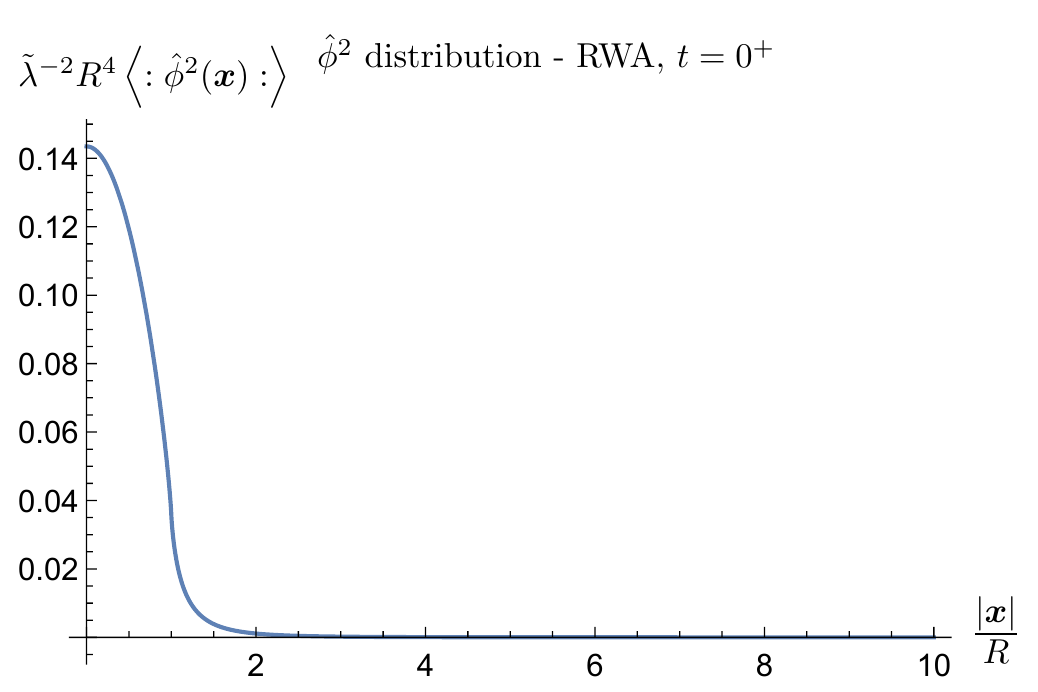}
\caption{$:\hat{\phi}^{2}:$ distribution immediately following $\delta$-coupling interaction under the RWA. Here $G(\bm{\zeta})=\Theta(1-\bm{\zeta})$, i.e. spherically symmetric with a sudden cutoff. Note the polynomially decaying tail for $\abs{\bm{x}}>R$. }\label{fig3}
\end{figure}

In contrast in figures \ref{fig1} and \ref{fig3} the renormalized energy density and normal ordered $\phi^{2}$ distributions are plotted respectively, for the RWA model, at time $t=0^{+}$. Particularly noteworthy are the non-zero values for $\abs{\bm{x}}>R$, demonstrating acausal behaviour in physically measurable quantities. Moreover, these acausal tails decay only polynomially, severely limiting the situations when the RWA can be treated as local in this regime.

\subsection{Perturbative evolution and long time regime}

The previous section demonstrated that for extremely short interaction times the RWA's Hamiltonian non-locality is reflected in the non-locality of physically measurable field quantities. However, one may perhaps expect the RWA to work well when considering long interaction times. We will see that this is not quite the case, and what the effects of the RWA on the causality of the model are in the long time regime.

In this case we consider an extended switching function and hence the RWA Hamiltonian we use is 
\begin{align}
\begin{split}
\hat{H}_{\text{I}}^{\textsc{rwa}}(t)&=\lambda\chi(t)\int\dd^{3}\bm{y}\,\frac{1}{R^{3}}G\left(\frac{\bm{y}}{R}\right)\int\frac{\dd^{3}\bm{k}}{(2\pi)^{3/2}\sqrt{2\omega}}\\
&\left(e^{-\ii(\omega-\Omega)t+\ii\bm{k}\cdot\bm{y}}\hat{a}_{\bm{k}}^{\vphantom{\dagger}}\hat{\sigma}^{+}+e^{\ii(\omega-\Omega)t-\ii\bm{k}\cdot\bm{y}}\hat{a}_{\bm{k}}^{\dagger}\hat{\sigma}^{-}\right)
\end{split}\nonumber\\
&=\chi(t)\left(\hat{\alpha}(t)\hat{\sigma}^{+}(t)+\hat{\alpha}^{\dagger}(t)\hat{\sigma}^{-}(t)\right),
\end{align}
where we take $\chi(t)=\Theta(t+T/2)-\Theta(t-T/2)$, i.e. an interaction of duration $T$ switched on at $t=-T/2$. Here $\hat{\alpha}$ and its conjugate are defined by:
\begin{align}
\tilde{F}(\bm{k})&\coloneqq \int \dd^{3}\bm{y}\,\frac{1}{R^{3}}G\left(\frac{\bm{y}}{R}\right)e^{\ii\bm{k}\cdot\bm{y}},\\
\hat{\alpha}(t)&\coloneqq \lambda\int\frac{\dd^{3}\bm{k}}{(2\pi)^{3/2}\sqrt{2\omega}}\tilde{F}(\bm{k})e^{-\ii\omega t}\hat{a}_{\bm{k}}^{\vphantom{\dagger}}.
\end{align}

Under these conditions the corresponding time evolution operator becomes, up to second order in the Dyson expansion,
\begin{align}
&\hat{U}^{\textsc{rwa}}(t)
=\hat{\openone}-\ii\int\limits_{-\infty}^{\infty}\dd t_{1}\,\chi(t_{1})\left(\hat{\alpha}(t_{1})\hat{\sigma}^{+}(t)+\hat{\alpha}^{\dagger}(t_{1})\hat{\sigma}^{-}(t)\right)\nonumber\\
&-\int\limits_{-\infty}^{\infty}\dd t_{1}\int\limits_{-\infty}^{t_{1}}\dd t_{2}\, \chi(t_{1})\chi(t_{2})
\left(
\vphantom{\hat{\alpha}(t_{1})\hat{\alpha}^{\dagger}(t_{2})\hat{\Pi}_{e}+\hat{\alpha}^{\dagger}(t_{1})\hat{\alpha}(t_{2})\hat{\Pi}_{g}}
\hat{\alpha}(t_{1})\hat{\alpha}^{\dagger}(t_{2})\hat{\Pi}_{e}e^{\ii\Omega(t_{1}-t_{2})}\right.\nonumber\\
+&\left.\vphantom{\hat{\alpha}(t_{1})\hat{\alpha}^{\dagger}(t_{2})\hat{\Pi}_{e}+\hat{\alpha}^{\dagger}(t_{1})\hat{\alpha}(t_{2})\hat{\Pi}_{g}}\hat{\alpha}^{\dagger}(t_{1})\hat{\alpha}(t_{2})\hat{\Pi}_{g}e^{-\ii\Omega(t_{1}-t_{2})}
\right)+\mathcal{O}(\lambda^{3}).
\end{align}
Here $\hat{\Pi}_{g,e}$ are the detector projection operators onto the ground and excited state respectively.

In contrast, the full model has a Hamiltonian
\begin{align}
\hat{H}_{\text{I}}^{\vphantom{\textsc{rwa}}}(t)&=\chi(t)\hat{\sigma}_{x}(t)\left(\hat{\alpha}(t)+\hat{\alpha}^{\dagger}(t)\right),
\end{align}
and the second order Dyson expansion of the time evolution operator yields
\begin{align}
&\hat{U}^{\textsc{full}}=\hat{\openone}-\ii\int\limits_{-\infty}^{\infty}\dd t_{1}\,\chi(t_{1})\hat{\sigma}_{x}(t_{1})\left(\hat{\alpha}(t_{1})+\hat{\alpha}^{\dagger}(t_{1})\right)\nonumber\\
&-\int\limits_{-\infty}^{\infty}\dd t_{1}\int\limits_{-\infty}^{t_{1}}\dd t_{2}\,\chi(t_{1})\chi( t_{2})\hat{\sigma}_{x}(t_{1})\hat{\sigma}_{x}( t_{2})
\left(\hat{\alpha}(t_{1})+\hat{\alpha}^{\dagger}(t_{1})\right)\nonumber\\
&\times\left(\hat{\alpha}( t_{2})+\hat{\alpha}^{\dagger}( t_{2})\right)+\mathcal{O}(\lambda^{3}).
\end{align}

If in the long time regime, it were satisfied that  \mbox{$||\hat{U}^{\textsc{rwa}}-\hat{U}^{\textsc{full}}||\rightarrow 0$}, then the rotating wave approximation would be guaranteed to work. Let us analyze this perturbatively (2nd order) and particularize for the initial state $\ket{\psi}=\ket{\varphi}\ket{0}$:
\begin{align}
\begin{split}
&\left(\hat{U}^{\textsc{rwa}}-\hat{U}^{\textsc{full}}\right)\ket{\varphi}\ket{0}\\
=&\bigg(\ii\lambda\int\frac{\dd^{3}\bm{k}}{(2\pi)^{3/2}\sqrt{2\omega}}\tilde{F}^{*}(\bm{k})\hat{a}_{\bm{k}}^{\dagger}\hat{\sigma}^{+}\int\limits_{-\infty}^{\infty}\dd t_{1}\,\chi(t_{1})e^{\ii(\omega+\Omega)t_{1}}\\
+&\lambda^{2}\int\frac{\dd^{3}\bm{k}\dd^{3}\bm{k}'}{(2\pi)^{3}\sqrt{4\omega\omega'}}\tilde{F}^{*}(\bm{k})\tilde{F}^{*}(\bm{k}')\hat{a}_{\bm{k}}^{\dagger}\hat{a}_{\bm{k}'}^{\dagger}\!\!\int\limits_{-\infty}^{\infty}\!\!\dd t_{1}\!\!\!\int\limits_{-\infty}^{t_{1}}\!\!\dd t_{2}\,\chi(t_{1})\\
\times&\chi(t_{2})\!\left(\hat{\Pi}_{e}e^{\ii(\omega+\Omega)t_{1}+\ii(\omega'-\Omega)t_{2}}+\hat{\Pi}_{g}e^{\ii(\omega-\Omega)t_{1}}e^{\ii(\omega'+\Omega)t_{2}}\right)\\
+&\int\limits_{-\infty}^{\infty}\dd t_{1}\int\limits_{-\infty}^{t_{1}}\dd t_{2}\left[\hat{\alpha}(t_{1}),\hat{\alpha}^{\dagger}(t_{2})\right]\hat{\Pi}_{g}e^{-\ii\Omega(t_{1}-t_{2})}
\bigg)\ket{\varphi}\ket{0}
\end{split}\label{eqq38}
\end{align}
should converge to zero for long times ($\hat{\Pi}_{g,e}$ are the usual ground and excited state detector projection operators). We will see that this is not the case.

For the first term in \eqref{eqq38} the time integral in the $T\rightarrow\infty$ limit will become a Dirac delta of a positive argument $\delta(\omega+\Omega)$, which after integration over $\bm k$ yields zero. Therefore, the difference in predictions \eqref{eqq38} vanishes to order $\mathcal{O}(\lambda)$ as $T\rightarrow \infty$; as dictated by the RWA.

However, the quadratic features a double integral over a semi-infinite domain, which will not yield a delta-like contribution of an always positive argument. Instead its contribution is governed by the expression \eqref{c4} in appendix \ref{sec:ac}. This ensures a persistent non-zero difference between the RWA and the exact prediction, of order $\mathcal{O}(\lambda^2)$, even when $T\rightarrow\infty$. Notice that this implies that as the coupling strength approaches non-perturbative regimes, the RWA validity becomes more and more  questionable.

We show below how this difference between the two models manifests in the field's observables' expectation values, with a focus on causality.

\begin{widetext}
As in section \ref{RWA:sec3}, we consider an initial state given by $(a_{g}\ket{g}+a_{e}\ket{e})\otimes\ket{0}$ then, as shown in appendix \ref{RWA:secb}, the second order expectation values are 

\begingroup
\allowdisplaybreaks

\begin{align}
\expect{:\hat{T}_{\mu\nu}(\bm{x},t):}_{\textsc{rwa}}&=\frac{\lambda^{2}}{4(2\pi)^{6}}\abs{a_{e}}^{2}\left[J_{\mu,e}^{1} (J_{\nu,e}^{1})^{*}+(J_{\mu,e}^{1})^{*}J_{\nu,e}^{1}-\eta_{\mu\nu}(J_{\gamma,e}^{1})^{*}J_{e}^{1\, \gamma}\right]+\mathcal{O}(\lambda^{3}),\label{prwat}\\
\expect{:\hat{\phi}^{2}(\bm{x},t):}_{\textsc{rwa}}&=\frac{\lambda^{2}}{2(2\pi)^{6}}\abs{a_{e}}^{2}\abs{M_{e}^{1}}^{2}+\mathcal{O}(\lambda^{3}),\label{prwaphi}
\end{align}
and 
\begin{align}
\begin{split}
\expect{:\hat{T}_{\mu\nu}(\bm{x},t):}_{\text{Full}}&=\frac{\lambda^{2}}{4(2\pi)^{6}}\sum_{i\in\{e,g\}}\abs{a_{i}}^{2}\left[J_{\mu,i}^{1}(J_{\nu,i}^{1})^{*}+(J_{\mu,i}^{1})^{*}J_{\nu,i}^{1}-\frac{\eta_{\mu\nu}}{2}\left(J_{\gamma,i}^{1}(J_{i}^{1\gamma })^{*}+(J_{\gamma,i}^{1 })^{*}J_{i}^{1 \gamma }\right)\vphantom{J_{\mu,i}^{1\vphantom{*}}J_{\nu,i}^{1*}+J_{\mu,i}^{1*}J_{\nu,i}^{1\vphantom{*}}-\frac{\eta_{\mu\nu}}{2}\left(J_{\gamma,i}^{1\vphantom{*}}J_{i}^{1\gamma *}+J_{\gamma,i}^{1 *}J_{i}^{1 \gamma \vphantom{*}}\right)+J_{\mu\nu,i}^{2\vphantom{*}}+J_{\mu\nu,i}^{2*}-\frac{\eta_{\mu\nu}}{2}\left(J_{\gamma,i}^{2\gamma\vphantom{*}}+J_{\gamma,i}^{2\gamma *}\right)}\right.\\
+&\left.\vphantom{J_{\mu,i}^{1\vphantom{*}}J_{\nu,i}^{1*}+J_{\mu,i}^{1*}J_{\nu,i}^{1\vphantom{*}}-\frac{\eta_{\mu\nu}}{2}\left(J_{\gamma,i}^{1\vphantom{*}}J_{i}^{1\gamma *}+J_{\gamma,i}^{1 *}J_{i}^{1 \gamma \vphantom{*}}\right)+J_{\mu\nu,i}^{2\vphantom{*}}+J_{\mu\nu,i}^{2*}-\frac{\eta_{\mu\nu}}{2}\left(J_{\gamma,i}^{2\gamma\vphantom{*}}+J_{\gamma,i}^{2\gamma *}\right)}
J_{\mu\nu,i}^{2}+(J_{\mu\nu,i}^{2})^{*}-\frac{\eta_{\mu\nu}}{2}\left(J_{\gamma,i}^{2\gamma}+(J_{\gamma,i}^{2\gamma })^{*}\right)\right]+\mathcal{O}(\lambda^{3}),
\end{split}\\
\expect{:\hat{\phi}^{2}(\bm{x},t):}_{\text{Full}}&=\frac{\lambda^{2}}{4(2\pi)^{6}}\sum_{i\in\{e,g\}}\abs{a_{i}}^{2} \left(2\abs{M_{i}^{1}}^{2}-M_{i}^{2}-(M_{i}^{2})^{*}\right)+\mathcal{O}(\lambda^{3}),
\end{align}
where 
\begin{align}
J_{\mu,e}^{1}(\bm{x},t)&\coloneqq \int\frac{\dd^{3}\bm{k}}{\omega}\,k_{\mu} \tilde{F}(\bm{k})e^{\ii\omega t-\ii\bm{k}\cdot\bm{x}}\int\limits_{-\infty}^{\infty}\dd t_{1}\,\chi(t_{1})e^{-\ii(\omega-\Omega)t_{1}},\\
\begin{split}
J_{\mu\nu,e}^{2}(\bm{x},t)&\coloneqq \int\frac{\dd^{3}\bm{k}\dd^{3}\bm{k}'}{\omega\omega'}\,k_{\mu}k'_{\nu} \tilde{F}(\bm{k})\tilde{F}(\bm{k}')e^{\ii\omega t-\ii\bm{k}\cdot\bm{x}}e^{\ii\omega' t-\ii\bm{k}'\cdot\bm{x}}\\
&\times\int\limits_{-\infty}^{\infty}\dd t_{1}\int\limits_{-\infty}^{t_{1}} \dd t_{2}\,\chi(t_{1})\chi(t_{2}) \left(e^{-\ii(\omega+\Omega)t_{1}-\ii(\omega'-\Omega)t_{2}}+e^{-\ii(\omega-\Omega)t_{2}-\ii(\omega'+\Omega)t_{1}}\right),
\end{split}\label{eqc44}\\
M_{e}^{1}(\bm{x},t)&\coloneqq \int\frac{\dd^{3}\bm{k}}{\omega}\tilde{F}(\bm{k})e^{\ii\omega t-\ii\bm{k}\cdot\bm{x}}\int\limits_{-\infty}^{\infty}\dd t_{1}\,\chi(t_{1})e^{-\ii(\omega-\Omega)t_{1}},\\
M_{e}^{2}(\bm{x},t)&=\int\frac{\dd^{3}\bm{k}\dd^{3}\bm{k}'}{\omega\omega'}\tilde{F}(\bm{k})\tilde{F}(\bm{k}')e^{\ii\omega t-\ii\bm{k}\cdot\bm{x}}e^{\ii\omega' t-\ii\bm{k}'\cdot\bm{x}}\\
&\times\int\limits_{-\infty}^{\infty}\dd t_{1}\int\limits_{-\infty}^{t_{1}} \dd t_{2}\,\chi(t_{1})\chi(t_{2})\left(e^{-\ii(\omega+\Omega)t_{1}-\ii(\omega'-\Omega)t_{2}}+e^{-\ii(\omega-\Omega)t_{2}-\ii(\omega'+\Omega)t_{1}}\right),\label{eqc47}
\end{align}
with $J_{\mu,g}^{1},J_{\mu\nu,g}^{2},M_{g}^{1}$ and $M_{g}^{2}$ differing from those above by a swap $\Omega\rightarrow -\Omega$ and $\tilde{F}(\bm{k})$ defined in equation \eqref{eqq16}. In the equations above the repeated Greek subindex and superindex pairs  follow Einstein's summation convention.
\endgroup

\end{widetext}

\subsubsection*{Numerical evaluations}

As with the $\delta$-coupling case we consider the situation of a spherically symmetric detector spatial distribution \eqref{eq31f}, i.e. a hard sphere with radius $R$. We also use sudden switching, i.e.
\begin{align}
\chi(t)=\Theta(t)\Theta(T-t),
\end{align}
which means that the interaction starts at $t=0$ and we evaluate as if the interaction stops at time $t=T$, hence $T$ represents the duration of the interaction. Here, instead of the energy density and $\hat{\phi}^{2}$ distributions at $t=0^{+}$, we consider the distributions at $t=T=150\, R$. This lies within the RWA criterion $T\Omega\gg 1$ as we take $\Omega=4\,R^{-1}$, such that $T\Omega=600$. Note that the detector is initially assumed to be excited. 

\begin{figure}[!h]
\includegraphics[width=0.9\columnwidth]{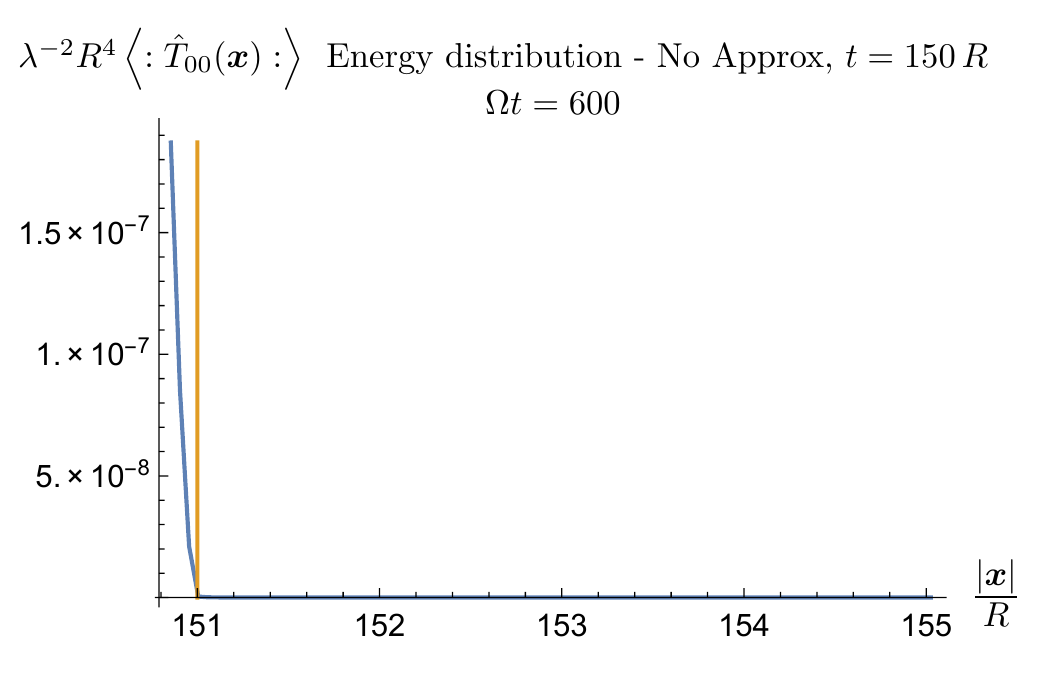}
\caption{Energy density distribution from a second order perturbative interaction where $\chi(t)=\Theta(t)\Theta(T-t)$ with no approximations and $T=150\,R$. Here $G(\bm{\zeta})=\Theta(1-\bm{\zeta})$, i.e. spherically symmetric with a sudden cutoff. Note that the interaction has no non-local field consequences, i.e. no effect beyond $\abs{\bm{x}}>151\,R$. The vertical line at $\abs{\bm{x}}=151\,R$ indicates the locality limit.}\label{fig6}
\end{figure}

\begin{figure}[!h]
\includegraphics[width=0.9\columnwidth]{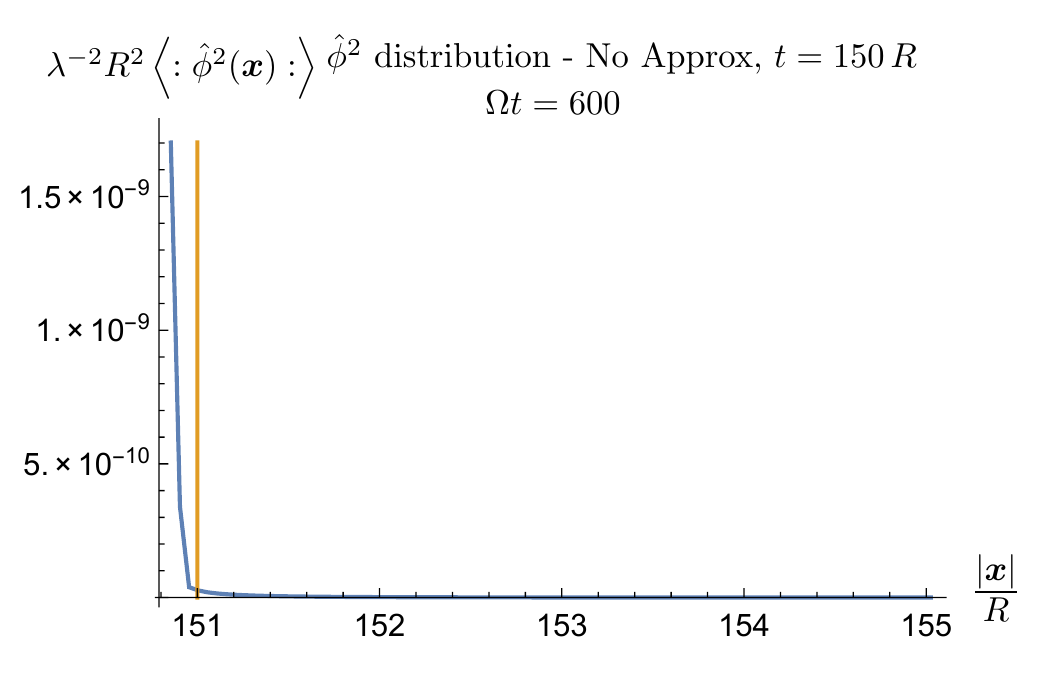}
\caption{$:\hat{\phi}^{2}:$ density distribution from a second order perturbative interaction where $\chi(t)=\Theta(t)\Theta(T-t)$ with no approximations and $T=150\,R$. Here $G(\bm{\zeta})=\Theta(1-\bm{\zeta})$, i.e. spherically symmetric with a sudden cutoff. Note that the interaction has no non-local field consequences, i.e. no effect beyond $\abs{\bm{x}}>151\,R$ (to numerical precision). The vertical line at $\abs{\bm{x}}=151\,R$ indicates the locality limit.}\label{fig8}
\end{figure}

In figures \ref{fig6} and \ref{fig8} the normal ordered energy density and $\phi^{2}$ distributions are plotted respectively for the full model at $t=T=150\,R$. The field observable expectations should be zero outside of the lightcone of the detector. Hence, from the support of the switching and smearing functions chosen, the field expectations should vanish for $\abs{\bm{x}}>151\,R$. This is the case for the full model prediction, as can be seen in the aforementioned figures. 

\begin{figure}[!h]
\includegraphics[width=0.9\columnwidth]{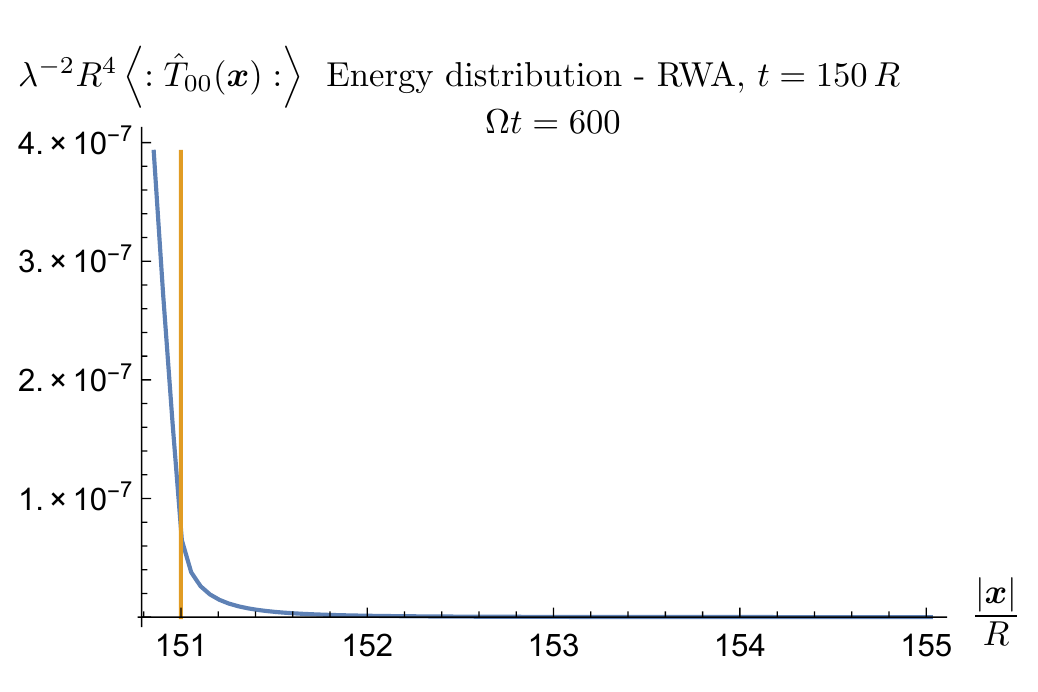}
\caption{Energy density distribution from a second order perturbative interaction where $\chi(t)=\Theta(t)\Theta(T-t)$ under the RWA and $T=150\,R$. Here $G(\bm{\zeta})=\Theta(1-\bm{\zeta})$, i.e. spherically symmetric with a sudden cutoff. Note the polynomial decaying tail for $\abs{\bm{x}}>151\,R$. The vertical line at $\abs{\bm{x}}=151\,R$ indicates the locality limit.}\label{fig5}
\end{figure}

\begin{figure}[!h]
\includegraphics[width=0.9\columnwidth]{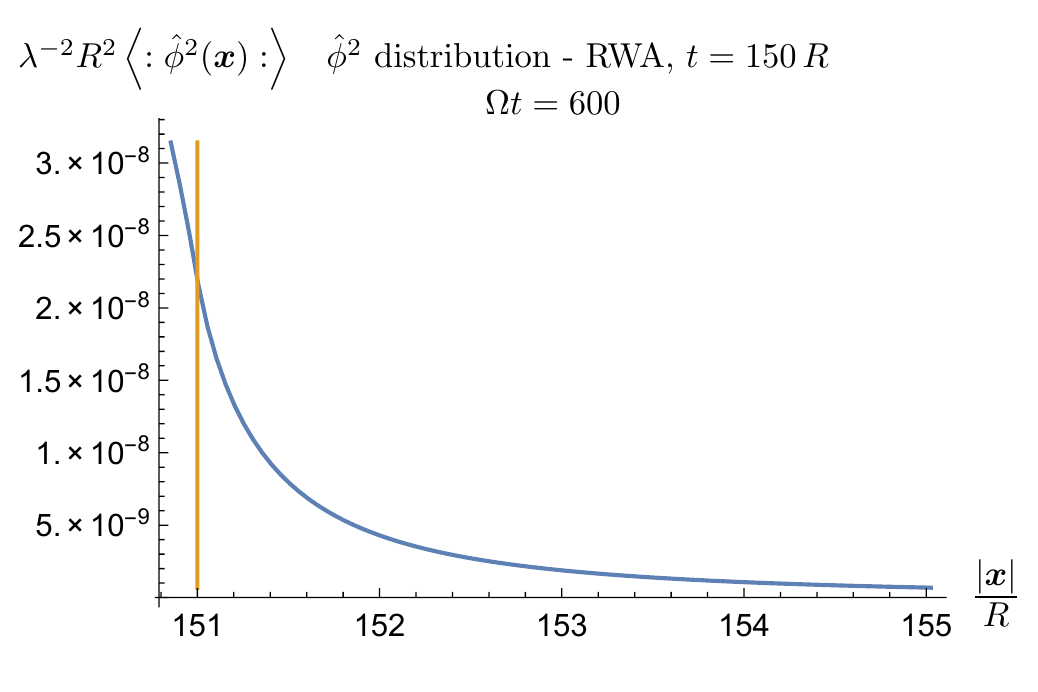}
\caption{$:\hat{\phi}^{2}:$ density distribution from a second order perturbative interaction where $\chi(t)=\Theta(t)\Theta(T-t)$ under the RWA and $T=150\,R$. Here $G(\bm{\zeta})=\Theta(1-\bm{\zeta})$, i.e. spherically symmetric with a sudden cutoff. Note the polynomial decaying tail for $\abs{\bm{x}}>151$. The vertical line at $\abs{\bm{x}}=151\,R$ indicates the locality limit.}\label{fig7}
\end{figure}

In figures \ref{fig5} and \ref{fig7} the normal ordered energy density and $\phi^{2}$ distributions are plotted respectively for the RWA model at $t=T=150\,R$. In this case the violations of causality in physically measurable quantities is apparent, especially so in figure \ref{fig7}, with an obvious polynomial tail extending well beyond $\abs{\bm{x}}=151\,R$. 

The results presented above satisfied the RWA criterion $T\Omega=600\gg1$, and yet causality violations have not been lessened. This could be perhaps more surprising than the $\delta$-coupling case, as it is usually stated that ``the RWA corrects itself over long times''. However, due to second order effects coming from the nested integration in time appearing in terms such as \eqref{eqc44}, as discussed in the previous section and in appendix \ref{sec:ac}, this is not the case. It is worth noting that figures \ref{fig6} and \ref{fig5} appear very similar over large scales, especially when far from the leading edge of the causal sphere. It is also equally important to note that figures \ref{fig8} and \ref{fig7} are wildly different. This can be attributed to the longer range effects of the qubit and the leading edge of the causal sphere on $\phi^{2}$. However, theoretically for sufficiently long times the two figures should begin to converge when far from the qubit and light cone surface. Nevertheless, they will always be different near the light-cone no matter how long the interaction time. Particularly, the faster-than-light tails that the RWA wrongfully predicts will not disappear for large $T$ (See more details in appendix \ref{sec:ac}).

\section{Communication in the RWA: faster-than-light signalling}

From the perspective of a relativistic quantum information theorist the non-localities in the field state make little impression if they do not translate into  causality violations during exchanges of information. For example, does the RWA allow for superluminal signalling between 2 detectors that communicate via `exchanging field quanta'? This section answers this question by considering two detectors coupling to the field at different times, communicating with each other through that interaction. We will see the emergence and behaviour of superluminal signalling when the RWA is assumed.

The leading order communication between two particle detectors has been formalized in \cite{PhysRevD.92.104019}. We will follow a similar scheme here comparing the RWA with the full model in a much more detailed way.

In the case of two detectors the RWA Hamiltonian is naturally extended to 
\begin{align}
\begin{split}
\hat{H}_{\text{I}}^{\textsc{rwa}}(t)&=\lambda_{\textsc{a}}\chi_{\textsc{a}}(t)\int\dd^{3}\bm{y}\,\frac{1}{R_{\textsc{a}}^{3}}G^{\vphantom{*}}_{\textsc{a}}\left(\frac{\bm{y}}{R_{\textsc{a}}}\right)\int\frac{\dd^{3}\bm{k}}{(2\pi)^{3/2}\sqrt{2\omega}}\\
&\left(e^{-\ii(\omega-\Omega_{\textsc{a}})t+\ii\bm{k}\cdot\bm{y}}\hat{a}_{\bm{k}}^{\vphantom{\dagger}}\hat{\sigma}_{\textsc{a}}^{+}+e^{\ii(\omega-\Omega_{\textsc{a}})t-\ii\bm{k}\cdot\bm{y}}\hat{a}_{\bm{k}}^{\dagger}\hat{\sigma}_{\textsc{a}}^{-}\right)\\
&+\lambda_{\textsc{b}}\chi_{\textsc{b}}(t)\int\dd^{3}\bm{y}\,\frac{1}{R_{\textsc{b}}^{3}}G^{\vphantom{*}}_{\textsc{b}}\left(\frac{\bm{y}}{R_{\textsc{b}}}\right)\int\frac{\dd^{3}\bm{k}}{(2\pi)^{3/2}\sqrt{2\omega}}\\
&\left(e^{-\ii(\omega-\Omega_{\textsc{b}})t+\ii\bm{k}\cdot\bm{y}}\hat{a}_{\bm{k}}^{\vphantom{\dagger}}\hat{\sigma}_{\textsc{b}}^{+}+e^{\ii(\omega-\Omega_{\textsc{b}})t-\ii\bm{k}\cdot\bm{y}}\hat{a}_{\bm{k}}^{\dagger}\hat{\sigma}_{\textsc{b}}^{-}\right),
\end{split}
\end{align}
where $\lambda_{\textsc{a},\textsc{b}},\chi_{\textsc{a},\textsc{b}},G^{\vphantom{*}}_{\textsc{a},\textsc{b}}$ are the interaction strength, switching function and spatial smearing functions of the two detectors respectively; and $\hat{\sigma}^{\pm}_{\textsc{a},\textsc{b}}$ are the usual ladder operators acting on detectors A and B respectively with their associated energy gaps $\Omega_{\textsc{a},\textsc{b}}$ respectively. Similarly for the full Hamiltonian:
\begin{align}
\begin{split}
\hat{H}_{\text{I}}^{\textsc{full}}&=\frac{\lambda_{\textsc{a}}\chi_{\textsc{a}}(t)}{R_{\textsc{a}}^{3}}\hat{\sigma}_{\textsc{a},x}(t)\int\dd^{3}\bm{y}\,G^{\vphantom{*}}_{\textsc{a}}\left(\frac{\bm{y}}{R_{\textsc{a}}}\right)\int\frac{\dd^{3}\bm{k}}{(2\pi)^{3/2}\sqrt{2\omega}}\\
&\left(e^{-\ii\omega t+\ii\bm{k}\cdot\bm{y}}\hat{a}_{\bm{k}}^{\vphantom{\dagger}}+e^{\ii\omega t-\ii\bm{k}\cdot\bm{y}}\hat{a}_{\bm{k}}^{\dagger}\right)\\
&+\frac{\lambda_{\textsc{b}}\chi_{\textsc{b}}(t)}{R_{\textsc{b}}^{3}}\hat{\sigma}_{\textsc{b},x}(t)\int\dd^{3}\bm{y}\,G^{\vphantom{*}}_{\textsc{b}}\left(\frac{\bm{y}}{R_{\textsc{b}}}\right)\int\frac{\dd^{3}\bm{k}}{(2\pi)^{3/2}\sqrt{2\omega}}\\
&\left(e^{-\ii\omega t+\ii\bm{k}\cdot\bm{y}}\hat{a}_{\bm{k}}^{\vphantom{\dagger}}+e^{\ii\omega t-\ii\bm{k}\cdot\bm{y}}\hat{a}_{\bm{k}}^{\dagger}\right),
\end{split}
\end{align}
where we recall that
\begin{equation}
    \hat{\sigma}_{\kappa,x}(t)\coloneqq e^{\ii\Omega_{\kappa} t}\hat \sigma^+_{\kappa} +e^{-\ii\Omega_{\kappa} t}\hat \sigma^-_{\kappa},
\end{equation}
acting on the subspace of states of detector $\kappa$. In order to compress notation we can encompass the Hamiltonians of both cases with the following expression:  
\begin{align}
\begin{split}
\hat{H}=&\chi_{\textsc{a}}(t)\left(\hat{\sigma}^{+}_{\textsc{a}}(t)\hat{\psi}_{\textsc{a}}^{\vphantom{\dagger}}+\hat{\sigma}_{\textsc{a}}^{-}(t)\hat{\psi}_{\textsc{a}}^{\dagger}\right)\\
+&\chi_{\textsc{b}}(t)\left(\hat{\sigma}^{+}_{\textsc{b}}(t)\hat{\psi}_{\textsc{b}}^{\vphantom{\dagger}}+\hat{\sigma}_{\textsc{b}}^{-}(t)\hat{\psi}_{\textsc{b}}^{\dagger}\right),
\end{split}
\end{align}
where
\begin{align}
\tilde{F}^{\vphantom{*}}_{\kappa}(\bm{k})&\coloneqq \lambda_{\kappa}\int \dd^{3}\bm{y}\,\frac{1}{R_{\kappa}^{3}}G^{\vphantom{*}}_{\kappa}\left(\frac{\bm{y}}{R_{\kappa}}\right)e^{\ii\bm{k}\cdot\bm{y}},\end{align}
\begin{align}
\hat{\alpha}_{\kappa}^{\vphantom{\dagger}}(t)&\coloneqq \int\frac{\dd^{3}\bm{k}}{(2\pi)^{3/2}\sqrt{2\omega}}\tilde{F}^{\vphantom{*}}_{\kappa}(\bm{k})e^{-\ii\omega t}\hat{a}_{\bm{k}}^{\vphantom{\dagger}},\end{align}
\begin{align}
    \hat{\psi}_{\kappa}^{\vphantom{\dagger}}(t)&\coloneqq\begin{cases}
\hat{\alpha}_{\kappa}^{\vphantom{\dagger}} & \text{if RWA},\\
\hat{\alpha}_{\kappa}^{\vphantom{\dagger}}+\hat{\alpha}_{\kappa}^{\dagger} & \text{otherwise}.
\end{cases}\label{eq48}
\end{align}
With this notation we only need to perform one formal second order Dyson expansion of the time evolution operator in order to investigate the possibilities of superluminal signalling. The corresponding second order Dyson expansion of the time evolution operator takes the form

\begin{widetext}
\begin{align}
\begin{split}
\hat{U}(t)&=\hat{\openone}-\ii\int\limits_{-\infty}^{\infty}\dd t_{1}\left(\chi_{\textsc{a}}(t_{1})\left(\hat{\sigma}_{\textsc{a}}^{+}(t_{1})\hat{\psi}_{\textsc{a}}^{\vphantom{\dagger}}(t_{1})+\hat{\sigma}_{\textsc{a}}^{-}(t_{1})\hat{\psi}_{\textsc{a}}^{\dagger}(t_{1})\right)+
\chi_{\textsc{b}}(t_{1})\left(\hat{\sigma}_{\textsc{b}}^{+}(t_{1})\hat{\psi}_{\textsc{b}}^{\vphantom{\dagger}}(t_{1})+\hat{\sigma}_{\textsc{b}}^{-}(t_{1})\hat{\psi}_{\textsc{b}}^{\dagger}(t_{1})\right)
\right)\\
&-\int\limits_{-\infty}^{\infty}\dd t_{1}\int\limits_{-\infty}^{t_{1}}\dd t_{2}
\left(\chi_{\textsc{a}}(t_{1})\left(\hat{\sigma}_{\textsc{a}}^{+}(t_{1})\hat{\psi}_{\textsc{a}}^{\vphantom{\dagger}}(t_{1})+\hat{\sigma}_{\textsc{a}}^{-}(t_{1})\hat{\psi}_{\textsc{a}}^{\dagger}(t_{1})\right)+
\chi_{\textsc{b}}(t_{1})\left(\hat{\sigma}_{\textsc{b}}^{+}(t_{1})\hat{\psi}_{\textsc{b}}^{\vphantom{\dagger}}(t_{1})+\hat{\sigma}_{\textsc{b}}^{-}(t_{1})\hat{\psi}_{\textsc{b}}^{\dagger}(t_{1})\right)
\right)\\
&
\left(\chi_{\textsc{a}}(t_{2})\left(\hat{\sigma}_{\textsc{a}}^{+}(t_{2})\hat{\psi}_{\textsc{a}}^{\vphantom{\dagger}}(t_{2})+\hat{\sigma}_{\textsc{a}}^{-}(t_{2})\hat{\psi}_{\textsc{a}}^{\dagger}(t_{2})\right)+
\chi_{\textsc{b}}(t_{2})\left(\hat{\sigma}_{\textsc{b}}^{+}(t_{2})\hat{\psi}_{\textsc{b}}^{\vphantom{\dagger}}(t_{2})+\hat{\sigma}_{\textsc{b}}^{-}(t_{2})\hat{\psi}_{\textsc{b}}^{\dagger}(t_{2})\right)
\right)+\mathcal{O}(\lambda^{3}).
\end{split}
\end{align}


As usual we assume the initial field state is the vacuum and we consider the initial state to be a completely uncorrelated state i.e. $\hat\rho= \hat{\rho}_{\textsc{a}}\otimes\hat{\rho}_{\textsc{b}}\otimes\ketbra{0}{0}$. For brevity we will also define $\hat{\rho}^{0}=\hat{\rho}_{\textsc{a}}\otimes\hat{\rho}_{\textsc{b}}$. Additionally, since we are investigating the causality of the signalling, we set up the detectors' switching and smearing functions to be compactly supported and their domains to be spacelike separated. With the extra assumption that the supports of the switching functions are non-overlapping in the frame $(t,\bm x)$, we can denest the time integrals in the same fashion as in \cite{PhysRevD.92.104019}, and we can assume WLOG that $\chi_{\textsc{b}}$ switches on and off before $\chi_{\textsc{a}}$ switches on in that frame. This plays a large role in simplifying the time ordered integral above.

Following the application of the time evolution operator we trace out the field and detector 2 and focus our attention on the reduced density matrix terms that involve communication, i.e. the $\lambda_{\textsc{a}}\lambda_{\textsc{b}}$ dependent terms. This leads us to

\begin{align}
\begin{split}
\hat{\rho}^{1}(t)&=\Tr_{\textsc{b}}\left(\hat{\rho}^{0}\right)+\mathcal{O}(\lambda_{\textsc{a}}^{2})+\mathcal{O}(\lambda_{\textsc{b}}^{2})+\int\limits_{-\infty}^{\infty}\dd t_{1}\int\limits_{-\infty}^{\infty}\dd t_{2}\bigg\{\chi_{\textsc{a}}(t_{1})\chi_{\textsc{b}}(t_{2})\\
&\Tr_{\textsc{b}}\bigg(\hat{\sigma}_{\textsc{a}}^{+}(t_{1})\hat{\rho}^{0}\hat{\sigma}_{\textsc{b}}^{+}(t_{2})\expect{\left[\hat{\psi}^{\vphantom{\dagger}}_{\textsc{b}}(t_{2}),\hat{\psi}^{\vphantom{\dagger}}_{\textsc{a}}(t_{1})\right]}
+\hat{\sigma}_{\textsc{a}}^{+}(t_{1})\hat{\rho}^{0}\hat{\sigma}_{\textsc{b}}^{-}(t_{2})\expect{\left[\hat{\psi}^{\dagger}_{\textsc{b}}(t_{2}),\hat{\psi}_{\textsc{a}}^{\vphantom{\dagger}}(t_{1})\right]}\\
+&\hat{\sigma}_{\textsc{a}}^{-}(t_{1})\hat{\rho}^{0}\hat{\sigma}_{\textsc{b}}^{+}(t_{2})\expect{\left[\hat{\psi}^{\vphantom{\dagger}}_{\textsc{b}}(t_{2}),\hat{\psi}_{\textsc{a}}^{\dagger}(t_{1})\right]}
+\hat{\sigma}_{\textsc{a}}^{-}(t_{1})\hat{\rho}^{0}\hat{\sigma}_{\textsc{b}}^{-}(t_{2})\expect{\left[\hat{\psi}^{\dagger}_{\textsc{b}}(t_{2}),\hat{\psi}_{\textsc{a}}^{\dagger}(t_{1})\right]}\bigg)\\
+&\Tr_{\textsc{b}}\bigg(
\hat{\sigma}_{\textsc{b}}^{+}(t_{2})\hat{\rho}^{0}\hat{\sigma}_{\textsc{a}}^{+}(t_{1})\expect{\left[\hat{\psi}^{\vphantom{\dagger}}_{\textsc{a}}(t_{1}),\hat{\psi}^{\vphantom{\dagger}}_{\textsc{b}}(t_{2})\right]}
+\hat{\sigma}_{\textsc{b}}^{+}(t_{2})\hat{\rho}^{0}\hat{\sigma}_{\textsc{a}}^{-}(t_{1})\expect{\left[\hat{\psi}^{\dagger}_{\textsc{a}}(t_{1}),\hat{\psi}_{\textsc{b}}^{\vphantom{\dagger}}(t_{2})\right]}\\
+&\hat{\sigma}_{\textsc{b}}^{-}(t_{2})\hat{\rho}^{0}\hat{\sigma}_{\textsc{a}}^{+}(t_{1})\expect{\left[\hat{\psi}^{\vphantom{\dagger}}_{\textsc{a}}(t_{1}),\hat{\psi}_{\textsc{b}}^{\dagger}(t_{2})\right]}
+\hat{\sigma}_{\textsc{b}}^{-}(t_{2})\hat{\rho}^{0}\hat{\sigma}_{\textsc{a}}^{-}(t_{1})\expect{\left[\hat{\psi}^{\dagger}_{\textsc{a}}(t_{1}),\hat{\psi}_{\textsc{b}}^{\dagger}(t_{2})\right]}\bigg)+\mathcal{O}(\lambda_{i}^{3}),
\end{split}\label{eqq56}
\end{align}
where the expectation values are taken over the field vacuum.
\end{widetext}

At this point we can examine the differences between the RWA and the full model by referring to our definitions in \eqref{eq48}. In the RWA, only expectation values of the form $\expect{\psi\psi^{\dagger}}$ will be non-zero, meaning that only 1 of the 2 terms in the commutators above would actually contribute. In these cases we have
\begin{align}
\nonumber &\expect{\left[\hat{\psi}_{\kappa}(t_{1}),\hat{\psi}_{\xi}^{\dagger}(t_{2})\right]}\rwaeq\expect{\hat{\psi}_{\kappa}^{\vphantom{\dagger}}(t_{1})\hat{\psi}_{\xi}^{\dagger}(t_{2})}=\\
&\;\quad\lambda_{\kappa}\lambda_{\xi}\int\frac{\dd^{3} \bm{y}_{1}\dd^{3} \bm{y}_{2}}{R_{\kappa}^{3}R_{\xi}^{3}}G_{\kappa}\left(\frac{\bm{y}_{1}}{R_{\kappa}}\right)G_{\xi}\left(\frac{\bm{y}_{2}}{R_{\xi}}\right)\label{eq51}\\
&\quad\qquad\times\int\frac{\dd^{3}\bm{k}}{(2\pi)^{3}2\omega}e^{-\ii\omega(t_{1}-t_{2})}e^{\ii\bm{k}\cdot(\bm{y}_{1}-\bm{y}_{2})},\nonumber
\end{align}
where the indices $\kappa$ and $\xi$ take values in $\{\text{A},\text{B}\}$.


In contrast, for the full model none of the expectations of the commutators are zero. Since the $\hat\psi_\kappa$ are self-adjoint in the full model (see \eqref{eq48}), then all the commutators are of the form:
\begin{align}
\begin{split}
&\expect{\left[\hat{\psi}_{\kappa}(t_{1}),\hat{\psi}_{\xi}(t_{2})\right]}\fulleq\lambda_{\kappa}\lambda_{\xi}\\
&\int\frac{\dd^{3}\bm{y}_{1}\dd^{3}\bm{y}_{2}}{R_{\kappa}^{3}R_{\xi}^{3}}G_{\kappa}\left(\frac{\bm{y}_{1}}{R_{\kappa}}\right)G_{\xi}\left(\frac{\bm{y}_{2}}{R_{\xi}}\right)\\
&\int\frac{\dd^{3}\bm{k}}{(2\pi)^{3}2\omega}\left(e^{-\ii\omega(t_{1}-t_{2})}-e^{\ii\omega(t_{1}-t_{2})}\right) e^{\ii\bm{k}\cdot(\bm{y}_{1}-\bm{y}_{2})}.
\end{split}\label{eq52}
\end{align}

The difference between \eqref{eq51} and \eqref{eq52} is the fact that the sole exponential $e^{-\ii\omega\Delta t}$ in the RWA case is replaced by the difference  $e^{-\ii\omega\Delta t}-e^{\ii\omega\Delta t}$. To understand the implications of this difference, let us evaluate the following integral:
\begin{align}
&\int\frac{\dd^{3}\bm{k}}{(2\pi)^{3}2\omega}e^{\mp\ii\omega \Delta t}e^{\ii\bm{k}\cdot\bm{x}}=\int\frac{\dd\omega\dd z\,\omega}{2(2\pi)^{2}}e^{\mp\ii\omega\Delta t}e^{\ii\omega \abs{\bm{x}} z}\\
&=\int\frac{\dd\omega}{(2\pi)^{2}}e^{\mp\ii\omega\Delta t}\frac{\sin(\omega\abs{\bm{x}})}{d}\\
&=\int\frac{\dd\omega}{(2\pi)^{2}2\ii \abs{\bm{x}}}\left(e^{\mp\ii\omega(\Delta t\mp\abs{\bm{x}})}-e^{\mp\ii\omega(\Delta t\pm\abs{\bm{x}})}\right)\\
\begin{split}
&=\frac{1}{8\pi^{2}\abs{\bm{x}}}\left(\frac{\text{P.V.}}{\abs{\bm{x}}+t}+\frac{\text{P.V.}}{\abs{\bm{x}}-t}\right)\\
&\pm\frac{\ii}{8\pi\abs{\bm{x}}}\big(\delta(\abs{\bm{x}}+\Delta t)-\delta(\abs{\bm{x}}-\Delta t)\big),
\end{split}
\end{align}
where P.V. indicates principal value integral when read under an integral sign. 

As can be seen these integrals yield non-local polynomially decaying terms that are present in the RWA case and enable superluminal communication. However, in the non-approximated model, we have the difference between such terms, i.e. $e^{-\ii\omega\Delta t}-e^{\ii\omega\Delta t}$. In this case, the the polynomial tails cancel out and only delta functions on the light-cone remain, as is expected from a non-approximated (and therefore causal) interaction theory \cite{PhysRevD.92.104019}.

\subsection*{Quantifying signalling through channel capacity}

In this subsection we illustrate with plots for particular cases the effect of the causality-violating tails in signalling for the RWA model. For simplicity we use spherically symmetric detector distributions, the same as in \eqref{eq31f}, with detector A centred around $\bm{x}=0$ and detector B centred around $\bm{x}=\bm{d}$. The switching functions where chosen to have compact, non-overlapping supports:
\begin{align}
\chi_{\textsc{a}}(t_{1})&=\begin{cases}
1 &\text{if $13<t_{1}\Omega_{1}<23$}\\
0&\text{otherwise},\end{cases}\\
\chi_{\textsc{b}}(t_{2})&=\begin{cases}
1 &\text{if $0<t_{2}\Omega_{2}<10$}\\
0&\text{otherwise},\end{cases}
\end{align}
where $\Omega_{\textsc{a}}=\Omega_{\textsc{b}}=R^{-1}$. Also $R_{\textsc{a}}=R_{\textsc{b}}=R$ and the precise numerical values for the support are chosen to maximize visibility in the plots.

\begin{figure}[!h]
\includegraphics[width=0.9\columnwidth]{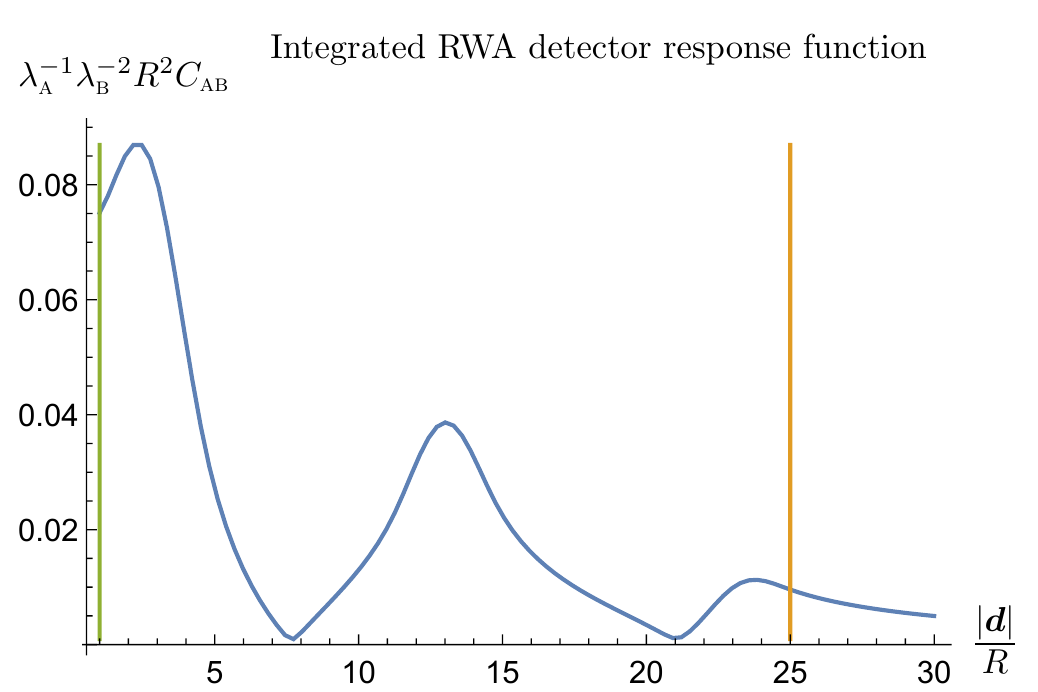}
\caption{Integrated RWA detector response function. Here we used $F_{\textsc{a}}(\bm{x})=\Theta(R_{\textsc{a}}-\abs{\bm{x}})$ and $F_{\textsc{b}}(\bm{x})=\Theta(R_{\textsc{b}}-\abs{\bm{x}-\bm{d}})$. In addition the detector interaction times where $\chi_{\textsc{b}}(t_{2})=1$ for $t_{2}\Omega_{\textsc{b}}\in(0,10)$ and zero otherwise; and $\chi_{\textsc{a}}(t_{1})=1$ for $t_{1}\Omega_{\textsc{a}}\in(13,23)$ and zero otherwise. Given that both detectors have a radius of $R$ the lightcone should only reach $\abs{\bm{d}}=25\,R$. The polynomial decay beyond this is a consequence of the RWA. The vertical line at $\abs{\bm{d}}=R$ indicates the superior limit of the strong Huygen's principle and the vertical line at $\abs{\bm{d}}=25\,R$ indicates the causal limit. Here $R_{\textsc{a}}=R_{\textsc{b}}=R$ and $\Omega_{\textsc{a}}=\Omega_{\textsc{b}}=R^{-1}$.}\label{fig9}
\end{figure}

In figure \ref{fig9} we plot, for the case of the RWA, the magnitude 
\begin{align}
C_{\textsc{a}\textsc{b}}&=\int\limits_{-\infty}^{\infty}\dd t_{1}\int\limits_{-\infty}^{\infty}\dd t_{2}\,\chi_{\textsc{a}}(t_{1})\chi_{\textsc{b}}(t_{2})\nonumber\\
&\times\expect{\left[\hat{\psi}_{\textsc{a}}^{\vphantom{\dagger}}(t_{1}),\hat{\psi}^{\dagger}_{\textsc{b}}(t_{2})\right]}e^{\ii\Omega(t_{1}-t_{2})},\label{eqa62}
\end{align}
 which is the coefficient of $\hat{\sigma}_{\textsc{b}}^{+}\hat{\rho}^{0}\hat{\sigma}_{\textsc{a}}^{-}$. This is a good estimator for a lower bound on the channel capacity between detectors A and B. As discussed in \cite{PhysRevD.92.104019}, when this quantity is non-zero there is communication between the operator of detector A and the operator of detector B (i.e. a local measurement on detector B can reveal information about the state of detector A through a simple protocol).

As we expect from \cite{PhysRevD.92.104019} the communication between 2 detectors arises from the commutators of the $\hat{\psi}_{\kappa}$ operators, as seen in \eqref{eqq56}. We see in figure \ref{fig9} how the non-locality in these commutators induce a non-vanishing signalling estimator $C_{\textsc{a}\textsc{b}}$ outside the causal contact  between A and B ($\abs{\bm{d}}>25\,R$), demonstrating  communication beyond the lightcone in the approximated model. The results from section \ref{sec2b} coincide exactly with figure \ref{fig9} in describing superluminal communication at $\abs{\bm{d}}>25\,R$ with polynomial decay.

\begin{figure}[!h]
\includegraphics[width=0.9\columnwidth]{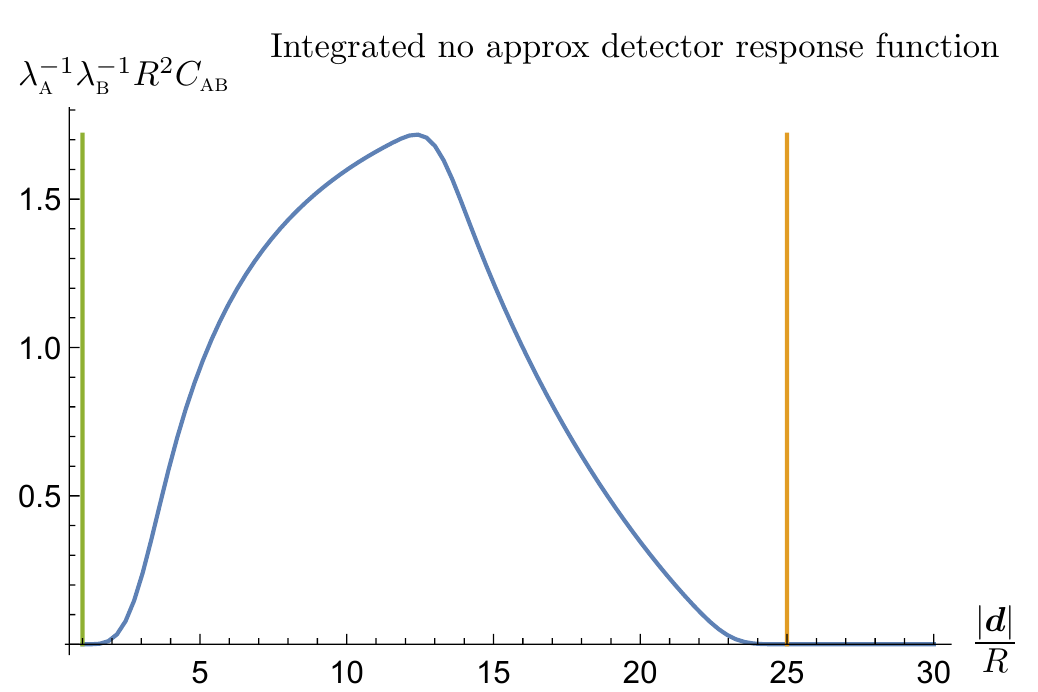}
\caption{Integrated no approx detector response function. Here we used $F_{\textsc{a}}(\bm{x})=\Theta(R_{\textsc{a}}-\abs{\bm{x}})$ and $F_{\textsc{b}}(\bm{x})=\Theta(R_{\textsc{b}}-\abs{\bm{x}-\bm{d}})$. In addition the detector interaction times where $\chi_{\textsc{b}}(t_{2})=1$ for $t_{2}\Omega_{\textsc{b}}\in(0,10)$ and zero otherwise; and $\chi_{\textsc{a}}(t_{1})=1$ for $t_{1}\Omega_{\textsc{a}}\in(13,23)$ and zero otherwise. Given that both detectors have a radius of $R$ the lightcone should only reach $\abs{\bm{d}}=25\,R$. Note how when considering the full model then causality is maintained. The vertical line at $\abs{\bm{d}}=R$ indicates the superior limit of the strong Huygen's principle and the vertical line at $\abs{\bm{d}}=25\,R$ indicates the causal limit. Here $R_{\textsc{a}}=R_{\textsc{b}}=R$ and $\Omega_{\textsc{a}}=\Omega_{\textsc{b}}=R^{-1}$.}\label{fig10}
\end{figure}

Conversely in figure \ref{fig10} we plot \eqref{eqa62} for a non-approximated model. These results are indeed consistent with causality. In fact a close look at $\abs{\bm{d}}<R$ verifies the strong Huygens's principle \cite{McLenaghan1974,McLenaghan2008,PhysRevLett.114.110505} at work.

\section{Discussion: The persistent violation of causality in the RWA}\label{sec:discuss}

From the results and plots above we have seen that the Hamiltonian non-locality introduced by the RWA translates into physically measurable non-causal effects such as non-causal field expectation values and superluminal communication between two particle detectors. Remarkably this is true regardless of how long the interaction lasts.

For very short interactions, we have seen that for the $\delta$-coupling, when considering spherically symmetric smearings of compact support of the form \eqref{eq31f} and under the assumption $\abs{\bm{x}}\gg R$ the expressions \eqref{eq25} and \eqref{eq26} yield 
\begin{align}
\expect{:\hat{T}_{00}(\bm{x},0):}_{{\textsc{rwa}}}&\sim \frac{4\lambda^{2}\sin^{2}\left(R^{2}\right)}{9\pi^{2}R^{4}\abs{\bm{x}}^{6}},\\
\expect{:\hat{\phi}^{2}(\bm{x},0):}_{\textsc{rwa}}&\sim\frac{2\lambda^{2}\sin^{2}\left(R^{2}\right)}{9\pi^{2}R^{4}\abs{\bm{x}}^{4}}.
\end{align}
This behaviour is perhaps unsurprising given $\hat{H}_{\text{I}}^{\textsc{rwa}}$ has a $1/r^{2}$ non-locality, combined with the quadratic nature of $\hat{\phi}^{2}$ should result in a $1/r^{4}$ non-locality. As for the stress-energy tensor, it is composed of $\partial_{\mu}\hat{\phi}\partial_{\nu}\hat{\phi}$, i.e. the two derivative operators act of the cannonical commutation relations to produce a $1/r^{6}$ non-locality. 

Remarkably, even for long timescales, when considering perturbative evolution under the assumption $\abs{\bm{x}}\gg t,R$ (events far ahead of the lightcone) and when considering spherically symmetric smearings of compact support of the form \eqref{eq31f}, the expectation values \eqref{prwat} and \eqref{prwaphi} asymptote to 
\begin{align}
\expect{:\hat{T}_{00}(\bm{x},t):}_{{\textsc{rwa}}}&\sim\frac{16 \lambda^{2} \sin^{2}\left(\frac{t \Omega }{2}\right)}{9\pi^{2}\abs{\bm{x}}^{6}\Omega^{2}},\\
\expect{:\hat{\phi}^{2}(\bm{x},t):}_{\textsc{rwa}}&\sim\frac{8\lambda^{2} \sin^{2}\left(\frac{t\Omega}{2}\right)}{9\pi^{2}\Omega^{2}\abs{\bm{x}}^{4}}.
\end{align}
Hence, perhaps not expected under the usual `RWA works for long times' belief, the asymptotic behaviour of the expectation values is the same for long interaction times as it is for very short interaction times. The satisfaction of the RWA's criterion does not improve the causality violation in any way. 

By applying the same asymptotic analysis and assumptions ($\abs{\bm{d}}\gg t,R_{\textsc{a}},R_{\textsc{b}}$, and smearings of the form \eqref{eq31f})  to the case of 2 detector communication \eqref{eq51}, we find that 
\begin{align}
\expect{\hat{\psi}^{\vphantom{\dagger}}_{\textsc{a}}(t_{1})\hat{\psi}^{\dagger}_{\textsc{b}}(t_{2})}\sim  \frac{1}{\abs{\bm{d}}^{2}},
\end{align}
where $\abs{\bm{d}}$ is the inter-detector spatial distance in the detector's comoving frame. This should not be a surprise, given that the communication capacity is given by the commutator of the respective field operators and \eqref{eq13} tells us that this commutation relations will decay as $1/r^{2}$.

Whilst the presence of a polynomially decaying non-locality should be a deal breaker for the RWA models we expect the behaviour of `resonant-rotating' terms to be more significant when considering situations well within the bulk of the light cone, and where relativistic considerations are not too important for the phsyics described. By considering fixed spatial points away from surface of the `light sphere', i.e. far from any causal considerations the RWA will pointwise converge to the full model, as shown in the appendix \ref{sec:ac}. One such example of this would be to consider cavity setups where the interaction time-scales are larger than the light crossing time of the cavity itself. However, as the realm of relativistic quantum information and ultra fast optical experiments expands \cite{Forn-Diaz2016}, the usefulness of the RWA diminishes and will become unsuitable for modelling experimental situations.

\section{Conclusions}

We have studied in detail the causality violations of the rotating wave approximation (RWA). Over the course of this manuscript we have followed up on the results of Compagno et al. \cite{Compagno_1989,Compagno_1990} in demonstrating the non-local physical effects of a RWA detector acting on a vacuum field, greatly extending their results and including an asymptotic study of detector response in several different regimes. We have also extended the results of Clerk \& Sipe \cite{Clerk1998} by finding the exact asymptotic behaviour of the RWA interaction Hamiltonian's non-locality. Our work found that the light-matter interaction assumes a $1/r^{2}$ non-locality when subject to the RWA, a non-locality that extends to the unitary time evolution operator by means of $1/r^{4}$ and $1/r^{6}$ non-localities for $\hat{\phi}^{2}(\bm{x})$ and $\hat{T}_{00}(\bm{x})$ field expectation values. This polynomial decay is independent of time, demonstrating that waiting for long times does not fix the causality violations of the RWA when looking at field observables. 

Additionally, we have also studied the fundamental relativistic quantum information scenario consisting of 2 detectors communicating through their coupling with a quantum field. In this situation the RWA predicts superluminal signalling, introducing a potentially severe $1/r^{2}$ non-locality, which becomes particularly important in vacuum field experiments, such as entanglement harvesting \cite{VALENTINI1991321,Reznik,Pozas2015,PhysRevD.94.064074}.  Again, no matter how long we wait, there are always polynomial tails that allow for faster-than-light signalling in the RWA.

The RWA may provide a certain simplification to the mathematical description of the physics as discussed at the end of appendix \ref{RWA:secb}; however, the non-localities introduced by RWA make it incompatible with any setup with relativistic considerations are relevant (such is the case in relativistic quantum information). Furthermore, these considerations are  becoming more relevant with the improvement of fast switching light-matter interaction experimental technologies \cite{Forn-Diaz2016}.

\acknowledgments
E.M-M acknowledges support of the NSERC Discovery program as well as his Ontario Early Researcher Award.

\onecolumngrid

\appendix 


\section{RWA Hamiltonian non-locality integrals}\label{seca1a}
In section \ref{sec2b} we made use of equations \eqref{eq12_0} and \eqref{eq12} to demonstrate that the RWA Hamiltonian has non-local interaction terms. Equation \eqref{eq12_0} is standard whilst equation \eqref{eq12} requires a couple of careful considerations. For brevity let $\bm{r}\coloneqq\bm{y}-\bm{z}$,

\begin{align}
\int\dd^{3}\bm{k}\frac{e^{\ii\bm{k}\cdot\bm{r}}}{\omega}=\int\limits_{0}^{\infty}\dd\omega\int\limits_{0}^{2\pi}\dd\phi\int\limits_{-1}^{1}\dd z\,\omega^{2}\frac{e^{\ii\omega r z}}{\omega}
=2\pi\int\dd\omega\, \omega \frac{e^{\ii\omega r}-e^{-\ii\omega r}}{\ii\omega r}
&=\frac{2\pi}{\ii r}\int\dd\omega\left(e^{\ii\omega r}-e^{-\ii\omega r}\right).
\end{align}
At this point we introduce a soft UV cutoff as a regularizator to facilitate the $\omega$ integral. This cutoff takes the form of $e^{-\varepsilon\omega}$, where following $\omega$ integration we will take $\varepsilon\rightarrow 0$.

\begin{align}
\frac{2\pi}{\ii r}\int\dd\omega\left(e^{\ii\omega r}-e^{-\ii\omega r}\right)&=\lim_{\varepsilon\rightarrow 0}\frac{2\pi}{\ii r}\int\dd\omega\left(e^{\omega(\ii r-\varepsilon)}-e^{\omega(-\ii r-\varepsilon)}\right)\\
&=\lim_{\varepsilon\rightarrow 0}\frac{2\pi}{\ii r}\left(-\frac{1}{\ii r-\varepsilon}+\frac{1}{-\ii r-\varepsilon}\right)\\
&=\lim_{\varepsilon\rightarrow 0}\frac{2\pi}{\ii r}\left(-\frac{2\ii r}{-r^{2}-\varepsilon^{2}}\right)=\frac{4\pi}{r^{2}}.
\end{align}
This leaves us with equation \eqref{eq12}, 
\begin{align}
\int\dd^{3}\bm{k}\frac{e^{\ii\bm{k}\cdot(\bm{y}-\bm{z})}}{\omega}&=\frac{4\pi}{\abs{\bm{y}-\bm{z}}^{2}}.
\end{align}

\section{RWA $\delta$-switching unitary time evolution operator}\label{seca2}
In section \ref{RWA:sec3} we stated that the time evolution operator generated by the RWA Hamiltonian under a $\delta$-switching, after considering that it will be acting on the vacuum (i.e. the time evolution operator restricted to that particular state of the field), is given by equation \eqref{eq21}. Its derivation follows:

\begin{align}
\hat{U}&=\mathcal{T}\exp\left(-\ii\int\hat{H}_{\text{I}}\dd t\right)\\
&=\exp\left(-\ii\left(\hat{\alpha}\hat{\sigma}^{+}+\hat{\alpha}^{\dagger}\hat{\sigma}^{-}\right)\right)\\
&=\sum_{n=0}^{\infty}\frac{(-\ii)^{2n}}{(2n)!}\left(\hat{\sigma}^{+}\hat{\alpha}+\hat{\sigma}^{-}\hat{\alpha}^{\dagger}\right)^{2n}+\sum_{n=0}^{\infty}\frac{(-\ii)^{2n+1}}{(2n+1)!}\left(\hat{\sigma}^{+}\hat{\alpha}+\hat{\sigma}^{-}\hat{\alpha}^{\dagger}\right)^{2n+1}\\
&=\sum_{n=0}^{\infty}\frac{(-1)^{n}}{(2n)!}\left(\hat{\Pi}_{e}(\hat{\alpha}\hat{\alpha}^{\dagger})^{n}+\hat{\Pi}_{g}(\hat{\alpha}^{\dagger}\hat{\alpha})^{n}\right)-\ii\sum_{n=0}^{\infty}\frac{(-1)^{n}}{(2n+1)!}\left(\hat{\sigma}^{+}(\hat{\alpha}\hat{\alpha}^{\dagger})^{n}\hat{\alpha}+\hat{\sigma}^{-}\hat{\alpha}^{\dagger}(\hat{\alpha}\hat{\alpha}^{\dagger})^{n}\right)\\
&=\sum_{n=0}^{\infty}\frac{(-1)^{n}}{(2n)!}\left(\hat{\Pi}_{e}(\hat{\alpha}^{\dagger}\hat{\alpha}+\K^{2}\hat{\openone})^{n}+\hat{\Pi}_{g}(\hat{\alpha}^{\dagger}\hat{\alpha})^{n}\right)-\ii\sum_{n=0}^{\infty}\frac{(-1)^{n}}{(2n+1)!}\left(\hat{\sigma}^{+}(\hat{\alpha}^{\dagger}\hat{\alpha}+\K^{2}\hat{\openone})^{n}\hat{\alpha}+\hat{\sigma}^{-}\hat{\alpha}^{\dagger}(\hat{\alpha}^{\dagger}\hat{\alpha}+\K^{2}\hat{\openone})^{n}\right).\label{eqb5}
\end{align}
Here $\hat{\Pi}_{g}\coloneqq \ket{g}\!\bra{g}, \hat{\Pi}_{e}\coloneqq \ket{e}\!\bra{e}$ refer to projection operators on the detector Hilbert space. Note that all the field operators $\hat{\alpha}$ are evaluated at $t=0$. Further note that 
\begin{align}
K^{2}\hat{\openone} \coloneqq \left[\hat{\alpha}(0),\hat{\alpha}^{\dagger}(0)\right]=\tilde{\lambda}^{2}\int\frac{\dd^{3}\bm{k}}{(2\pi)^{3}2\omega}\big|{\tilde{F}(\bm{k})}\big|^{2}\hat\openone,
\end{align}
Acting with \eqref{eqb5} on the vacuum we can cancel all terms that annihilate it and therefore
\begin{align}
\hat{U}\ket{0}=\left[\hat{\Pi}_{g}+\sum_{n=0}^{\infty}\frac{(-1)^{n}}{(2n)!}\hat{\Pi}_{e}\K^{2n}-\ii\sum_{n=0}^{\infty}\frac{(-1)^{n}}{(2n+1)!}\hat{\sigma}^{-}\hat{\alpha}^{\dagger}\K^{2n}\right]\ket{0}=\left[\hat{\Pi}_{g}+\hat{\Pi}_{e}\cos\K-\ii\frac{\hat{\sigma}^{-}\hat{\alpha}^{\dagger}(0)}{\K}\sin\K\right]\ket{0},
\end{align}
where in the final step the time dependence of $\hat{\alpha}$ is explicitly shown for clarity.

\section{Field expectations under perturbative expansions}\label{RWA:secb}
Here we present a derivation of the expectation values  $\langle\hat{T}_{\mu\nu}\rangle$ and $\langle\hat{\phi}^{2}\rangle$ when using second order perturbation theory both for the full model and under the RWA.

\subsection{Full model expectations}

Without the RWA approximation the interaction Hamiltonian is
\begin{align}
\hat{H}_{\text{I}}(t)&=\lambda\chi(t)\hat{\sigma}_{x}(t)\int\dd^{3}\bm{y}\,\frac{1}{R^{3}}G\left(\frac{\bm{y}}{R}\right)\int\frac{\dd^{3}\bm{k}}{(2\pi)^{3/2}\sqrt{2\omega}}\left(e^{-\ii\omega t+\ii\bm{k}\cdot\bm{y}}\hat{a}_{\bm{k}}^{\vphantom{\dagger}}+e^{\ii\omega t-\ii\bm{k}\cdot\bm{y}}\hat{a}_{\bm{k}}^{\dagger}\right),
\end{align}
where, in order to simplify, we can define
\begin{align}
\tilde{F}(\bm{k})&\coloneqq \int \dd^{3}\bm{y}\,\frac{1}{R^{3}}G\left(\frac{\bm{y}}{R}\right)e^{\ii\bm{k}\cdot\bm{y}},\\
\hat{\alpha}(t)&\coloneqq \lambda\int\frac{\dd^{3}\bm{k}}{(2\pi)^{3/2}\sqrt{2\omega}}\tilde{F}(\bm{k})e^{-\ii\omega t}\hat{a}_{\bm{k}}^{\vphantom{\dagger}},
\end{align}
then
\begin{align}
\hat{H}_{\text{I}}(t)&=\chi(t)\hat{\sigma}_{x}(t)\left(\hat{\alpha}+\hat{\alpha}^{\dagger}\right).
\end{align}

The corresponding second order time evolution operator becomes
\begin{align}
\hat{U}&=\hat{\openone}-\ii\int\limits_{-\infty}^{\infty}\dd t_{1}\,\chi(t_{1})\hat{\sigma}_{x}(t_{1})\left(\hat{\alpha}(t_{1})+\hat{\alpha}^{\dagger}(t_{1})\right)\nonumber\\
&-\int\limits_{-\infty}^{\infty}\dd t_{1} \int\limits_{-\infty}^{t_{1}}\dd t_{2}\,\chi(t_{1})\chi( t_{2})\hat{\sigma}_{x}(t_{1})\hat{\sigma}_{x}( t_{2})
\left(\hat{\alpha}(t_{1})+\hat{\alpha}^{\dagger}(t_{1})\right)
\left(\hat{\alpha}( t_{2})+\hat{\alpha}^{\dagger}( t_{2})\right)+\mathcal{O}(\lambda^{3}),
\end{align}
where the interaction time is encoded in the shape and support of $\chi(t)$.

Taking into account that
\begin{align}
\left[\hat{\alpha}(t_{1}),\hat{\alpha}^{\dagger}( t_{2})\right]&=\lambda^{2}\int\frac{\dd^{3}\bm{k}}{(2\pi)^{3}2\omega}\abs{\tilde{F}(\bm{k})}^{2}e^{-\ii\omega(t_{1}- t_{2})},\label{b24}
\end{align}
 $\hat{U}$ acting on the vacuum can be simplified to
\begin{align}
\hat{U}\ket{0}&=\left[\hat{\openone}-\ii\int\limits_{-\infty}^{\infty}\dd t_{1}\,\chi(t_{1})\hat{\sigma}_{x}(t_{1})\hat{\alpha}^{\dagger}(t_{1})-\int\limits_{-\infty}^{\infty}\dd t_{1} \int\limits_{-\infty}^{t_{1}}\dd t_{2}\,\chi(t_{1})\chi( t_{2})\left(\hat{\Pi}_{e}e^{\ii\Omega(t_{1}- t_{2})}+\Pi_{g}e^{-\ii\Omega(t_{1}- t_{2})}\right)\right.\nonumber\\
&\times\left.\left(\hat{\alpha}^{\dagger}(t_{1})\hat{\alpha}^{\dagger}( t_{2})+\lambda^{2}\int\frac{\dd^{3}\bm{k}}{(2\pi)^{3}2\omega}e^{-\ii\omega(t_{1}- t_{2})}\abs{\tilde{F}(\bm{k})}^{2}\right)\right]\ket{0}.
\end{align}
This yields components with 0, 1 and 2 excitations. By  taking the expectation values and using that
 \begin{equation}
     \left[\hat{a}_{\bm{k}}^{\vphantom{\dagger}},\hat{\alpha}^{\dagger}(t_{1})\right]=\lambda\frac{e^{\ii\omega t_{1}}\tilde{F}^{*}(\bm{k})}{(2\pi)^{3/2}\sqrt{2\omega}},
 \end{equation}
 we can write
\begin{align}
\begin{split}
&\expect{:\hat{T}_{\mu\nu}(\bm{x},t):}_{\text{Full}}=\lambda^{2}\int\frac{\dd^{3}\bm{k}\dd^{3}\bm{k}'}{(2\pi)^{6}4\omega\omega'}\left(k_{\mu}k'_{\nu}-\frac{\eta_{\mu\nu}}{2}k_{\gamma}k'^{\gamma}\right)\\
&\Bigg[
e^{-\ii(\omega-\omega')t+\ii(\bm{k}-\bm{k}')\cdot\bm{x}}\int\limits_{-\infty}^{\infty}\dd t_{1}\int\limits_{-\infty}^{\infty}\dd t_{1}'\,\chi(t_{1})\chi(t_{1}')\left(\hat{\Pi}_{e}e^{\ii\Omega(t_{1}-t_{1}')}+\hat{\Pi}_{g}e^{-\ii\Omega(t_{1}-t_{1}')}\right)\tilde{F}(\bm{k}')\tilde{F}^{*}(\bm{k})e^{-\ii\omega' t_{1}+\ii\omega t_{1}'}\\
&+e^{\ii(\omega-\omega')t-\ii(\bm{k}-\bm{k}')\cdot\bm{x}}\int\limits_{-\infty}^{\infty}\dd t_{1}\int\limits_{-\infty}^{\infty}\dd t_{1}'\,\chi(t_{1})\chi(t_{1}')\left(\hat{\Pi}_{e}e^{\ii\Omega(t_{1}-t_{1}')}+\hat{\Pi}_{g}e^{-\ii\Omega(t_{1}-t_{1}')}\right)\tilde{F}^{*}(\bm{k}')\tilde{F}(\bm{k})e^{-\ii\omega t_{1}+\ii\omega' t_{1}'}\\
&+e^{-\ii(\omega+\omega')t+\ii(\bm{k}+\bm{k}')\cdot\bm{x}}\!\int\limits_{-\infty}^{\infty}\!\dd t_{1}\! \int\limits_{-\infty}^{t_{1}}\!\dd t_{2}\,\chi(t_{1})\chi(t_{2})\left(\hat{\Pi}_{e}e^{\ii\Omega(t_{1}-t_{2})}+\hat{\Pi}_{g}e^{-\ii\Omega(t_{1}-t_{2})}\right)\tilde{F}^{*}(\bm{k}')\tilde{F}^{*}(\bm{k})\left(e^{\ii\omega t_{1}+\ii\omega' t_{2}}+e^{\ii\omega' t_{1}+\ii\omega t_{2}}\right)\\
&+e^{\ii(\omega+\omega')t-\ii(\bm{k}+\bm{k}')\cdot\bm{x}}\!\int\limits_{-\infty}^{\infty}\!\dd t_{1}\!\! \int\limits_{-\infty}^{t_{1}}\!\dd t_{2}\,\chi(t_{1})\chi(t_{2})\left(\hat{\Pi}_{e}e^{-\ii\Omega(t_{1}-t_{2})}+\hat{\Pi}_{g}e^{\ii\Omega(t_{1}-t_{2})}\right)\tilde{F}(\bm{k}')\tilde{F}(\bm{k})\left(e^{-\ii\omega t_{1}-\ii\omega' t_{2}}+e^{-\ii\omega' t_{1}-\ii\omega t_{2}}\right)\Bigg],
\end{split}\\
\begin{split}
&\expect{:\hat{\phi}^{2}(\bm{x},t):}_{\text{Full}}=\lambda^{2}\int\frac{\dd^{3}\bm{k}\dd^{3}\bm{k}'}{(2\pi)^{6}4\omega\omega'}\\
&\Bigg[
e^{-\ii(\omega-\omega')t+\ii(\bm{k}-\bm{k}')\cdot\bm{x}}\int\limits_{-\infty}^{\infty}\dd t_{1}\int\limits_{-\infty}^{\infty}\dd t_{1}'\,\chi(t_{1})\chi(t_{1}')\left(\hat{\Pi}_{e}e^{\ii\Omega(t_{1}-t_{1}')}+\hat{\Pi}_{g}e^{-\ii\Omega(t_{1}-t_{1}')}\right)\tilde{F}(\bm{k}')\tilde{F}^{*}(\bm{k})e^{-\ii\omega' t_{1}+\ii\omega t_{1}'}\\
&+e^{\ii(\omega-\omega')t-\ii(\bm{k}-\bm{k}')\cdot\bm{x}}\int\limits_{-\infty}^{\infty}\dd t_{1}\int\limits_{-\infty}^{\infty}\dd t_{1}'\,\chi(t_{1})\chi(t_{1}')\left(\hat{\Pi}_{e}e^{\ii\Omega(t_{1}-t_{1}')}+\hat{\Pi}_{g}e^{-\ii\Omega(t_{1}-t_{1}')}\right)\tilde{F}^{*}(\bm{k}')\tilde{F}(\bm{k})e^{-\ii\omega t_{1}+\ii\omega' t_{1}'}\\
&-e^{-\ii(\omega+\omega')t+\ii(\bm{k}+\bm{k}')\cdot\bm{x}}\!\int\limits_{-\infty}^{\infty}\!\dd t_{1}\! \int\limits_{-\infty}^{t_{1}}\!\dd t_{2}\,\chi(t_{1})\chi(t_{2})\left(\hat{\Pi}_{e}e^{\ii\Omega(t_{1}-t_{2})}+\hat{\Pi}_{g}e^{-\ii\Omega(t_{1}-t_{2})}\right)\tilde{F}^{*}(\bm{k}')\tilde{F}^{*}(\bm{k})\left(e^{\ii\omega t_{1}+\ii\omega' t_{2}}+e^{\ii\omega' t_{1}+\ii\omega t_{2}}\right)\\
&-e^{\ii(\omega+\omega')t-\ii(\bm{k}+\bm{k}')\cdot\bm{x}}\!\int\limits_{-\infty}^{\infty}\!\dd t_{1}\!\! \int\limits_{-\infty}^{t_{1}}\!\dd t_{2}\,\chi(t_{1})\chi(t_{2})\left(\hat{\Pi}_{e}e^{-\ii\Omega(t_{1}-t_{2})}+\hat{\Pi}_{g}e^{\ii\Omega(t_{1}-t_{2})}\right)\tilde{F}(\bm{k}')\tilde{F}(\bm{k})\left(e^{-\ii\omega t_{1}-\ii\omega' t_{2}}+e^{-\ii\omega' t_{1}-\ii\omega t_{2}}\right)\Bigg].
\end{split}
\end{align}
In the equations above the contributions to the expectations from 1 excitation states are those of the form $\tilde{F}\tilde{F}^{*}$, where as the remainder, i.e. $\tilde{F}\tilde{F}$ and $\tilde{F}^{*}\tilde{F}^{*}$, are contributions from the superposition of 0 and 2 excitation states.\\
Here we assumed that $t$ is larger than the maximum $t$ in the support of $\chi(t)$, i.e. they represent the evolution of the stress-energy density after the detector's interaction.

In order to simplify this rather long expression and further compare with the RWA, we define the following:
\begin{align}
J_{\mu,e}^{1}(\bm{x},t)&\coloneqq \int\frac{\dd^{3}\bm{k}}{\omega}k_{\mu} \tilde{F}(\bm{k})e^{\ii\omega t-\ii\bm{k}\cdot\bm{x}}\int\limits_{-\infty}^{\infty}\dd t_{1}\,\chi(t_{1})e^{-\ii(\omega-\Omega)t_{1}},\label{eq12c11}\\
\begin{split}
J_{\mu\nu,e}^{2}(\bm{x},t)&\coloneqq \int\frac{\dd^{3}\bm{k}\dd^{3}\bm{k}'}{\omega\omega'}\,k_{\mu}k'_{\nu} \tilde{F}(\bm{k})\tilde{F}(\bm{k}')e^{\ii\omega t-\ii\bm{k}\cdot\bm{x}}e^{\ii\omega' t-\ii\bm{k}'\cdot\bm{x}}\\
&\times\int\limits_{-\infty}^{\infty}\dd t_{1} \int\limits_{-\infty}^{t_{1}}\dd t_{2}\,\chi(t_{1})\chi(t_{2}) \left(e^{-\ii(\omega+\Omega)t_{1}-\ii(\omega'-\Omega)t_{2}}+e^{-\ii(\omega-\Omega)t_{2}-\ii(\omega'+\Omega)t_{1}}\right),
\end{split}\label{b31}\\
M_{e}^{1}(\bm{x},t)&\coloneqq \int\frac{\dd^{3}\bm{k}}{\omega}\tilde{F}(\bm{k})e^{\ii\omega t-\ii\bm{k}\cdot\bm{x}}\int\limits_{-\infty}^{\infty}\dd t_{1}\,\chi(t_{1})e^{-\ii(\omega-\Omega)t_{1}},\label{eq12c13}\\
M_{e}^{2}(\bm{x},t)&=\int\frac{\dd^{3}\bm{k}\dd^{3}\bm{k}'}{\omega\omega'}\tilde{F}(\bm{k})\tilde{F}(\bm{k}')e^{\ii\omega t-\ii\bm{k}\cdot\bm{x}}e^{\ii\omega' t-\ii\bm{k}'\cdot\bm{x}}\nonumber\\
&\times\int\limits_{-\infty}^{\infty}\dd t_{1} \int\limits_{-\infty}^{t_{1}}\dd t_{2}\,\chi(t_{1})\chi(t_{2})\left(e^{-\ii(\omega+\Omega)t_{1}-\ii(\omega'-\Omega)t_{2}}+e^{-\ii(\omega-\Omega)t_{2}-\ii(\omega'+\Omega)t_{1}}\right),\label{b33}
\end{align}
with $J_{\mu,g}^{1},J_{\mu\nu,g}^{2},M_{g}^{1}$ and $M_{g}^{2}$ differing from those above by a swap $\Omega\rightarrow -\Omega$. This way

\begin{align}
\expect{:\hat{T}_{\mu\nu}(\bm{x},t):}_{\text{Full}}&=\frac{\lambda^{2}}{4(2\pi)^{6}}\sum_{i\in\{e,g\}}\hat{\Pi}_{i}\left(J_{\mu,i}^{1\vphantom{*}}J_{\nu,i}^{1*}+J_{\mu,i}^{1*}J_{\nu,i}^{1\vphantom{*}}-\frac{\eta_{\mu\nu}}{2}\left(J_{\gamma,i}^{1\vphantom{*}}J_{i}^{1\gamma *}+J_{\gamma,i}^{1 *}J_{i}^{1 \gamma \vphantom{*}}\right)+J_{\mu\nu,i}^{2\vphantom{*}}+J_{\mu\nu,i}^{2*}\right.\nonumber\\
&\left.-\frac{\eta_{\mu\nu}}{2}\left(J_{\gamma,i}^{2\gamma\vphantom{*}}+J_{\gamma,i}^{2\gamma *}\right)\right)+\mathcal{O}(\lambda^{3}),\\
\expect{:\hat{\phi}^{2}(\bm{x},t):}_{\text{Full}}&=\frac{\lambda^{2}}{4(2\pi)^{6}}\sum_{i\in\{e,g\}}\hat{\Pi}_{i} \left(2\abs{M_{i}^{1}}^{2}-M_{i}^{2\vphantom{*}}-M_{i}^{2*}\right)+\mathcal{O}(\lambda^{3}).
\end{align}

The projection operators meant that if we consider an initial state given by $\ket{0}\otimes(a_{g}\ket{g}+a_{e}\ket{e})$ then the equations above simplify to

\begin{align}
\expect{:\hat{T}_{\mu\nu}(\bm{x},t):}_{\text{Full}}&=\frac{\lambda^{2}}{4(2\pi)^{6}}\sum_{i\in\{e,g\}}\abs{a_{i}}^{2}\left(J_{\mu,i}^{1\vphantom{*}}J_{\nu,i}^{1*}+J_{\mu,i}^{1*}J_{\nu,i}^{1\vphantom{*}}-\frac{\eta_{\mu\nu}}{2}\left(J_{\gamma,i}^{1\vphantom{*}}J_{i}^{1\gamma *}+J_{\gamma,i}^{1 *}J_{i}^{1 \gamma \vphantom{*}}\right)+J_{\mu\nu,i}^{2\vphantom{*}}+J_{\mu\nu,i}^{2*}\right.\nonumber\\
&\left.-\frac{\eta_{\mu\nu}}{2}\left(J_{\gamma,i}^{2\gamma\vphantom{*}}+J_{\gamma,i}^{2\gamma *}\right)\right)+\mathcal{O}(\lambda^{3}),\label{b36}\\
\expect{:\hat{\phi}^{2}(\bm{x},t):}_{\text{Full}}&=\frac{\lambda^{2}}{4(2\pi)^{6}}\sum_{i\in\{e,g\}}\abs{a_{i}}^{2} \left(2\abs{M_{i}^{1}}^{2}-M_{i}^{2\vphantom{*}}-M_{i}^{2*}\right)+\mathcal{O}(\lambda^{3}).\label{b37}
\end{align}

\subsection{RWA expectations}
The RWA interaction Hamiltonian is (see \eqref{eq7})
\begin{align}
\hat{H}_{\text{I}}(t)&=\lambda\chi(t)\int\dd^{3}\bm{y}\,\frac{1}{R^{3}}G\left(\frac{\bm{y}}{R}\right)\int\frac{\dd^{3}\bm{k}}{(2\pi)^{3/2}\sqrt{2\omega}}\left(e^{-\ii(\omega-\Omega)t+\ii\bm{k}\cdot\bm{y}}\hat{a}_{\bm{k}}^{\vphantom{\dagger}}\hat{\sigma}^{+}+e^{\ii(\omega-\Omega)t-\ii\bm{k}\cdot\bm{y}}\hat{a}_{\bm{k}}^{\dagger}\hat{\sigma}^{-}\right),
\end{align}
where, in order to simplify, we can define
\begin{align}
\tilde{F}(\bm{k})&\coloneqq \int \dd^{3}\bm{y}\,\frac{1}{R^{3}}G\left(\frac{\bm{y}}{R}\right)e^{\ii\bm{k}\cdot\bm{y}},\\
\hat{\alpha}(t)&\coloneqq \lambda\int\frac{\dd^{3}\bm{k}}{(2\pi)^{3/2}\sqrt{2\omega}}\tilde{F}(\bm{k})e^{-\ii(\omega-\Omega)t}\hat{a}_{\bm{k}}^{\vphantom{\dagger}},
\end{align}
then
\begin{align}
\hat{H}_{\text{I}}(t)&=\chi(t)\left(\hat{\alpha}(t)\hat{\sigma}^{+}+\hat{\alpha}^{\dagger}(t)\hat{\sigma}^{-}\right).
\end{align}
The corresponding second order time evolution operator becomes
\begin{align}
\hat{U}&=\hat{\openone}-\ii\int\limits_{-\infty}^{\infty}\dd t_{1}\,\chi(t_{1})\left(\hat{\alpha}(t_{1})\hat{\sigma}^{+}+\hat{\alpha}^{\dagger}(t_{1})\hat{\sigma}^{-}\right)\nonumber\\
&-\int\limits_{-\infty}^{\infty}\dd t_{1} \int\limits_{-\infty}^{t_{1}}\dd t_{2}\, \chi(t_{1})\chi(t_{2})\left(\hat{\alpha}(t_{1})\hat{\sigma}^{+}+\hat{\alpha}^{\dagger}(t_{1})\hat{\sigma}^{-}\right)\left(\hat{\alpha}(t_{2})\hat{\sigma}^{+}+\hat{\alpha}^{\dagger}(t_{2})\hat{\sigma}^{-}\right)+\mathcal{O}(\lambda^{3})\\
&=\hat{\openone}-\ii\int\limits_{-\infty}^{\infty}\dd t_{1}\,\chi(t_{1})\left(\hat{\alpha}(t_{1})\hat{\sigma}^{+}+\hat{\alpha}^{\dagger}(t_{1})\hat{\sigma}^{-}\right)-\int\limits_{-\infty}^{\infty}\dd t_{1} \int\limits_{-\infty}^{t_{1}}\dd t_{2}\, \chi(t_{1})\chi(t_{2})
\left(\hat{\alpha}(t_{1})\hat{\alpha}^{\dagger}(t_{2})\hat{\Pi}_{e}+\hat{\alpha}^{\dagger}(t_{1})\hat{\alpha}(t_{2})\hat{\Pi}_{g}
\right)+\mathcal{O}(\lambda^{3}),
\end{align}
where $\hat{\Pi}_{g,e}$ are the projection operators onto the ground and excited states of the detector respectively. Note the interaction duration is encoded in the shape and support of $\chi(t)$.

For compactness we define
\begin{align}
\hat{\xi}&\coloneqq-\ii\int\limits_{-\infty}^{\infty}\dd t_{1} \,\chi(t_{1})\hat{\alpha}^{\dagger}(t_{1}).
\end{align}
Using that
\begin{align}
\left[\hat{\alpha}(t_{1}),\hat{\alpha}^{\dagger}(t_{2})\right]&=\lambda^{2}\int\frac{\dd^{3}\bm{k}}{(2\pi)^{3}2\omega}\abs{\tilde{F}(\bm{k})}^{2}e^{-\ii(\omega-\Omega)t_{1}}e^{\ii(\omega-\Omega)t_{2}},
\end{align}
the time evolution operator acting on the vacuum state simplifies to
\begin{align}
\hat{U}\ket{0}&=\left(\hat{\openone}+\hat{\xi}\hat{\sigma}^{-}-\int\limits_{-\infty}^{\infty}\dd t_{1} \int\limits_{-\infty}^{t_{1}}\dd t_{2}\,\chi(t_{1})\chi(t_{2})\hat{\Pi}_{e}\int\frac{\dd^{3}\bm{k}}{(2\pi)^{3}2\omega}\abs{F(\bm{k})}^{2}e^{-\ii(\omega-\Omega)t_{1}}e^{\ii(\omega-\Omega)t_{2}} \right)\ket{0}.
\end{align}
For computational purposes we only need to focus on the $\hat{\xi}$ term, given it is the only one with a field excitation. Since we only have 0 and 1 field excitations,  and using that
\begin{equation}
\left[\hat{a}_{\bm{k}},\hat{\xi}\right]=-\ii\lambda\int\limits_{-\infty}^{\infty}\dd t_{1} e^{\ii(\omega-\Omega)t_{1}}\chi(t_{1})\frac{\tilde{F}^{*}(\bm{k})}{(2\pi)^{3/2}\sqrt{2\omega}},
\end{equation}
the stress energy tensor and $\hat\phi^{2}$ expectations  reduce to
\begin{align}
\expect{:\hat{T}_{\mu\nu}(\bm{x},t):}_{\textsc{rwa}}=&\hat{\Pi}_{e}\int\frac{\dd^{3}\bm{k}\dd^{3}\bm{k}'}{(2\pi)^{3}\sqrt{4\omega\omega'}}\left(k_{\mu}k'_{\nu}-\frac{\eta_{\mu\nu}}{2}k_{\gamma}k'^{\gamma}\right)\nonumber\\
&\times\left(e^{-\ii (\omega-\omega')t+\ii(\bm{k}-\bm{k}')\cdot\bm{x}}\left[\hat{\xi}^{\dagger},\hat{a}_{\bm{k}'}^{\dagger}\right]\left[\hat{a}_{\bm{k}}^{\vphantom{\dagger}},\hat{\xi}\right]
+e^{\ii (\omega-\omega')t-\ii(\bm{k}-\bm{k}')\cdot\bm{x}}\left[\hat{\xi}^{\dagger},\hat{a}_{\bm{k}}^{\dagger}\right]\left[\hat{a}_{\bm{k}'}^{\vphantom{\dagger}},\hat{\xi}\right]
\right)+\mathcal{O}(\lambda^{3}),\label{eqc11}\\
&\!\!\!\!\!\!\!\!\!\!\!\!\!\!\!\!\!\!\!\!\!\!\!\!\!\!\!\!\!\!\!\!\!\!\!\!\!\!\!\!\!\!\!\!\!\!\!\!\!\!\!\!\!\!\expect{:\hat{\phi}^{2}(\bm{x},t):}_{\textsc{rwa}}\!\!\!\!\!=\!\hat{\Pi}_{e}\!\int\!\frac{\dd^{3}\bm{k}\dd^{3}\bm{k}'}{(2\pi)^{3}\sqrt{4\omega\omega'}}\left(e^{-\ii (\omega-\omega')t+\ii(\bm{k}-\bm{k}')\cdot\bm{x}}\left[\hat{\xi}^{\dagger},\hat{a}_{\bm{k}'}^{\dagger}\right]\left[\hat{a}_{\bm{k}}^{\vphantom{\dagger}},\hat{\xi}\right]
\!+e^{\ii (\omega-\omega')t-\ii(\bm{k}-\bm{k}')\cdot\bm{x}}\left[\hat{\xi}^{\dagger},\hat{a}_{\bm{k}}^{\dagger}\right]\left[\hat{a}_{\bm{k}'}^{\vphantom{\dagger}},\hat{\xi}\right]
\right)\!+\!\mathcal{O}(\lambda^{3}).\label{eqc12}
\end{align}
Here we assumed that $t$ is larger than the maximum $t$ in the support of $\chi(t)$, i.e. post interaction. Also note that these results require the initial state of the detector to have some excited state component. If the initial state is the ground state then the expectation of the stress-energy density and the field amplitude squared (normal ordered) are exactly zero. In order to simplify \eqref{eqc11} and \eqref{eqc12} we use \eqref{eq12c11} and \eqref{eq12c13}, where the expectation values then become
\begin{align}
\expect{:\hat{T}_{\mu\nu}(\bm{x},t):}_{\textsc{rwa}}&=\frac{\lambda^{2}}{4(2\pi)^{6}}\hat{\Pi}_{e}\left[J_{\mu,e}^{1\vphantom{*}} J_{\nu,e}^{1*}+J_{\mu,e}^{1*}J_{\nu,e}^{1\vphantom{*}}-\eta_{\mu\nu}J_{\gamma,e}^{1*}J_{e}^{1\, \gamma}\right]+\mathcal{O}(\lambda^{3}),\\
\expect{:\hat{\phi}^{2}(\bm{x},t):}_{\textsc{rwa}}&=\frac{\lambda^{2}}{2(2\pi)^{6}}\hat{\Pi}_{e}\abs{M_{e}^{1}}^{2}+\mathcal{O}(\lambda^{3}).
\end{align}

The projection operators meant that if we consider an initial state given by $\ket{0}\otimes(a_{g}\ket{g}+a_{e}\ket{e})$ then the equations above simplify to
\begin{align}
\expect{:\hat{T}_{\mu\nu}(\bm{x},t):}_{\textsc{rwa}}&=\frac{\lambda^{2}}{4(2\pi)^{6}}\abs{a_{e}}^{2}\left[J_{\mu,e}^{1\vphantom{*}} J_{\nu,e}^{1*}+J_{\mu,e}^{1*}J_{\nu,e}^{1\vphantom{*}}-\eta_{\mu\nu}J_{\gamma,e}^{1*}J_{e}^{1\, \gamma}\right]+\mathcal{O}(\lambda^{3}),\\
\expect{:\hat{\phi}^{2}(\bm{x},t):}_{\textsc{rwa}}&=\frac{\lambda^{2}}{2(2\pi)^{6}}\abs{a_{e}}^{2}\abs{M_{e}^{1}}^{2}+\mathcal{O}(\lambda^{3}).
\end{align}

\subsection{Why RWA?}
As shown above the implementation of the RWA avoids the need to calculate $J^{2}_{\mu\nu,e}$ when determining the stress-energy expectations. The great advantage to this is that a 2D semi-infinite integral can be avoided, i.e. $0<\omega<\infty$ and $0<\omega'<\infty$. Unlike the $J^{1}_{\mu,e}$ terms that can be separated, the terms $J^{2}_{\mu\nu,e}$ (as shown in \eqref{c4}) contain a denominator that cannot be separated.


\section{RWA in large time limit}\label{sec:ac}

In the discussion of the main text (section \ref{sec:discuss}) the field expectations are Laurent expanded in the limit $\abs{\bm{x}}\gg t,R$, showing that the RWA continues to violate causality in the long time limit. Here we  demonstrate why this occurs.

\subsection{Persistence of RWA causality violations}

Consider equation \eqref{b33}, with 
\begin{align}
\chi(t)=\Theta(t+T)-\Theta(t-T).
\end{align}

Again, a handwavy argument can be put together along the lines of that as $T\rightarrow \infty$ then the $t_{1}$ integral will resemble $\delta(\omega+\Omega)$, and the  $t_{2}$ integral will resemble $\delta(\omega-\Omega)$ given the Fourier transform definition of the Dirac $\delta$. This in turn would mean that in the long time limit the contribution from that integral would be zero once one integrates over $\bm k$ since the argument of the delta is always strictly positive, and hence one can just throw away the contribution from those counter-rotating terms. In the same fashion the emergent $\delta(\omega-\Omega)$ would allow one to keep only one frequency in the field (the so-called single mode approximation) for the integrals involving de-excitation probabilities (e.g. \eqref{eq12c11} and \eqref{eq12c13}) in the same long time limit.

This may be true if we keep the position at which we evaluate the observables fixed and we take the limit of large $T$. However this will not be true if we take the limit of long times and long spatial separation simultaneously as to evaluate field observables near the light-cone of the detector. In this particular situation it is important to consider the terms outside the integrals, i.e. $e^{\ii\omega t-\ii\bm{k}\cdot\bm{x}}e^{\ii\omega' t-\ii\bm{k}'\cdot\bm{x}}$, which are evaluated at $t=T$ and since as $T\rightarrow \infty$ these terms will begin to oscillate wildly such as to unravel the integral definition of the Dirac delta introducing polynomial decays in $\omega$, eliminating the foundations on which the RWA (and the single mode approxiamtion) and SMA are based. That is to say, the long time limit of the integrals of $J^{2}_{\mu\nu,e}$ do not converge uniformly to zero when considering the external exponentials.\\

Mathematically, consider the following expression (central to $J_{\mu\nu,e}^{2}$ and $M_{e}^{2}$)
\begin{align}
e^{\ii\omega T-\ii\bm{k}\cdot\bm{x}}e^{\ii\omega' T-\ii\bm{k}'\cdot\bm{x}}
\int\limits_{-\infty}^{\infty}\dd t_{1} \int\limits_{-\infty}^{t_{1}}\dd t_{2}\,\chi(t_{1})\chi(t_{2}) \left(e^{-\ii(\omega+\Omega)t_{1}-\ii(\omega'-\Omega)t_{2}}+e^{-\ii(\omega-\Omega)t_{2}-\ii(\omega'+\Omega)t_{1}}\right).\label{c2}
\end{align}
First note that when evaluating the expectations of $\hat{T}_{00}$ or $\hat{\phi}^{2}$ that we can swap $\omega\leftrightarrow \omega'$ in the second term of \eqref{c2} without affecting the result of \eqref{b36} and \eqref{b37} (although not for the off diagonal stress energy terms). We perform this swap to simplify the equations in this derivation, i.e. the expression becomes
\begin{align}
I\coloneqq 2e^{\ii\omega T-\ii\bm{k}\cdot\bm{x}}e^{\ii\omega' T-\ii\bm{k}'\cdot\bm{x}}
\int\limits_{-\infty}^{\infty}\dd t_{1} \int\limits_{-\infty}^{t_{1}}\dd t_{2}\,\chi(t_{1})\chi(t_{2}) e^{-\ii(\omega+\Omega)t_{1}-\ii(\omega'-\Omega)t_{2}}.
\end{align}
When we perform the integrals in question we obtain
\begin{align}
I=e^{\ii\omega T-\ii\bm{k}\cdot\bm{x}}e^{\ii\omega' T-\ii\bm{k}'\cdot\bm{x}}
\frac{4\ii}{\Omega-\omega'}\left(\frac{e^{\ii(\omega'-\Omega)T}\sin[(\omega+\Omega)T]}{\omega+\Omega}-\frac{\sin[(\omega+\omega')T]}{\omega+\omega'}\right).\label{c4}
\end{align}
Here we note that the sinc functions in the brackets usually pointwise converge to delta functions (as $T\rightarrow \infty$) and since $\omega,\omega',\Omega>0$ then these will naturally be zero, making $J^{2}_{\mu\nu,e}\rightarrow 0$ and therefore seemingly demonstrating that the RWA predictions tend to the full model predictions in the infinite time limit. However, we must consider the exponentials outside the brackets. Since we are looking at the violations of causality near the surface of the light cone and the interaction lasts from $-T$ to $T$, we must set $\abs{\bm{x}}\approx 2T$, which is the leading edge of the detectors perturbation on the field. 

Therefore, when we consider the integrals in momentum space in equations \eqref{eq12c11} and \eqref{eq12c13}, the  oscillatory terms outside of the  $t_1,t_2$ integrals,  near the lightcone, go as $ e^{2\ii\omega T}- e^{-2\ii\omega T}$ (this difference between two exponentials emerges from $e^{\ii \bm k\cdot \bm x}$ after integrating the angular variables in momentum space). Therefore the terms of \eqref{c4} near the lightcone, approximately oscillate as:
\begin{align}
I=\underbrace{e^{\ii\omega T-\ii\bm{k}\cdot\bm{x}}}_{e^{-\ii\omega T}-e^{3\ii\omega T}}\overbrace{e^{\ii\omega' T-\ii\bm{k}'\cdot\bm{x}}}^{e^{-\ii\omega' T}-e^{3\ii\omega' T}}
\frac{4\ii}{\Omega-\omega'}\left(\underbrace{\frac{e^{\ii(\omega'-\Omega)T}\sin[(\omega+\Omega)T]}{\omega+\Omega}}_{e^{\ii(\omega+\omega') T}-e^{-\ii(\omega-\omega'+2\Omega) T}}
-\underbrace{\frac{\sin[(\omega+\omega')T]}{\omega+\omega'}}_{e^{\ii(\omega+\omega')T}-e^{-\ii(\omega+\omega')T}}\right).\label{aeeq5}
\end{align}
A quick inspection reveals that within $I$ there will be terms that oscillate slowly (or not at all), i.e. if $\abs{\bm{x}}=2T+\epsilon$, terms of the form $e^{\ii\omega \epsilon }$ will appear that oscillate slowly with respect to the significant sections of the smearing Fourier transforms (i.e. $\tilde{F}(\bm{k})$) and other terms within the $\bm{k},\bm{k}'$ integrals. This of course means that even if $T\rightarrow \infty$ the polynomial decay remains. With more rigorous working it can be shown that the non-locality introduced by the RWA Hamiltonian persists, polynomially decaying from the surface of the light cone/sphere, similar to the plots shown in the manuscript. This of course should not be surprising considering the explicit non-locality of the interaction Hamiltonian. 

Note, however, that if $\abs{\bm{x}}\ll 2T$ then the arguments above no longer hold and we can take the pointwise limit of Dirac $\delta$, i.e. for $\abs{\bm{x}}\ll 2T$: RWA$\rightarrow$UdW, as described in the next section.

\subsection{RWA convergence to the full model}

The derivation above showed that the second order counter-rotating terms do not vanish for long times near the light cone. One can ask under what conditions there are points where the second order counter-rotating terms do vanish.

Consider the counter-rotating contributions to the expectation of $\hat \phi^2$ \eqref{b37} which are given by the real part of \eqref{b33}. Consider a simple switching of duration $T$: 
\begin{align}
\chi(t)=\Theta(t+T)-\Theta(t-T),
\end{align}
and a spherically symmetric detector smearing. Then
\begin{align}
\begin{split}
M_{e}^{2}&=\frac{2(2\pi)^{2}}{\abs{\bm{x}}^{2}}\int\dd\omega\,\dd\omega'\,F(\omega)F(\omega')\bigg\{\\
&\frac{1}{(\omega'-\Omega)}\left[\frac{e^{\ii(\omega+\omega')\abs{\bm{x}}}-e^{\ii(\omega+\omega')(2T+\abs{\bm{x}})}}{(\omega+\omega')}-\frac{e^{2\ii(\omega'-\Omega)T+\ii(\omega+\omega')\abs{\bm{x}}}-e^{\ii(\omega+\omega')(2T+\abs{\bm{x}})}}{(\omega+\Omega)}\right]\\
&-\frac{1}{(\omega'-\Omega)}\left[\frac{e^{\ii(\omega-\omega')\abs{\bm{x}}}-e^{2\ii(\omega+\omega')T+\ii(\omega-\omega')\abs{\bm{x}}}}{(\omega+\omega')}-\frac{e^{2\ii(\omega'-\Omega)T+\ii(\omega-\omega')\abs{\bm{x}}}-e^{2\ii(\omega+\omega')T+\ii(\omega-\omega')\abs{\bm{x}}}}{(\omega+\Omega)}\right]\\
&-\frac{1}{(\omega'-\Omega)}\left[\frac{e^{-\ii(\omega-\omega')\abs{\bm{x}}}-e^{2\ii(\omega+\omega')T-\ii(\omega-\omega')\abs{\bm{x}}}}{(\omega+\omega')}-\frac{e^{2\ii(\omega'-\Omega)T-\ii(\omega-\omega')\abs{\bm{x}}}-e^{2\ii(\omega+\omega')T-\ii(\omega-\omega')\abs{\bm{x}}}}{(\omega+\Omega)}\right]\\
&+\frac{1}{(\omega'-\Omega)}\left[\frac{e^{-\ii(\omega+\omega')\abs{\bm{x}}}-e^{\ii(\omega+\omega')(2T-\abs{x})}}{(\omega+\omega')}-\frac{e^{2\ii(\omega'-\Omega)T-\ii(\omega+\omega')\abs{\bm{x}}}-e^{\ii(\omega+\omega')(2T-\abs{\bm{x}})}}{(\omega+\Omega)}\right]\bigg\}.
\end{split}
\end{align}

This can be decomposed in the following way:
\begin{equation}
    M_{e}^{2}=M_{e,\text{residual}}^{2}(T,\bm x)+M_{e,\text{uniform}}^{2}(T,\bm x),
\end{equation}
where

\begin{align}
M_{e,\text{residual}}^{2}&=\frac{2(2\pi)^{2}}{\abs{\bm{x}}^{2}}\int\dd\omega\dd\omega'\,\tilde{F}(\omega)\tilde{F}(\omega')\bigg\{\frac{e^{\ii(\omega+\omega')\abs{\bm{x}}}-e^{\ii(\omega-\omega')\abs{\bm{x}}}-e^{-\ii(\omega-\omega')\abs{\bm{x}}}+e^{-\ii(\omega+\omega')\abs{\bm{x}}}-e^{\ii(\omega+\omega')(2T-\abs{\bm{x}})}}{(\omega'-\Omega)(\omega+\omega')}\nonumber\\
&+\frac{e^{\ii(\omega+\omega')(2T-\abs{\bm{x}})}}{(\omega'-\Omega)(\omega+\Omega)}\bigg\},\\
\nonumber M_{e,\text{uniform}}^{2}&=\frac{2(2\pi)^{2}}{\abs{\bm{x}}^{2}}\int\dd\omega\dd\omega'\,\tilde{F}(\omega)\tilde{F}(\omega')\bigg\{\frac{-e^{\ii(\omega+\omega')(2T+\abs{\bm{x}})}+e^{\ii\omega(2T+\abs{\bm{x}})+\ii\omega'(2T-\abs{\bm{x}})}+e^{\ii\omega(2T-\abs{\bm{x}})+\ii\omega'(2T+\abs{\bm{x}})}}{(\omega'-\Omega)(\omega+\omega')}\\
&-\frac{e^{-2\ii\Omega T+\ii\omega\abs{\bm{x}}+\ii\omega'(2T+\abs{\bm{x}})}
-e^{\ii(\omega+\omega')(2T+\abs{\bm{x}})}
-e^{-2\ii\Omega T+\ii\omega\abs{\bm{x}}+\ii\omega'(2T-\abs{\bm{x}})}
+e^{\ii\omega(2T+\abs{\bm{x}})+\ii\omega'(2T-\abs{\bm{x}})}
}{(\omega'-\Omega)(\omega+\Omega)}\nonumber\\
&-\frac{
e^{-2\ii\Omega T-\ii\omega \abs{\bm{x}}+\ii\omega'(2T-\abs{\bm{x}})}
-e^{-2\ii\Omega T-\ii\omega\abs{\bm{x}}+\ii\omega'(2T+\abs{\bm{x}})}
+e^{\ii\omega(2T-\abs{\bm{x}})+\ii\omega'(2T+\abs{\bm{x}})}
}
{(\omega'-\Omega)(\omega+\Omega)}\bigg\}.
\end{align}

The complex exponentials in $M_{e,\text{uniform}}^{2}$  always become highly oscillatory in the limit of very long interaction $T\rightarrow\infty$ (irrespective of the choice of $\bm{x}$). This means that the handwavy argument that the contribution of the counter-rotating terms goes to zero in such a limit is correct for these terms: 
\begin{equation}
\lim_{T\to\infty}M_{e,\text{uniform}}^{2} =0.
\end{equation}
However, for the term $M_{e,\text{Residual}}^{2}$ there are values of $\bm x$ for which some of the complex exponentials in the integral become slowly oscillatory and provide finite contributions even in the limit of very long interaction times $T\to\infty$. In particular,  when $\bm x \approx 2T$ (around the boundary of the lightcone of the detector's interaction) some of the complex exponentials in the integral become slowly oscillatory and will not cancel in the long $T$ limit.

As a conclusion, we see that as we evaluate the field observables deeper into the lightcone of the detector's interaction (long interaction times but evaluating in the timelike are far from the detector's interaction lightcone front) the rotating wave approximation becomes more accurate. When we look at values closer to detector's lightcone boundary the approximation fails no matter how long the interaction time is. 


\section{RWA signalling - 2 detector perturbative expansion}
Here we go step by step over the 2 detector perturbative expansion, resulting in the reduced density matrix for 1 of the 2 detectors with the field completely traced out.

\begin{align}
\tilde{F}(\bm{k})&\coloneqq \int \dd^{3}\bm{y}\,\frac{1}{R^{3}}G\left(\frac{\bm{y}}{R}\right)e^{\ii\bm{k}\cdot\bm{y}}.
\end{align}
The Unruh-DeWitt interaction Hamiltonian has the form
\begin{align}
\hat{H}=\lambda\chi(t)\int\dd^{3}\bm{x}\,G(\bm{x})\hat{\sigma}_{x}(t)\int\frac{\dd^{3}\bm{k}}{(2\pi)^{3/2}\sqrt{2\omega}}\left(e^{-\ii\omega t+\ii\bm{k}\cdot\bm{x}}\hat{a}_{\bm{k}}^{\vphantom{\dagger}}+e^{\ii\omega t-\ii\bm{k}\cdot\bm{x}}\hat{a}_{\bm{k}}^{\dagger}\right).
\end{align}
In order to proceed we define the following, if Unruh-DeWitt coupling:
\begin{align}
\hat{\psi}_{i}=\lambda_{i}\int\frac{\dd^{3}\bm{k}}{(2\pi)^{3/2}\sqrt{2\omega}}\left(\tilde{F}_{i}(\bm{k})e^{-\ii\omega t}\hat{a}_{\bm{k}}^{\vphantom{\dagger}}+\tilde{F}_{i}^{*}(\bm{k})e^{\ii\omega t}\hat{a}_{\bm{k}}^{\dagger}\right),
\end{align}
if RWA coupling:
\begin{align}
\hat{\psi}_{i}=\lambda_{i}\int\frac{\dd^{3}\bm{k}}{(2\pi)^{3/2}\sqrt{2\omega}}\tilde{F}_{i}(\bm{k})e^{-\ii\omega t}\hat{a}_{\bm{k}}^{\vphantom{\dagger}},
\end{align}
c.f. \eqref{eq48}.

This way the interaction Hamiltonian becomes
\begin{align}
\hat{H}=\chi_{\textsc{a}}(t)\left(\hat{\sigma}^{+}_{\textsc{a}}(t)\hat{\psi}_{\textsc{a}}^{\vphantom{\dagger}}+\hat{\sigma}_{\textsc{a}}^{-}(t)\hat{\psi}_{\textsc{a}}^{\dagger}\right)+\chi_{\textsc{b}}(t)\left(\hat{\sigma}^{+}_{\textsc{b}}(t)\hat{\psi}_{\textsc{b}}^{\vphantom{\dagger}}+\hat{\sigma}_{\textsc{b}}^{-}(t)\hat{\psi}_{\textsc{b}}^{\dagger}\right).
\end{align}

The time evolution operator then looks like
\begin{align}
\begin{split}
\hat{U}(t)&=\hat{\openone}-\ii\int\limits_{-\infty}^{\infty}\dd  t_{1} \left(\chi_{\textsc{a}}( t_{1} )\left(\hat{\sigma}_{\textsc{a}}^{+}( t_{1} )\hat{\psi}_{\textsc{a}}^{\vphantom{\dagger}}( t_{1} )+\hat{\sigma}_{\textsc{a}}^{-}( t_{1} )\hat{\psi}_{\textsc{a}}^{\dagger}( t_{1} )\right)+
\chi_{\textsc{b}}( t_{1} )\left(\hat{\sigma}_{\textsc{b}}^{+}( t_{1} )\hat{\psi}_{\textsc{b}}^{\vphantom{\dagger}}( t_{1} )+\hat{\sigma}_{\textsc{b}}^{-}( t_{1} )\hat{\psi}_{\textsc{b}}^{\dagger}( t_{1} )\right)
\right)\\
&-\int\limits_{-\infty}^{\infty}\dd  t_{1}  \int\limits_{-\infty}^{ t_{1} }\dd  t_{2} 
\left(\chi_{\textsc{a}}( t_{1} )\left(\hat{\sigma}_{\textsc{a}}^{+}( t_{1} )\hat{\psi}_{\textsc{a}}^{\vphantom{\dagger}}( t_{1} )+\hat{\sigma}_{\textsc{a}}^{-}( t_{1} )\hat{\psi}_{\textsc{a}}^{\dagger}( t_{1} )\right)+
\chi_{\textsc{b}}( t_{1} )\left(\hat{\sigma}_{\textsc{b}}^{+}( t_{1} )\hat{\psi}_{\textsc{b}}^{\vphantom{\dagger}}( t_{1} )+\hat{\sigma}_{\textsc{b}}^{-}( t_{1} )\hat{\psi}_{\textsc{b}}^{\dagger}( t_{1} )\right)
\right)\\
&
\left(\chi_{\textsc{a}}( t_{2} )\left(\hat{\sigma}_{\textsc{a}}^{+}( t_{2} )\hat{\psi}_{\textsc{a}}^{\vphantom{\dagger}}( t_{2} )+\hat{\sigma}_{\textsc{a}}^{-}( t_{2} )\hat{\psi}_{\textsc{a}}^{\dagger}( t_{2} )\right)+
\chi_{\textsc{b}}( t_{2} )\left(\hat{\sigma}_{\textsc{b}}^{+}( t_{2} )\hat{\psi}_{\textsc{b}}^{\vphantom{\dagger}}( t_{2} )+\hat{\sigma}_{\textsc{b}}^{-}( t_{2} )\hat{\psi}_{\textsc{b}}^{\dagger}( t_{2} )\right)
\right)+\mathcal{O}(\lambda^{3}).
\end{split}
\end{align}

By assuming the initial field state is the vacuum and the initial detector states is $\hat{\rho}_{0}$, the reduced detector density matrix becomes

\begin{align}
\begin{split}
\hat{\rho}_{q}(t)&=\hat{\rho}_{0}+\int\limits_{-\infty}^{\infty}\dd  t_{1} \int\limits_{-\infty}^{\infty}\dd  t_{2} \bigg\{\\
&\chi_{\textsc{a}}( t_{1} )\chi_{\textsc{a}}( t_{2} )\bigg(\hat{\sigma}_{\textsc{a}}^{+}\hat{\rho}_{0}\hat{\sigma}_{\textsc{a}}^{+}e^{\ii\Omega_{\textsc{a}}( t_{1} + t_{2} )}\expect{\hat{\psi}^{\vphantom{\dagger}}_{\textsc{a}}( t_{2} )\hat{\psi}^{\vphantom{\dagger}}_{\textsc{a}}( t_{1} )}
+\hat{\sigma}_{\textsc{a}}^{+}\hat{\rho}_{0}\hat{\sigma}_{\textsc{a}}^{-}e^{\ii\Omega_{\textsc{a}}( t_{1} - t_{2} )}\expect{\hat{\psi}^{\dagger}_{\textsc{a}}( t_{2} )\hat{\psi}_{\textsc{a}}^{\vphantom{\dagger}}( t_{1} )}\\
+&\hat{\sigma}_{\textsc{a}}^{-}\hat{\rho}_{0}\hat{\sigma}_{\textsc{a}}^{+}e^{-\ii\Omega_{\textsc{a}}( t_{1} - t_{2} )}\expect{\hat{\psi}^{\vphantom{\dagger}}_{\textsc{a}}( t_{2} )\hat{\psi}_{\textsc{a}}^{\dagger}( t_{1} )}
+\hat{\sigma}_{\textsc{a}}^{-}\hat{\rho}_{0}\hat{\sigma}_{\textsc{a}}^{-}e^{-\ii\Omega_{\textsc{a}}( t_{1} + t_{2} )}\expect{\hat{\psi}^{\dagger}_{\textsc{a}}( t_{2} )\hat{\psi}_{\textsc{a}}^{\dagger}( t_{1} )}\bigg)\\
+&\chi_{\textsc{a}}( t_{1} )\chi_{\textsc{b}}( t_{2} )\bigg(\hat{\sigma}_{\textsc{a}}^{+}\hat{\rho}_{0}\hat{\sigma}_{\textsc{b}}^{+}e^{\ii(\Omega_{\textsc{b}} t_{2} +\Omega_{\textsc{a}} t_{1} )}\expect{\hat{\psi}^{\vphantom{\dagger}}_{\textsc{b}}( t_{2} )\hat{\psi}^{\vphantom{\dagger}}_{\textsc{a}}( t_{1} )}
+\hat{\sigma}_{\textsc{a}}^{+}\hat{\rho}_{0}\hat{\sigma}_{\textsc{b}}^{-}e^{-\ii(\Omega_{\textsc{b}} t_{2} -\Omega_{\textsc{a}} t_{1} )}\expect{\hat{\psi}^{\dagger}_{\textsc{b}}( t_{2} )\hat{\psi}_{\textsc{a}}^{\vphantom{\dagger}}( t_{1} )}\\
+&\hat{\sigma}_{\textsc{a}}^{-}\hat{\rho}_{0}\hat{\sigma}_{\textsc{b}}^{+}e^{\ii(\Omega_{\textsc{b}} t_{2} -\Omega_{\textsc{a}} t_{1} )}\expect{\hat{\psi}^{\vphantom{\dagger}}_{\textsc{b}}( t_{2} )\hat{\psi}_{\textsc{a}}^{\dagger}( t_{1} )}
+\hat{\sigma}_{\textsc{a}}^{-}\hat{\rho}_{0}\hat{\sigma}_{\textsc{b}}^{-}e^{-\ii(\Omega_{\textsc{b}} t_{2} +\Omega_{\textsc{a}} t_{1} )}\expect{\hat{\psi}^{\dagger}_{\textsc{b}}( t_{2} )\hat{\psi}_{\textsc{a}}^{\dagger}( t_{1} )}\bigg)\\
+&\chi_{\textsc{b}}( t_{1} )\chi_{\textsc{a}}( t_{2} )\bigg(\hat{\sigma}_{\textsc{b}}^{+}\hat{\rho}_{0}\hat{\sigma}_{\textsc{a}}^{+}e^{\ii(\Omega_{\textsc{a}} t_{2} +\Omega_{\textsc{b}} t_{1} )}\expect{\hat{\psi}^{\vphantom{\dagger}}_{\textsc{a}}( t_{2} )\hat{\psi}^{\vphantom{\dagger}}_{\textsc{b}}( t_{1} )}
+\hat{\sigma}_{\textsc{b}}^{+}\hat{\rho}_{0}\hat{\sigma}_{\textsc{a}}^{-}e^{-\ii(\Omega_{\textsc{a}} t_{2} -\Omega_{\textsc{b}} t_{1} )}\expect{\hat{\psi}^{\dagger}_{\textsc{a}}( t_{2} )\hat{\psi}_{\textsc{b}}^{\vphantom{\dagger}}( t_{1} )}\\
+&\hat{\sigma}_{\textsc{b}}^{-}\hat{\rho}_{0}\hat{\sigma}_{\textsc{a}}^{+}e^{\ii(\Omega_{\textsc{a}} t_{2} -\Omega_{\textsc{b}} t_{1} )}\expect{\hat{\psi}^{\vphantom{\dagger}}_{\textsc{a}}( t_{2} )\hat{\psi}_{\textsc{b}}^{\dagger}( t_{1} )}
+\hat{\sigma}_{\textsc{b}}^{-}\hat{\rho}_{0}\hat{\sigma}_{\textsc{a}}^{-}e^{-\ii(\Omega_{\textsc{a}} t_{2} +\Omega_{\textsc{b}} t_{1} )}\expect{\hat{\psi}^{\dagger}_{\textsc{a}}( t_{2} )\hat{\psi}_{\textsc{b}}^{\dagger}( t_{1} )}\bigg)\\
+&\chi_{\textsc{b}}( t_{1} )\chi_{\textsc{b}}( t_{2} )\bigg(\hat{\sigma}_{\textsc{b}}^{+}\hat{\rho}_{0}\hat{\sigma}_{\textsc{b}}^{+}e^{\ii\Omega_{\textsc{b}}( t_{1} + t_{2} )}\expect{\hat{\psi}^{\vphantom{\dagger}}_{\textsc{b}}( t_{2} )\hat{\psi}^{\vphantom{\dagger}}_{\textsc{b}}( t_{1} )}
+\hat{\sigma}_{\textsc{b}}^{+}\hat{\rho}_{0}\hat{\sigma}_{\textsc{b}}^{-}e^{\ii\Omega_{\textsc{b}}( t_{1} - t_{2} )}\expect{\hat{\psi}^{\dagger}_{\textsc{b}}( t_{2} )\hat{\psi}_{\textsc{b}}^{\vphantom{\dagger}}( t_{1} )}\\
+&\hat{\sigma}_{\textsc{b}}^{-}\hat{\rho}_{0}\hat{\sigma}_{\textsc{b}}^{+}e^{-\ii\Omega_{\textsc{b}}( t_{1} - t_{2} )}\expect{\hat{\psi}^{\vphantom{\dagger}}_{\textsc{b}}( t_{2} )\hat{\psi}_{\textsc{b}}^{\dagger}( t_{1} )}
+\hat{\sigma}_{\textsc{b}}^{-}\hat{\rho}_{0}\hat{\sigma}_{\textsc{b}}^{-}e^{-\ii\Omega_{\textsc{b}}( t_{1} + t_{2} )}\expect{\hat{\psi}^{\dagger}_{\textsc{b}}( t_{2} )\hat{\psi}_{\textsc{b}}^{\dagger}( t_{1} )}\bigg)\bigg\}\\
-&\int\limits_{-\infty}^{\infty}\dd  t_{1}  \int\limits_{-\infty}^{ t_{1} }\dd  t_{2} \bigg\{\\
&\chi_{\textsc{a}}( t_{1} )\chi_{\textsc{a}}( t_{2} )\bigg(\hat{\Pi}_{e}^{1}\hat{\rho}_{0}e^{\ii\Omega_{\textsc{a}}( t_{1} - t_{2} )}\expect{\hat{\psi}_{\textsc{a}}^{\vphantom{\dagger}}( t_{1} )\hat{\psi}_{\textsc{a}}^{\dagger}( t_{2} )}
+\hat{\Pi}_{g}^{1}\hat{\rho}_{0}e^{-\ii\Omega_{\textsc{a}}( t_{1} - t_{2} )}\expect{\hat{\psi}^{\dagger}_{\textsc{a}}( t_{1} )\hat{\psi}_{\textsc{a}}^{\vphantom{\dagger}}( t_{2} )}
+\hat{\rho}_{0}\hat{\Pi}_{e}^{1}e^{-\ii\Omega_{\textsc{a}}( t_{1} - t_{2} )}\expect{\hat{\psi}_{\textsc{a}}^{\vphantom{\dagger}}( t_{2} )\hat{\psi}_{\textsc{a}}^{\dagger}( t_{1} )}\\
+&\hat{\rho}_{0}\hat{\Pi}_{g}^{1}e^{\ii\Omega_{\textsc{a}}( t_{1} - t_{2} )}\expect{\hat{\psi}_{\textsc{a}}^{\dagger}( t_{2} )\psi_{\textsc{a}}^{\vphantom{\dagger}}( t_{1} )}\bigg)\\
+&\chi_{\textsc{b}}( t_{1} )\chi_{\textsc{a}}( t_{2} )\bigg(
\hat{\sigma}_{\textsc{b}}^{+}\hat{\sigma}_{\textsc{a}}^{+}\hat{\rho}_{0}e^{\ii(\Omega_{\textsc{a}} t_{2} +\Omega_{\textsc{b}} t_{1} )}\expect{\hat{\psi}^{\vphantom{\dagger}}_{\textsc{b}}( t_{1} )\hat{\psi}_{\textsc{a}}^{\vphantom{\dagger}}( t_{2} )}
+\hat{\sigma}_{\textsc{b}}^{-}\hat{\sigma}_{\textsc{a}}^{+}\hat{\rho}_{0}e^{\ii(\Omega_{\textsc{a}} t_{2} -\Omega_{\textsc{b}} t_{1} )}\expect{\hat{\psi}^{\dagger}_{\textsc{b}}( t_{1} )\hat{\psi}_{\textsc{a}}^{\vphantom{\dagger}}( t_{2} )}\\
+&\hat{\sigma}_{\textsc{b}}^{+}\hat{\sigma}_{\textsc{a}}^{-}\hat{\rho}_{0}e^{-\ii(\Omega_{\textsc{a}} t_{2} -\Omega_{\textsc{b}} t_{1} )}\expect{\hat{\psi}^{\vphantom{\dagger}}_{\textsc{b}}( t_{1} )\hat{\psi}_{\textsc{a}}^{\dagger}( t_{2} )}
+\hat{\sigma}_{\textsc{b}}^{-}\hat{\sigma}_{\textsc{a}}^{-}\hat{\rho}_{0}e^{-\ii(\Omega_{\textsc{a}} t_{2} +\Omega_{\textsc{b}} t_{1} )}\expect{\hat{\psi}^{\dagger}_{\textsc{b}}( t_{1} )\hat{\psi}_{\textsc{a}}^{\dagger}( t_{2} )}
+\hat{\rho}_{0}\hat{\sigma}_{\textsc{a}}^{-}\hat{\sigma}_{\textsc{b}}^{-}e^{-\ii(\Omega_{\textsc{a}} t_{2} +\Omega_{\textsc{b}} t_{1} )}\expect{\hat{\psi}_{\textsc{a}}^{\dagger}( t_{2} )\hat{\psi}_{\textsc{b}}^{\dagger}( t_{1} )}\\
+&\hat{\rho}_{0}\hat{\sigma}_{\textsc{a}}^{-}\hat{\sigma}_{\textsc{b}}^{+}e^{-\ii(\Omega_{\textsc{a}} t_{2} -\Omega_{\textsc{b}} t_{1} )}\expect{\hat{\psi}_{\textsc{a}}^{\dagger}( t_{2} )\hat{\psi}_{\textsc{b}}^{\vphantom{\dagger}}( t_{1} )}
+\hat{\rho}_{0}\hat{\sigma}_{\textsc{a}}^{+}\hat{\sigma}_{\textsc{b}}^{-}e^{\ii(\Omega_{\textsc{a}} t_{2} -\Omega_{\textsc{b}} t_{1} )}\expect{\hat{\psi}_{\textsc{a}}^{\vphantom{\dagger}}( t_{2} )\hat{\psi}_{\textsc{b}}^{\dagger}( t_{1} )}
+\hat{\rho}_{0}\hat{\sigma}_{\textsc{a}}^{+}\hat{\sigma}_{\textsc{b}}^{+}e^{\ii(\Omega_{\textsc{a}} t_{2} +\Omega_{\textsc{b}} t_{1} )}\expect{\hat{\psi}_{\textsc{a}}^{\vphantom{\dagger}}( t_{2} )\hat{\psi}_{\textsc{b}}^{\vphantom{\dagger}}( t_{1} )}
\bigg)\\
+&\chi_{\textsc{a}}( t_{1} )\chi_{\textsc{b}}( t_{2} )\bigg(
\hat{\sigma}_{\textsc{a}}^{+}\hat{\sigma}_{\textsc{b}}^{+}\hat{\rho}_{0}e^{\ii(\Omega_{\textsc{a}} t_{1} +\Omega_{\textsc{b}} t_{2} )}\expect{\hat{\psi}^{\vphantom{\dagger}}_{\textsc{a}}( t_{1} )\hat{\psi}_{\textsc{b}}^{\vphantom{\dagger}}( t_{2} )}
+\hat{\sigma}_{\textsc{a}}^{-}\hat{\sigma}_{\textsc{b}}^{+}\hat{\rho}_{0}e^{-\ii(\Omega_{\textsc{a}} t_{1} -\Omega_{\textsc{b}} t_{2} )}\expect{\hat{\psi}^{\dagger}_{\textsc{a}}( t_{1} )\hat{\psi}_{\textsc{b}}^{\vphantom{\dagger}}( t_{2} )}\\
+&\hat{\sigma}_{\textsc{a}}^{+}\hat{\sigma}_{\textsc{b}}^{-}\hat{\rho}_{0}e^{\ii(\Omega_{\textsc{a}} t_{1} -\Omega_{\textsc{b}} t_{2} )}\expect{\hat{\psi}^{\vphantom{\dagger}}_{\textsc{a}}( t_{1} )\hat{\psi}_{\textsc{b}}^{\dagger}( t_{2} )}
+\hat{\sigma}_{\textsc{a}}^{-}\hat{\sigma}_{\textsc{b}}^{-}\hat{\rho}_{0}e^{-\ii(\Omega_{\textsc{a}} t_{1} +\Omega_{\textsc{b}} t_{2} )}\expect{\hat{\psi}^{\dagger}_{\textsc{a}}( t_{1} )\hat{\psi}_{\textsc{b}}^{\dagger}( t_{2} )}
+\hat{\rho}_{0}\hat{\sigma}_{\textsc{b}}^{-}\hat{\sigma}_{\textsc{a}}^{-}e^{-\ii(\Omega_{\textsc{a}} t_{1} +\Omega_{\textsc{b}} t_{2} )}\expect{\hat{\psi}_{\textsc{b}}^{\dagger}( t_{2} )\hat{\psi}_{\textsc{a}}^{\dagger}( t_{1} )}\\
+&\hat{\rho}_{0}\hat{\sigma}_{\textsc{b}}^{-}\hat{\sigma}_{\textsc{a}}^{+}e^{\ii(\Omega_{\textsc{a}} t_{1} -\Omega_{\textsc{b}} t_{2} )}\expect{\hat{\psi}_{\textsc{b}}^{\dagger}( t_{2} )\hat{\psi}_{\textsc{a}}^{\vphantom{\dagger}}( t_{1} )}
+\hat{\rho}_{0}\hat{\sigma}_{\textsc{b}}^{+}\hat{\sigma}_{\textsc{a}}^{-}e^{-\ii(\Omega_{\textsc{a}} t_{1} -\Omega_{\textsc{b}} t_{2} )}\expect{\hat{\psi}_{\textsc{b}}^{\vphantom{\dagger}}( t_{2} )\hat{\psi}_{\textsc{a}}^{\dagger}( t_{1} )}
+\hat{\rho}_{0}\hat{\sigma}_{\textsc{b}}^{+}\hat{\sigma}_{\textsc{a}}^{+}e^{\ii(\Omega_{\textsc{a}} t_{1} +\Omega_{\textsc{b}} t_{2} )}\expect{\hat{\psi}_{\textsc{b}}^{\vphantom{\dagger}}( t_{2} )\hat{\psi}_{\textsc{a}}^{\vphantom{\dagger}}( t_{1} )}
\bigg)\\
+&\chi_{\textsc{b}}( t_{1} )\chi_{\textsc{b}}( t_{2} )\bigg(\hat{\Pi}_{e}^{2}\hat{\rho}_{0}e^{\ii\Omega_{\textsc{b}}( t_{1} - t_{2} )}\expect{\hat{\psi}_{\textsc{b}}^{\vphantom{\dagger}}( t_{1} )\hat{\psi}_{\textsc{b}}^{\dagger}( t_{2} )}
+\hat{\Pi}_{g}^{2}\hat{\rho}_{0}e^{-\ii\Omega_{\textsc{b}}( t_{1} - t_{2} )}\expect{\hat{\psi}^{\dagger}_{\textsc{b}}( t_{1} )\hat{\psi}_{\textsc{b}}^{\vphantom{\dagger}}( t_{2} )}
+\hat{\rho}_{0}\hat{\Pi}_{e}^{2}e^{-\ii\Omega_{\textsc{b}}( t_{1} - t_{2} )}\expect{\hat{\psi}_{\textsc{b}}^{\vphantom{\dagger}}( t_{2} )\hat{\psi}_{\textsc{b}}^{\dagger}( t_{1} )}\\
+&\hat{\rho}_{0}\hat{\Pi}_{g}^{2}e^{\ii\Omega_{\textsc{b}}( t_{1} - t_{2} )}\expect{\hat{\psi}_{\textsc{b}}^{\dagger}( t_{2} )\psi_{\textsc{b}}^{\vphantom{\dagger}}( t_{1} )}\bigg)\bigg\}+\mathcal{O}(\lambda_{i}^{3})
\end{split}
\end{align}

Now in order to gauge the non-local effects we consider the following scenario, $\chi_{\textsc{b}}$ occurs before $\chi_{\textsc{a}}$ and in our frame of reference their supports do not overlap, same as in \cite{PhysRevD.92.104019}. This allows us to eliminate the terms $\chi_{\textsc{b}}( t_{1} )\chi_{\textsc{a}}( t_{2} )$ from the ordered integral and allows us to compare terms from the integrals that proportional to $\lambda_{\textsc{a}}\lambda_{\textsc{b}}$, note that our choice of switching means that the time-ordering becomes trivial for $\lambda_{\textsc{a}}\lambda_{\textsc{b}}$ terms. Furthermore we trace out the second detector and inspect the first detector's density matrix,

\begin{align}
\begin{split}
\hat{\rho}_{\textsc{a}}(t)&=\hat{\rho}_{\textsc{a}}^{0}+\mathcal{O}(\lambda_{\textsc{a}}^{2})+\mathcal{O}(\lambda_{\textsc{b}}^{2})+\int\limits_{-\infty}^{\infty}\dd  t_{1}\int\limits_{-\infty}^{\infty} \dd  t_{2} \bigg\{\chi_{\textsc{a}}( t_{1} )\chi_{\textsc{b}}( t_{2} )\\
&\Tr_{\textsc{b}}\bigg(\hat{\sigma}_{\textsc{a}}^{+}\hat{\rho}_{0}\hat{\sigma}_{\textsc{b}}^{+}\expect{\left[\hat{\psi}^{\vphantom{\dagger}}_{\textsc{b}}( t_{2} ),\hat{\psi}^{\vphantom{\dagger}}_{\textsc{a}}( t_{1} )\right]}
+\hat{\sigma}_{\textsc{a}}^{+}\hat{\rho}_{0}\hat{\sigma}_{\textsc{b}}^{-}\expect{\left[\hat{\psi}^{\dagger}_{\textsc{b}}( t_{2} ),\hat{\psi}_{\textsc{a}}^{\vphantom{\dagger}}( t_{1} )\right]}
+\hat{\sigma}_{\textsc{a}}^{-}\hat{\rho}_{0}\hat{\sigma}_{\textsc{b}}^{+}\expect{\left[\hat{\psi}^{\vphantom{\dagger}}_{\textsc{b}}( t_{2} ),\hat{\psi}_{\textsc{a}}^{\dagger}( t_{1} )\right]}\\
+&\hat{\sigma}_{\textsc{a}}^{-}\hat{\rho}_{0}\hat{\sigma}_{\textsc{b}}^{-}\expect{\left[\hat{\psi}^{\dagger}_{\textsc{b}}( t_{2} ),\hat{\psi}_{\textsc{a}}^{\dagger}( t_{1} )\right]}\bigg)\\
+&\Tr_{\textsc{b}}\bigg(
\hat{\sigma}_{\textsc{b}}^{+}\hat{\rho}_{0}\hat{\sigma}_{\textsc{a}}^{+}\expect{\left[\hat{\psi}^{\vphantom{\dagger}}_{\textsc{a}}( t_{1} ),\hat{\psi}^{\vphantom{\dagger}}_{\textsc{b}}( t_{2} )\right]}
+\hat{\sigma}_{\textsc{b}}^{+}\hat{\rho}_{0}\hat{\sigma}_{\textsc{a}}^{-}\expect{\left[\hat{\psi}^{\dagger}_{\textsc{a}}( t_{1} ),\hat{\psi}_{\textsc{b}}^{\vphantom{\dagger}}( t_{2} )\right]}
+\hat{\sigma}_{\textsc{b}}^{-}\hat{\rho}_{0}\hat{\sigma}_{\textsc{a}}^{+}\expect{\left[\hat{\psi}^{\vphantom{\dagger}}_{\textsc{a}}( t_{1} ),\hat{\psi}_{\textsc{b}}^{\dagger}( t_{2} )\right]}\\
+&\hat{\sigma}_{\textsc{b}}^{-}\hat{\rho}_{0}\hat{\sigma}_{\textsc{a}}^{-}\expect{\left[\hat{\psi}^{\dagger}_{\textsc{a}}( t_{1} ),\hat{\psi}_{\textsc{b}}^{\dagger}( t_{2} )\right]}\bigg)+\mathcal{O}(\lambda_{i}^{3}).
\end{split}
\end{align}

In this last step we have used the cyclic property of the partial trace to arrange terms nicely. All that remains is to evaluate the commutators, all of which can be accomplished easily, as shown in \eqref{eq51} and \eqref{eq52}.

\twocolumngrid
\bibliography{bibliography}

\providecommand{\noopsort}[1]{}\providecommand{\singleletter}[1]{#1}%
  \newcommand{\noop}[1]{}
\begin{thebibliography}{26}%
\makeatletter
\providecommand \@ifxundefined [1]{%
 \@ifx{#1\undefined}
}%
\providecommand \@ifnum [1]{%
 \ifnum #1\expandafter \@firstoftwo
 \else \expandafter \@secondoftwo
 \fi
}%
\providecommand \@ifx [1]{%
 \ifx #1\expandafter \@firstoftwo
 \else \expandafter \@secondoftwo
 \fi
}%
\providecommand \natexlab [1]{#1}%
\providecommand \enquote  [1]{``#1''}%
\providecommand \bibnamefont  [1]{#1}%
\providecommand \bibfnamefont [1]{#1}%
\providecommand \citenamefont [1]{#1}%
\providecommand \href@noop [0]{\@secondoftwo}%
\providecommand \href [0]{\begingroup \@sanitize@url \@href}%
\providecommand \@href[1]{\@@startlink{#1}\@@href}%
\providecommand \@@href[1]{\endgroup#1\@@endlink}%
\providecommand \@sanitize@url [0]{\catcode `\\12\catcode `\$12\catcode
  `\&12\catcode `\#12\catcode `\^12\catcode `\_12\catcode `\%12\relax}%
\providecommand \@@startlink[1]{}%
\providecommand \@@endlink[0]{}%
\providecommand \url  [0]{\begingroup\@sanitize@url \@url }%
\providecommand \@url [1]{\endgroup\@href {#1}{\urlprefix }}%
\providecommand \urlprefix  [0]{URL }%
\providecommand \Eprint [0]{\href }%
\providecommand \doibase [0]{http://dx.doi.org/}%
\providecommand \selectlanguage [0]{\@gobble}%
\providecommand \bibinfo  [0]{\@secondoftwo}%
\providecommand \bibfield  [0]{\@secondoftwo}%
\providecommand \translation [1]{[#1]}%
\providecommand \BibitemOpen [0]{}%
\providecommand \bibitemStop [0]{}%
\providecommand \bibitemNoStop [0]{.\EOS\space}%
\providecommand \EOS [0]{\spacefactor3000\relax}%
\providecommand \BibitemShut  [1]{\csname bibitem#1\endcsname}%
\let\auto@bib@innerbib\@empty
\bibitem [{\citenamefont {{G{\"o}ppert-Mayer}}(1931)}]{1931AnP...401..273G}%
  \BibitemOpen
  \bibfield  {author} {\bibinfo {author} {\bibfnamefont {M.}~\bibnamefont
  {{G{\"o}ppert-Mayer}}},\ }\href {\doibase 10.1002/andp.19314010303}
  {\bibfield  {journal} {\bibinfo  {journal} {Annalen der Physik}\ }\textbf
  {\bibinfo {volume} {401}},\ \bibinfo {pages} {273} (\bibinfo {year}
  {1931})}\BibitemShut {NoStop}%
\bibitem [{\citenamefont {Pozas-Kerstjens}\ and\ \citenamefont
  {Mart\'{\i}n-Mart\'{\i}nez}(2016)}]{PhysRevD.94.064074}%
  \BibitemOpen
  \bibfield  {author} {\bibinfo {author} {\bibfnamefont {A.}~\bibnamefont
  {Pozas-Kerstjens}}\ and\ \bibinfo {author} {\bibfnamefont {E.}~\bibnamefont
  {Mart\'{\i}n-Mart\'{\i}nez}},\ }\href {\doibase 10.1103/PhysRevD.94.064074}
  {\bibfield  {journal} {\bibinfo  {journal} {Phys. Rev. D}\ }\textbf {\bibinfo
  {volume} {94}},\ \bibinfo {pages} {064074} (\bibinfo {year}
  {2016})}\BibitemShut {NoStop}%
\bibitem [{\citenamefont {Mart\'{\i}n-Mart\'{\i}nez}\ and\ \citenamefont
  {Rodriguez-Lopez}(2018)}]{Pablo}%
  \BibitemOpen
  \bibfield  {author} {\bibinfo {author} {\bibfnamefont {E.}~\bibnamefont
  {Mart\'{\i}n-Mart\'{\i}nez}}\ and\ \bibinfo {author} {\bibfnamefont
  {P.}~\bibnamefont {Rodriguez-Lopez}},\ }\href {\doibase
  10.1103/PhysRevD.97.105026} {\bibfield  {journal} {\bibinfo  {journal} {Phys.
  Rev. D}\ }\textbf {\bibinfo {volume} {97}},\ \bibinfo {pages} {105026}
  (\bibinfo {year} {2018})}\BibitemShut {NoStop}%
\bibitem [{\citenamefont {Scully}\ and\ \citenamefont
  {Zubairy}(1997)}]{Scully}%
  \BibitemOpen
  \bibfield  {author} {\bibinfo {author} {\bibfnamefont {M.~O.}\ \bibnamefont
  {Scully}}\ and\ \bibinfo {author} {\bibfnamefont {M.~S.}\ \bibnamefont
  {Zubairy}},\ }\href@noop {} {\emph {\bibinfo {title} {Quantum Optics}}}\
  (\bibinfo  {publisher} {Cambridge University Press},\ \bibinfo {year}
  {1997})\BibitemShut {NoStop}%
\bibitem [{\citenamefont {Dirac}(1927)}]{Dirac1927}%
  \BibitemOpen
  \bibfield  {author} {\bibinfo {author} {\bibfnamefont {P.~A.~M.}\
  \bibnamefont {Dirac}},\ }\href {\doibase 10.1098/rspa.1927.0039} {\bibfield
  {journal} {\bibinfo  {journal} {Proc. Royal Soc. A: Math., Phys. and Eng.
  Sci,}\ }\textbf {\bibinfo {volume} {114}},\ \bibinfo {pages} {243} (\bibinfo
  {year} {1927})}\BibitemShut {NoStop}%
\bibitem [{\citenamefont {Jonsson}\ \emph {et~al.}(2014)\citenamefont
  {Jonsson}, \citenamefont {Mart\'{\i}n-Mart\'{\i}nez},\ and\ \citenamefont
  {Kempf}}]{PhysRevA.89.022330}%
  \BibitemOpen
  \bibfield  {author} {\bibinfo {author} {\bibfnamefont {R.~H.}\ \bibnamefont
  {Jonsson}}, \bibinfo {author} {\bibfnamefont {E.}~\bibnamefont
  {Mart\'{\i}n-Mart\'{\i}nez}}, \ and\ \bibinfo {author} {\bibfnamefont
  {A.}~\bibnamefont {Kempf}},\ }\href {\doibase 10.1103/PhysRevA.89.022330}
  {\bibfield  {journal} {\bibinfo  {journal} {Phys. Rev. A}\ }\textbf {\bibinfo
  {volume} {89}},\ \bibinfo {pages} {022330} (\bibinfo {year}
  {2014})}\BibitemShut {NoStop}%
\bibitem [{\citenamefont
  {Mart\'{\i}n-Mart\'{\i}nez}(2015)}]{PhysRevD.92.104019}%
  \BibitemOpen
  \bibfield  {author} {\bibinfo {author} {\bibfnamefont {E.}~\bibnamefont
  {Mart\'{\i}n-Mart\'{\i}nez}},\ }\href {\doibase 10.1103/PhysRevD.92.104019}
  {\bibfield  {journal} {\bibinfo  {journal} {Phys. Rev. D}\ }\textbf {\bibinfo
  {volume} {92}},\ \bibinfo {pages} {104019} (\bibinfo {year}
  {2015})}\BibitemShut {NoStop}%
\bibitem [{\citenamefont {Fermi}(1932)}]{RevModPhys.4.87}%
  \BibitemOpen
  \bibfield  {author} {\bibinfo {author} {\bibfnamefont {E.}~\bibnamefont
  {Fermi}},\ }\href {\doibase 10.1103/RevModPhys.4.87} {\bibfield  {journal}
  {\bibinfo  {journal} {Rev. Mod. Phys.}\ }\textbf {\bibinfo {volume} {4}},\
  \bibinfo {pages} {87} (\bibinfo {year} {1932})}\BibitemShut {NoStop}%
\bibitem [{\citenamefont {Kirchhoff}(2009)}]{kirchhoff2009abhandlungen}%
  \BibitemOpen
  \bibfield  {author} {\bibinfo {author} {\bibfnamefont {M.}~\bibnamefont
  {Kirchhoff}},\ }\href {https://books.google.ca/books?id=XDIXZiXWIuEC} {\emph
  {\bibinfo {title} {Abhandlungen Ber Emission Und Absorption: I. Ueber Die
  Fraunhofer'schen Linien.(1859.) II. Ueber de}}}\ (\bibinfo  {publisher}
  {BiblioBazaar},\ \bibinfo {year} {2009})\BibitemShut {NoStop}%
\bibitem [{\citenamefont {Schmidt}\ \emph {et~al.}(2013)\citenamefont
  {Schmidt}, \citenamefont {Blatter},\ and\ \citenamefont
  {Keeling}}]{0953-4075-46-22-224020}%
  \BibitemOpen
  \bibfield  {author} {\bibinfo {author} {\bibfnamefont {S.}~\bibnamefont
  {Schmidt}}, \bibinfo {author} {\bibfnamefont {G.}~\bibnamefont {Blatter}}, \
  and\ \bibinfo {author} {\bibfnamefont {J.}~\bibnamefont {Keeling}},\ }\href
  {http://stacks.iop.org/0953-4075/46/i=22/a=224020} {\bibfield  {journal}
  {\bibinfo  {journal} {Journal of Physics B: Atomic, Molecular and Optical
  Physics}\ }\textbf {\bibinfo {volume} {46}},\ \bibinfo {pages} {224020}
  (\bibinfo {year} {2013})}\BibitemShut {NoStop}%
\bibitem [{\citenamefont {Meystre}(1992)}]{MEYSTRE1992243}%
  \BibitemOpen
  \bibfield  {author} {\bibinfo {author} {\bibfnamefont {P.}~\bibnamefont
  {Meystre}},\ }\href {\doibase 10.1016/0370-1573(92)90141-L} {\bibfield
  {journal} {\bibinfo  {journal} {Physics Reports}\ }\textbf {\bibinfo {volume}
  {219}},\ \bibinfo {pages} {243} (\bibinfo {year} {1992})}\BibitemShut
  {NoStop}%
\bibitem [{\citenamefont {Compagno}\ \emph {et~al.}(1989)\citenamefont
  {Compagno}, \citenamefont {Palma}, \citenamefont {Passante},\ and\
  \citenamefont {Persico}}]{Compagno_1989}%
  \BibitemOpen
  \bibfield  {author} {\bibinfo {author} {\bibfnamefont {G.}~\bibnamefont
  {Compagno}}, \bibinfo {author} {\bibfnamefont {G.~M.}\ \bibnamefont {Palma}},
  \bibinfo {author} {\bibfnamefont {R.}~\bibnamefont {Passante}}, \ and\
  \bibinfo {author} {\bibfnamefont {F.}~\bibnamefont {Persico}},\ }\href
  {http://stacks.iop.org/0295-5075/9/i=3/a=005} {\bibfield  {journal} {\bibinfo
   {journal} {EPL (Europhysics Letters)}\ }\textbf {\bibinfo {volume} {9}},\
  \bibinfo {pages} {215} (\bibinfo {year} {1989})}\BibitemShut {NoStop}%
\bibitem [{\citenamefont {Compagno}\ \emph {et~al.}(1990)\citenamefont
  {Compagno}, \citenamefont {Passante},\ and\ \citenamefont
  {Persico}}]{Compagno_1990}%
  \BibitemOpen
  \bibfield  {author} {\bibinfo {author} {\bibfnamefont {G.}~\bibnamefont
  {Compagno}}, \bibinfo {author} {\bibfnamefont {R.}~\bibnamefont {Passante}},
  \ and\ \bibinfo {author} {\bibfnamefont {F.}~\bibnamefont {Persico}},\ }\href
  {\doibase 10.1080/09500349014551511} {\bibfield  {journal} {\bibinfo
  {journal} {Journal of Modern Optics}\ }\textbf {\bibinfo {volume} {37}},\
  \bibinfo {pages} {1377} (\bibinfo {year} {1990})}\BibitemShut {NoStop}%
\bibitem [{\citenamefont {Clerk}\ and\ \citenamefont {Sipe}(1998)}]{Clerk1998}%
  \BibitemOpen
  \bibfield  {author} {\bibinfo {author} {\bibfnamefont {A.~A.}\ \bibnamefont
  {Clerk}}\ and\ \bibinfo {author} {\bibfnamefont {J.~E.}\ \bibnamefont
  {Sipe}},\ }\href {\doibase 10.1023/A:1018717823725} {\bibfield  {journal}
  {\bibinfo  {journal} {Foundations of Physics}\ }\textbf {\bibinfo {volume}
  {28}},\ \bibinfo {pages} {639} (\bibinfo {year} {1998})}\BibitemShut
  {NoStop}%
\bibitem [{\citenamefont {Landulfo}(2016)}]{PhysRevD.93.104019}%
  \BibitemOpen
  \bibfield  {author} {\bibinfo {author} {\bibfnamefont {A.~G.~S.}\
  \bibnamefont {Landulfo}},\ }\href {\doibase 10.1103/PhysRevD.93.104019}
  {\bibfield  {journal} {\bibinfo  {journal} {Phys. Rev. D}\ }\textbf {\bibinfo
  {volume} {93}},\ \bibinfo {pages} {104019} (\bibinfo {year}
  {2016})}\BibitemShut {NoStop}%
\bibitem [{\citenamefont {Jonsson}\ \emph {et~al.}(2015)\citenamefont
  {Jonsson}, \citenamefont {Mart\'{\i}n-Mart\'{\i}nez},\ and\ \citenamefont
  {Kempf}}]{PhysRevLett.114.110505}%
  \BibitemOpen
  \bibfield  {author} {\bibinfo {author} {\bibfnamefont {R.~H.}\ \bibnamefont
  {Jonsson}}, \bibinfo {author} {\bibfnamefont {E.}~\bibnamefont
  {Mart\'{\i}n-Mart\'{\i}nez}}, \ and\ \bibinfo {author} {\bibfnamefont
  {A.}~\bibnamefont {Kempf}},\ }\href {\doibase 10.1103/PhysRevLett.114.110505}
  {\bibfield  {journal} {\bibinfo  {journal} {Phys. Rev. Lett.}\ }\textbf
  {\bibinfo {volume} {114}},\ \bibinfo {pages} {110505} (\bibinfo {year}
  {2015})}\BibitemShut {NoStop}%
\bibitem [{\citenamefont {Blasco}\ \emph {et~al.}(2015)\citenamefont {Blasco},
  \citenamefont {Garay}, \citenamefont {Mart\'{\i}n-Benito},\ and\
  \citenamefont {Mart\'{\i}n-Mart\'{\i}nez}}]{PhysRevLett.114.141103}%
  \BibitemOpen
  \bibfield  {author} {\bibinfo {author} {\bibfnamefont {A.}~\bibnamefont
  {Blasco}}, \bibinfo {author} {\bibfnamefont {L.~J.}\ \bibnamefont {Garay}},
  \bibinfo {author} {\bibfnamefont {M.}~\bibnamefont {Mart\'{\i}n-Benito}}, \
  and\ \bibinfo {author} {\bibfnamefont {E.}~\bibnamefont
  {Mart\'{\i}n-Mart\'{\i}nez}},\ }\href {\doibase
  10.1103/PhysRevLett.114.141103} {\bibfield  {journal} {\bibinfo  {journal}
  {Phys. Rev. Lett.}\ }\textbf {\bibinfo {volume} {114}},\ \bibinfo {pages}
  {141103} (\bibinfo {year} {2015})}\BibitemShut {NoStop}%
\bibitem [{\citenamefont {Jonsson}\ \emph {et~al.}(2018)\citenamefont
  {Jonsson}, \citenamefont {Ried}, \citenamefont
  {Mart{\'{\i}}n-Mart{\'{\i}}nez},\ and\ \citenamefont {Kempf}}]{Jonsson_2018}%
  \BibitemOpen
  \bibfield  {author} {\bibinfo {author} {\bibfnamefont {R.~H.}\ \bibnamefont
  {Jonsson}}, \bibinfo {author} {\bibfnamefont {K.}~\bibnamefont {Ried}},
  \bibinfo {author} {\bibfnamefont {E.}~\bibnamefont
  {Mart{\'{\i}}n-Mart{\'{\i}}nez}}, \ and\ \bibinfo {author} {\bibfnamefont
  {A.}~\bibnamefont {Kempf}},\ }\href {\doibase 10.1088/1751-8121/aae78a}
  {\bibfield  {journal} {\bibinfo  {journal} {Journal of Physics A:
  Mathematical and Theoretical}\ }\textbf {\bibinfo {volume} {51}},\ \bibinfo
  {pages} {485301} (\bibinfo {year} {2018})}\BibitemShut {NoStop}%
\bibitem [{\citenamefont {Hotta}(2008)}]{Hotta08}%
  \BibitemOpen
  \bibfield  {author} {\bibinfo {author} {\bibfnamefont {M.}~\bibnamefont
  {Hotta}},\ }\href {\doibase 10.1103/PhysRevD.78.045006} {\bibfield  {journal}
  {\bibinfo  {journal} {Phys. Rev. D}\ }\textbf {\bibinfo {volume} {78}},\
  \bibinfo {pages} {045006} (\bibinfo {year} {2008})}\BibitemShut {NoStop}%
\bibitem [{\citenamefont {Pozas-Kerstjens}\ \emph {et~al.}(2017)\citenamefont
  {Pozas-Kerstjens}, \citenamefont {Louko},\ and\ \citenamefont
  {Mart\'{\i}n-Mart\'{\i}nez}}]{PhysRevD.95.105009}%
  \BibitemOpen
  \bibfield  {author} {\bibinfo {author} {\bibfnamefont {A.}~\bibnamefont
  {Pozas-Kerstjens}}, \bibinfo {author} {\bibfnamefont {J.}~\bibnamefont
  {Louko}}, \ and\ \bibinfo {author} {\bibfnamefont {E.}~\bibnamefont
  {Mart\'{\i}n-Mart\'{\i}nez}},\ }\href {\doibase 10.1103/PhysRevD.95.105009}
  {\bibfield  {journal} {\bibinfo  {journal} {Phys. Rev. D}\ }\textbf {\bibinfo
  {volume} {95}},\ \bibinfo {pages} {105009} (\bibinfo {year}
  {2017})}\BibitemShut {NoStop}%
\bibitem [{\citenamefont {McLenaghan}(1974)}]{McLenaghan1974}%
  \BibitemOpen
  \bibfield  {author} {\bibinfo {author} {\bibfnamefont {R.~G.}\ \bibnamefont
  {McLenaghan}},\ }\href {http://eudml.org/doc/75801} {\bibfield  {journal}
  {\bibinfo  {journal} {Annales de l'I.H.P. Physique théorique}\ }\textbf
  {\bibinfo {volume} {20}},\ \bibinfo {pages} {153} (\bibinfo {year}
  {1974})}\BibitemShut {NoStop}%
\bibitem [{\citenamefont {Czapor}\ and\ \citenamefont
  {McLenaghan}(2008)}]{McLenaghan2008}%
  \BibitemOpen
  \bibfield  {author} {\bibinfo {author} {\bibfnamefont {S.}~\bibnamefont
  {Czapor}}\ and\ \bibinfo {author} {\bibfnamefont {R.~G.}\ \bibnamefont
  {McLenaghan}},\ }\href
  {http://www.actaphys.uj.edu.pl/findarticle?series=Sup&vol=1&page=55}
  {\bibfield  {journal} {\bibinfo  {journal} {Acta. Phys. Pol. B Proc. Suppl.}\
  }\textbf {\bibinfo {volume} {1}},\ \bibinfo {pages} {55} (\bibinfo {year}
  {2008})}\BibitemShut {NoStop}%
\bibitem [{\citenamefont {Forn-D{\'i}az}\ \emph {et~al.}(2016)\citenamefont
  {Forn-D{\'i}az}, \citenamefont {Garc{\'i}a-Ripoll}, \citenamefont
  {Peropadre}, \citenamefont {Orgiazzi}, \citenamefont {Yurtalan},
  \citenamefont {Belyansky}, \citenamefont {Wilson},\ and\ \citenamefont
  {Lupascu}}]{Forn-Diaz2016}%
  \BibitemOpen
  \bibfield  {author} {\bibinfo {author} {\bibfnamefont {P.}~\bibnamefont
  {Forn-D{\'i}az}}, \bibinfo {author} {\bibfnamefont {J.~J.}\ \bibnamefont
  {Garc{\'i}a-Ripoll}}, \bibinfo {author} {\bibfnamefont {B.}~\bibnamefont
  {Peropadre}}, \bibinfo {author} {\bibfnamefont {J.-L.}\ \bibnamefont
  {Orgiazzi}}, \bibinfo {author} {\bibfnamefont {M.~A.}\ \bibnamefont
  {Yurtalan}}, \bibinfo {author} {\bibfnamefont {R.}~\bibnamefont {Belyansky}},
  \bibinfo {author} {\bibfnamefont {C.~M.}\ \bibnamefont {Wilson}}, \ and\
  \bibinfo {author} {\bibfnamefont {A.}~\bibnamefont {Lupascu}},\ }\href
  {https://doi.org/10.1038/nphys3905} {\bibfield  {journal} {\bibinfo
  {journal} {Nature Physics}\ }\textbf {\bibinfo {volume} {13}},\ \bibinfo
  {pages} {39 EP } (\bibinfo {year} {2016})}\BibitemShut {NoStop}%
\bibitem [{\citenamefont {Valentini}(1991)}]{VALENTINI1991321}%
  \BibitemOpen
  \bibfield  {author} {\bibinfo {author} {\bibfnamefont {A.}~\bibnamefont
  {Valentini}},\ }\href {\doibase 10.1016/0375-9601(91)90952-5} {\bibfield
  {journal} {\bibinfo  {journal} {Physics Letters A}\ }\textbf {\bibinfo
  {volume} {153}},\ \bibinfo {pages} {321 } (\bibinfo {year}
  {1991})}\BibitemShut {NoStop}%
\bibitem [{\citenamefont {Reznik}\ \emph {et~al.}(2005)\citenamefont {Reznik},
  \citenamefont {Retzker},\ and\ \citenamefont {Silman}}]{Reznik}%
  \BibitemOpen
  \bibfield  {author} {\bibinfo {author} {\bibfnamefont {B.}~\bibnamefont
  {Reznik}}, \bibinfo {author} {\bibfnamefont {A.}~\bibnamefont {Retzker}}, \
  and\ \bibinfo {author} {\bibfnamefont {J.}~\bibnamefont {Silman}},\ }\href
  {\doibase 10.1103/PhysRevA.71.042104} {\bibfield  {journal} {\bibinfo
  {journal} {Phys. Rev. A}\ }\textbf {\bibinfo {volume} {71}},\ \bibinfo
  {pages} {042104} (\bibinfo {year} {2005})}\BibitemShut {NoStop}%
\bibitem [{\citenamefont {Pozas-Kerstjens}\ and\ \citenamefont
  {Mart\'{\i}n-Mart\'{\i}nez}(2015)}]{Pozas2015}%
  \BibitemOpen
  \bibfield  {author} {\bibinfo {author} {\bibfnamefont {A.}~\bibnamefont
  {Pozas-Kerstjens}}\ and\ \bibinfo {author} {\bibfnamefont {E.}~\bibnamefont
  {Mart\'{\i}n-Mart\'{\i}nez}},\ }\href {\doibase 10.1103/PhysRevD.92.064042}
  {\bibfield  {journal} {\bibinfo  {journal} {Phys. Rev. D}\ }\textbf {\bibinfo
  {volume} {92}},\ \bibinfo {pages} {064042} (\bibinfo {year}
  {2015})}\BibitemShut {NoStop}%
\end{thebibliography}%

\end{document}